\documentclass[aps,pre,twoside,letterpaper,showpacs]{revtex4}  
\usepackage{amssymb} 
\usepackage{amsmath}
\usepackage{amsfonts}
\usepackage{graphicx}
\usepackage[T1]{fontenc}

\newcommand{\vsr}{v_{\scriptscriptstyle SR}}
\newcommand{\Vsr}{V_{\scriptscriptstyle SR}}
\newcommand{\Vself}{V_{\rm self}}
\newcommand{\Vimsc}{V_{\rm im}^{ {\scriptscriptstyle B}\,\rm sc}}
\newcommand{\dw}{\Delta_{\rm el}}

\newcommand{\xid}{\xi_{\scriptscriptstyle D}}
\newcommand{\ew}{\epsilon_{\scriptscriptstyle W}}

\newcommand{\kbun}{\delta\kappa_{\rm \scriptscriptstyle B}^{\scriptscriptstyle(1)}}
\newcommand{\Zeffba}{ Z_{\as}^{\rm eff\,{\scriptscriptstyle B} }}
\newcommand{\Zeffbap}{ Z_{\asp}^{\rm eff\, {\scriptscriptstyle B} }}
\newcommand{\Zeffwa}{ Z_{\as}^{\rm eff\,{\scriptscriptstyle W} }}
\newcommand{\soma}{\sum_{ \alpha}}
\newcommand{\vecr}{ {\bf r}}
\newcommand{\as}{{\scriptscriptstyle \alpha}}
\newcommand{\asp}{{\scriptscriptstyle \alpha'}}
\newcommand{\aspp}{{\scriptscriptstyle \alpha''}}
\newcommand{\gs}{{\scriptscriptstyle \gamma}}
\newcommand{\gsp}{{\scriptscriptstyle \gamma'}}
\newcommand{\gspu}{{\scriptscriptstyle \gamma'_1}}
\newcommand{\gspd}{{\scriptscriptstyle \gamma'_2}}
\newcommand{\roa}{\rho_{\scriptscriptstyle \alpha}}
\newcommand{\roap}{\rho_{\scriptscriptstyle \alpha'}}
\newcommand{\roab}{\rho_{\scriptscriptstyle \alpha}^{\rm\scriptscriptstyle B}}
\newcommand{\rogb}{\rho_{\scriptscriptstyle \gamma}^{\rm\scriptscriptstyle B}}
\newcommand{\rogpb}{\rho_{\scriptscriptstyle \gamma'}^{\rm \scriptscriptstyle B}}
\newcommand{\roapb}{\rho_{\scriptscriptstyle \alpha'}^{\rm \scriptscriptstyle B}}
\newcommand{\Fcc}{F^{\rm c  c}}
\newcommand{\Fr}{F_{\scriptscriptstyle{\rm R}}}
\newcommand{\Frt}{F_{\scriptscriptstyle{\rm RT}}}
\newcommand{\Ir}{I}
\newcommand{\Irbar}{\overline{I}}

\newcommand{\kbar}{{\overline{\kappa}}}
\newcommand{\kd}{\kappa_{\rm\scriptscriptstyle D}}
\newcommand{\vy}{{\bf y}}
\newcommand{\vq}{{\bf q}}
\newcommand{\vk}{{\bf k}}
\newcommand{\vl}{{\bf l}}
\newcommand{\modq}{\vert{\bf q}\vert}
\newcommand{\phit}{{\widetilde{\phi}}}
\newcommand{\phid}{{\phi}_{\rm\scriptscriptstyle D}}
\newcommand{\phidd}{{\phi}_{\rm\scriptscriptstyle D}^{2}}
\newcommand{\xt}{{\widetilde{x}}}
\newcommand{\bt}{{\widetilde{b}}}
\newcommand{\lt}{{\widetilde{l}}}
\newcommand{\rac}{\sqrt{1+\vq^2}}
\newcommand{\phitzero}{{\widetilde{\phi}}^{(0)}}
\newcommand{\cO}{{\cal O}}

\newcommand{\eps}{\varepsilon}
\newcommand{\Lb}{{\overline{L}}}

\newcommand{\hcc}{h^{\rm cc}}
\newcommand{\hct}{h^{\rm c-}}
\newcommand{\htc}{h^{\rm -c}}
\newcommand{\htt}{h^{\rm --}}
\newcommand{\hccb}{h^{\rm cc\, {\scriptscriptstyle B}}}
\newcommand{\hctb}{h^{\rm c-\, {\scriptscriptstyle B}}}
\newcommand{\htcb}{h^{\rm -c\, {\scriptscriptstyle B}}}
\newcommand{\httb}{h^{\rm --\, {\scriptscriptstyle B}}}

\newcommand{\kappab}{\kappa_{\rm {\scriptscriptstyle B}}}
\newcommand{\hcczero}{h^{{\rm cc}\,(0)}}
\newcommand{\hctzero}{h^{{\rm c-}\,(0)}}
\newcommand{\htczero}{h^{{\rm -c}\,(0)}}
\newcommand{\httzero}{h^{{\rm --}\,(0)}}
\newcommand{\Mb}{\overline{M}}
\newcommand{\ract}{\sqrt{t^2-1}}

\newcommand{\somg}{\sum_{\gamma}}
\newcommand{\somd}{\Sigma_2}
\newcommand{\somt}{\Sigma_3}
\newcommand{\somq}{\Sigma_4}

\newcommand{\Ei}{{\rm Ei}}

\begin{document}
\title{Charge renormalization and other exact coupling corrections in the dipolar effective interaction 
 in an electrolyte near
a dielectric  wall}

\author{J.-N. Aqua}
\affiliation{Laboratoire de Physique\\
\'Ecole Normale Sup\'erieure de Lyon
\\46 all\'ee d'Italie, 69364 Lyon, \textsc{France}}
\altaffiliation{Laboratoire associ\'e au Centre National de la Recherche Scientifique UMR 5672}
\altaffiliation{Present address : Institute for Physical Science and Technology, University of Maryland, College Park, Maryland 20910, \textsc{USA}}

\author{F. Cornu}
\affiliation{Laboratoire de Physique Th\'eorique \\
Universit\'e Paris-Sud, B\^{a}timent 210\\
91405 Orsay, \textsc{France}}
\altaffiliation{Laboratoire associ\'e au Centre National de la Recherche Scientifique - UMR 8627}

\date{\today}

\begin{abstract}

The aim  of the paper is to study the renormalizations of the  charge and of the screening length that appear in the large-distance behavior of
the effective pairwise interaction $w_{\as\asp}$  between two charges $e_{\as}$ and 
$e_{\asp}$ in a dilute electrolyte solution,  both along a dielectric wall and in the bulk. The electrolyte is described by the primitive model in the framework of classical statistical mechanics and the electrostatic response of the wall is characterized by its dielectric constant. In Ref.\cite{AquaCor03I} a graphic reorganization of resummed Mayer diagrammatics has been devised in order to exhibit the general structure of the  $1/y^3$ leading  tail of $w_{\as\asp}(x,x',y)$ for two charges located at distances $x$ and $x'$ from the wall and separated by a distance $y$ along the wall. When all species have the same closest approach distance $b$ to the wall, the coefficient of the  $1/y^3$ tail is the product $D_{\as}(x)D_{\asp}(x')$ of two effective dipoles. Here we use the same graphic reorganization in order to systematically investigate  the   exponential large-distance behavior of $w_{\as\asp}$ in the bulk. (We show that the reorganization also enables one to derive the basic screening rules in both cases.) Then, in a  regime of high dilution and weak coupling,  
 the exact analytical corrections to the leading tail of $w_{\as\asp}$, both in the bulk or along the wall, are calculated
at  first order in  the coupling parameter $\eps$ and in the limit where $b$ becomes negligible with respect to the Debye screening length. ($\eps$ is proportional to the so-called plasma parameter.) The structure of corrections to the terms of order $\eps$ is exhibited, and the scaling regime for the validity of the Debye limit is specified. In the vicinity of the wall, 
 we use the density profiles calculated in Ref.\cite{AquaCor01I} up to order $\eps$ and the method devised in  Ref.\cite{AquaCor01II} for the determination of the  corresponding correction in the auxiliary screened potential, which also appears in the linear-response theory. The first coupling correction to the effective dipole  $D_{\as}(x)$ is a function (not a mere exponential decay) determined by the nonuniformity of the density profiles  as well as by three- and four-body screened interactions in $w_{\as\asp}$.
Though the effective screening length (beyond the Debye value) in the direction perpendicular to the wall
is the same as in the bulk, the bare solvated charges are not renormalized by the
same quantity as in the bulk,
because of combined steric and electrostatic effects induced by the wall. 

\pacs{05.20.Jj, 05.70.Np.}

\end{abstract}  

\maketitle

{\it Corresponding author }: cornu@th.u-psud.fr.

\section{Introduction}

\subsection{Issue at stake}

The paper is devoted to the large-distance behavior of the pairwise effective interaction between two charges in
an electrolyte solution, which is confined to the region $x>0$ by a plane impenetrable 
 dielectric wall. The electrolyte solution is described by the usual
 {\it primitive model} \cite{McQuarrie} with $n_s$ species of charges which interact  via the Coulomb interaction.
 Every charged particle of species $\alpha$  is
represented as a hard sphere -- with diameter $\sigma_{\as}$ -- where the 
net  bare solvated charge $e_\as\equiv Z_\as e$ is concentrated  at the center of
the sphere. ($e$ is the abolute value of
the electron charge and $Z_\as$ may be positive or negative.) 
The solvent  is handled with as a continuous medium of
uniform dielectric constant $\epsilon_{\rm solv}$. 
The wall matter is characterized by a dielectric constant $\ew\not=\epsilon_{\rm solv}$, and the latter difference results into an electrostatic response of the wall to the moving charges in the electrolyte. 
Moreover, the excluded-volume sphere of every particle is assumed 
to be
made of a material with the same dielectric constant as that of the solvent. 
(Therefore 
$\epsilon=\epsilon_{\rm solv}$ when $x>0$ and 
$\epsilon=\epsilon_{\rm \scriptscriptstyle W}$ when $x<0$). In the framework of statistical mechanics, the effective pairwise interaction $w_{\as\asp}(\vecr,\vecr')$ between two
charges $e_\as$ and $e_\asp$ located at positions $\vecr$ and $\vecr'$, 
respectively, is defined 
from the pair correlation function $ h_{\as\asp}$ by (see e.g. Ref.\cite{HansenMcDonald})
\begin{equation}
\label{defw}
1+h_{\as\asp}\equiv \exp(-\beta w_{\as\asp}),
\end{equation}
where $\beta=1/k_{\scriptscriptstyle B} T$ is the inverse temperature, in which 
$k_{\scriptscriptstyle B} $ is   Boltzmann constant and $T$ is the 
absolute temperature.
($w_{\as\asp}$ is  also called potential of mean force, while 
$ h_{\as\asp}$  is known as the Ursell function.) 
In the vicinity of the wall, symmetries enforce that 
$w_{\as\asp}(\vecr,\vecr')=w_{\as\asp}(x,x',y)$, where  $x$ 
and $x'$ are the distances of $\vecr$ and $\vecr'$ from the wall and $y$ is
the norm of the projection $\vy$ of $\vecr-\vecr'$ onto the wall plane.
Along the wall, contrary to the bulk case,
$w_{\as\asp}(x,x',y)$ does not decay exponentially fast:  its leading behavior at large distances $y$ takes a dipolar form
$f_{\as\asp}(x,x')/y^3$, as a result of 
the deformation of screening clouds enforced
by the presence of the wall (see Ref\cite{revueMartin} for a review or e.g. Ref.\cite{AquaCor03I}).

An electrolyte solution can be considered as a dilute charge fluid
where  the closest approach distance between the center of a charge  with species $\alpha$ and the dielectric wall takes  the same value $b$ for all species. The reason is that the differences in the various ion diameters are negligible with respect to  all other characteristic lengths.  ($b_{\as}=b$ for all $\alpha$'s, whether $b_{\as}$ is only determined by the radius of the excluded-volume sphere of species $\alpha$ or  $b_{\as}$ involves some other more complicated microscopic mechanism for the short-distance repulsion from the wall. For instance, a layer of water molecules, with a thickness of molecular dimensions, may lie between the wall and the electrolyte solution, as it has been suggested for instance for another situation, the mercury-aqueous solution interface \cite{MottParsons&62}.) 
As a consequence, as shown in Ref.\cite{AquaCor03I}, called Paper I in the following, the coefficient $f_{\as\asp}(x,x')$ of the $1/y^3$ tail of $w_{\as\asp}(x,x',y)$ is a product of effective dipoles $D_{\as}(x)$ and $D_{\asp}(x')$,
\begin{equation}
\label{has}
w_{\as\asp}(x,x',y) \underset{y\rightarrow +\infty}{\sim}
  \frac{D_{\as}(x)D_{\asp}(x')}{y^3}.
\end{equation}
(Therefore the tail of $w_{\as\as}(x,x',y)$ between two particles of the same species $\alpha$ is repulsive when $x=x'$, as it is the case for identical point dipoles with the  same direction.)

The general result \eqref{has} arises from a  property about the screened potential $\phi$ defined as follows.
$(\delta q \delta q' /\epsilon_{\rm solv})\phi$
 is the immersion free
energy  between two {\it infinitesimal
external point }
charges $\delta q$ and $\delta q'$ calculated in the framework of the linear-response theory  as if the radii of  the excluded-volume spheres of the fluid charges were equal to zero \cite{AquaCor99}. (The effect of hard cores is briefly discussed after \eqref{eqintphi}.)
As shown in Paper I, when all particles have the same closest approach distance $b$ to the wall,
\begin{equation}
\label{tailphi}
\phi(x,x',y)
\underset{y\rightarrow +\infty}{\sim}
\frac{{\overline D}_{\phi}(x){\overline D}_{\phi}(x')}{y^3}.
\end{equation}
(In \eqref{tailphi}
${\overline D}_{\phi}(x)$  vanishes  for $x<b$.)
In the following, a quantity that is independent of charge species $\alpha$ is denoted by an overlined letter when it is analogous to another one that depends on $\alpha$, as it is the case for $D_{\as}(x)$ and ${\overline D}_{\phi}(x)$. 
Since $\phi$ obeys an  ``{inhomogeneous}'' Debye equation where the effective screening length depends on the distance $x$ from the  wall through the density profiles,
${\overline D}_{\phi}(x)$ has the same sign at any distance $x$ from the wall, contrarily to the effective dipole $D_\as(x)$, the sign of which may {\it a priori} vary with the distance $x$. Thus, the $1/y^3$ tail of $\phi(x,x',y)$ is  repulsive at all distances $x$ and $x'$ from the wall. Moreover, 
the $x$-dependent screening length
  tends to the  Debye length $\xid$ at large distances, and ${\overline D}_{\phi}(x)$ can be rewritten, for $x>b$, as
\begin{equation}
\label{structDphi}
{\overline D}_{\phi}(x)=-
\sqrt{\frac{2\ew}{\epsilon_{\rm solv}}}
\frac{e^{-\kd(x-b)}}{\kd}\left[1+C_{\phi}+
{\overline G}^{\rm exp}_{\phi}(x)\right],
\end{equation} 
where 
${\overline G}^{\rm exp}_{\phi}(x)$ tends to zero exponentially fast over a scale of order $1/\kd$. 
In Gauss units, the Debye length $\xid$   reads
\begin{equation}
  \label{defkd}
 \xid^{-1}\equiv \kd =\sqrt{ \frac{ 4 \pi \beta e^2}{\epsilon_{\rm solv}}\soma 
  Z_\as^2\roab},
\end{equation} 
 where $\roab $ is the bulk density of species $\alpha$. 
$C_{\phi}$ is a constant which vanishes, as well as 
${\overline G}^{\rm exp}_{\phi}(x)$, in the infinite-dilution and vanishing-coupling limit considered hereafter. The global minus sign in \eqref{structDphi}  has been introduced, because, in the latter limit and
in the case of a plain wall ($\ew=\epsilon_{\rm solv}$), ${\overline D}_{\phi}(x) $ is expected to have the same sign as the dipole $d(x)$ carried by the set made of a positive unit charge and its screening cloud repelled from the wall.
 The sign of $1+C_{\phi}$ depends on the temperature, on the composition of the electrolyte, on the value of the closest approach distance $b$ to the wall, and on the dielectric constants $\ew$ and $\epsilon_{\rm solv}$.

In an electrolyte solution, the $Z_\as$'s of all species $\alpha$'s are of unit order and the diameters $\sigma_\as$'s of excluded-volume spheres  also have the same typical value, denoted by  $\sigma$.  Moreover,  all densities $\roab$'s are of the same magnitude order. Thus, if the solution is highly diluted, the
Coulomb coupling between charges of any species separated by the mean interparticle distance $a$ is weak: the condition of low densities,
$\sigma/a\ll 1$,  implies that
$(\beta e^2/(\epsilon_{\rm solv}a))\ll 1$, if the temperature is  high enough for $\beta e^2/(\epsilon_{\rm solv}\sigma)$ to be far smaller than 1 or of unit order.  Detailed  scaling regimes are given in Section \ref{mainresults}.
In the corresponding limit, denoted by the superscript $(0)$ hereafter, where the fluid is infinitely diluted and extremely weakly-coupled, 
the large-distance behavior 
$w_{\as\asp}^{\rm as}(\vecr,\vecr')$ of the effective pairwise interaction $w_{\as\asp}(\vecr,\vecr')$ is the same  as if the charges $e_\as$ and $e_\asp$ were infinitesimal external point charges embedded in the infinitely-diluted and vanishingly-coupled fluid, 
\begin{equation}
\label{w0phi0}
w_{\as\asp}^{\rm as\, (0)}= \frac{e^2}{\epsilon_{\rm solv}}
Z_{\as} Z_{\asp} \phi^{\rm as\,(0)}.
\end{equation} 
Moreover, in this limit,
the density profiles are uniform at leading order and ${\overline D}^{(0)}_{\phi}(x)$ is given by \eqref{structDphi} where the
constant $C_{\phi}$ and the function ${\overline G}^{\rm exp}_{\phi}(x)$  vanish: $C_{\phi}^{(0)}=0$ and 
${\overline G}^{\rm exp\,(0)}_{\phi}(x)=0$
\cite{Russes,Janco82I}. Then, by virtue of \eqref{has} and \eqref{w0phi0}, 
\begin{equation}
\label{valueDzero}
D_{\as}^{(0)}(x)=\frac{ e}{\sqrt{\epsilon_{\rm solv}}}\,
  Z_\as {\overline D}^{(0)}_{\phi}(x)\quad\mbox{with}\quad
{\overline D}^{(0)}_{\phi}(x)= -
\sqrt{\frac{2\ew}{\epsilon_{\rm solv}}}
\frac{e^{-\kd(x-b)}}{\kd}.
\end{equation}

 As long as the dilution is high  enough,  the large-distance behavior 
$w_{\as\asp}^{\rm as}(\vecr,\vecr')$ of the effective pairwise interaction $w_{\as\asp}(\vecr,\vecr')$ is expected to have the same functional form as its expression
$w_{\as\asp}^{{\rm as}\,(0)}(\vecr,\vecr')$ in  the infinite-dilution and vanishing-coupling  limit. This assuption is supported by two reasons. First,
since densities are low, the functional form of the large-distance behavior
$w_{\as\asp}^{\rm as}(\vecr,\vecr')$ is ruled by the effect of long-range Coulomb interactions, whereas short-ranged hard-core repulsions are only involved in the values of the coefficients of this leading tail. Second, the leading Coulomb effects are due to the  large-distance nonintegrability of Coulomb interaction, and
the leading-order contribution from  any integral involving the Boltzman factor of either  the effective interaction $w_{\as\asp}$ or the bare Coulomb  interaction $e_{\as}e_{\asp} v$ is obtained by linearizing the latter exponential factors. This is the procedure that underlies the Debye-H\"uckel approximation  for bulk correlations, which was initially derived as a linearized  Poisson-Boltzmann theory, where $w_{\as\asp}$ is dealt with in a linear-response framework  as if charges $e_{\as}$ and $e_{\asp}$ were infinitesimal external charges \cite{Debye,McQuarrie}.
(In the Mayer diagrammatic approach of Debye-H\"uckel theory,  the linearized Boltzmann factor is that of the bare potential and  one must also resum the infinite series of the most divergent integrals that arise from this linearization (see e.g. Ref.\cite{HansenMcDonald}).) The second reason amounts to state that leading Coulomb effects are properly described in a  linearized mean-field scheme.

In other words, as long as the dilution and the temperature are  high enough, in $w_{\as\asp}^{\rm as}$ the many-body effects beyond the  linearized mean-field structure only result into the renormalization of charges and of the screening length with respect to their values in the infinite-dilution and vanishing-coupling limit, namely  with respect to the bare solvated charges $Z_\as$'s and the Debye screening length $\xid$. For instance,
in the bulk,  
   $w_{\as\asp}^{\rm {\scriptscriptstyle B}\, as}(\vecr,\vecr')$
behaves as the leading-order function
 $w_{\as\asp}^{\rm{ \scriptscriptstyle B} \, as\,(0)}(\vecr,\vecr')$
given by the Debye-H\"uckel theory for point charges, 
but  bare
charges $Z_{\as}$'s and the Debye length $\kd^{-1}$ are replaced by 
effective charges $Z_{\as}^{\rm {\scriptscriptstyle B}\,eff}$'s and the 
 screening length
$\kappab^{-1}$, respectively:
\begin{equation}
\label{defwbulkeff}
w_{\as\asp}^{\rm {\scriptscriptstyle B}}(|\vecr-\vecr'|)
\underset{|\vecr-\vecr'|\rightarrow +\infty}{\sim}  
\frac{e^2}{\epsilon_{\rm solv}}\,  \Zeffba\Zeffbap   
\frac{ e^{- \kappab|\vecr-\vecr'|}}{|\vecr-\vecr'|},
\end{equation}
where
\begin{equation}
\label{valuewzerobulk}
Z_{\as}^{\rm eff\,{\scriptscriptstyle B}\, (0) }=Z_\as 
\quad\mbox{\rm and}\quad  \kappab^{(0)}=\kd.
\end{equation}
(In the following, the superscript ${\rm B}$ signals all bulk quantities.) In the present paper we  show that 
the effective dipole $D_{\as}(x)$ in the large-distance pairwise interaction along a dielectric wall  takes the form
\begin{equation}
\label{comparaison}
D_{\as}(x)=-
\sqrt{\frac{2\ew}{\epsilon_{\rm solv}}}
\frac{ e}{\sqrt{\epsilon_{\rm solv}}}\,
 \Zeffwa\,
\frac{e^{-\kappa(x-b)}}{\kappa}
\left[1+  G^{\rm exp}_{\as}(x)\right],
\end{equation}
where $G^{\rm exp}_{\as}(x)$ is an
exponentially decaying function which tends to $0$ when $x$ goes to infinity. 
At distances from the wall larger than a few screening lengths, $D_{\as}(x)$ takes the same functional form as $D_{\as}^{(0)}(x)$, and 
many-body effects 
 reduce  to the introduction of effective charges $\Zeffwa$ and of a  screening length $\kappa^{-1}$.

 On the other hand,
in the bulk many-body effects upon effective charges and the screening length arise only from pair interactions between ions, whereas along the wall they involve also the electrostatic potential and the geometric constraint created by the wall. In order to investigate these differences, we  also
determine the bulk pairwise interaction
$w_{\as\asp}^{\rm {\scriptscriptstyle B} }$ up to the same order
in the coupling parameter as for the pairwise interaction along the wall.

\subsection{Main results}
\label{mainresults}

In the present paper  we determine
$w_{\as\asp}^{\rm {\scriptscriptstyle B}}(|\vecr-\vecr'|)$,
${\overline D}_{\phi}(x)$, and $D_{\as}(x)$ up to first-order in the dimensionless coupling  parameter
\begin{equation}
\label{defeps}
 \eps\equiv \frac{1}{2} \kd \frac{\beta e^2}{\epsilon_{\rm solv}} \ll 1.
\end{equation}
in regimes where $\eps\ll1$, $(\sigma/a)^3\ll1$ and $\kd b\ll1$.
$\eps\propto \left(\beta e^2/\epsilon_{\rm solv}a\right)^{3/2}$ and \eqref{defeps} implies that $\beta e^2/\epsilon_{\rm solv}\ll a\ll \xid$.
(Up to the factor $1/2$,  $\eps$ coincides with the so-called plasma parameter of an
electron gas.) As shown in Section \ref{section4}, contributions from steric effects involving $\sigma/a$  are corrections of higher order with respect to the terms of order $\eps$ in
some scaling regimes of high dilution  where $(\sigma/a)^3\ll \eps$. Moreover, as shown in Section \ref{titato}, the first corrections involving $\kd b$ appear only at order $\eps\ln(\kd b)$. In the first scaling regime, the  density vanishes while the temperature is fixed; then  $\beta e^2/(\epsilon_{\rm solv}\sigma)$ and 
$\beta e^2/(\epsilon_{\rm solv}b)$ are fixed, namely
\begin{equation}
\label{case1bis}
\left(\frac{\sigma}{a}\right)^3\propto \eps^{2}\quad \textrm{and}\quad \kd b\propto \eps\ll1\quad \mbox{\rm regime (1)}.
\end{equation}
In the second case, the density vanishes while the temperature goes to infinity, but not too  fast in order to ensure that 
$(\sigma/a)^3\ll \eps$; then both
$\beta e^2/(\epsilon_{\rm solv}\sigma)$ and 
$\beta e^2/(\epsilon_{\rm solv}b)$ vanish. These conditions can be summarized in the following way,
\begin{equation}
\label{subcase2bis}
\eps^2\ll \left(\frac{\sigma}{a}\right)^3\ll \eps
\ll\kd b \ll 1\quad \mbox{\rm regime (2)}.
\end{equation}
The expressions at leading order in $\eps$  and $\kd b$ in regime (2) can be obtained from those derived in regime (1) for a fixed ratio $(\eps/\kd b)\propto \beta e^2/(\epsilon_{\rm solv}b)$ by taking the limit where $\beta e^2/(\epsilon_{\rm solv}b)$  goes to zero while $\kd b$ is kept fixed.
We notice that, when the solvent is water, the Bjerrum length
$\beta e^2/\epsilon_{\rm solv}$ at room temperature is about $7$ Angstr\oe m and, for concentrations  around $10^{-4}$ mol/liter, $\eps$ is of order $10^{-2}$ and $(\sigma/a)^3$ is of order $\eps^2$ for $\sigma\sim 5$ Angstr\oe m. 
 For the sake of conciseness,  both regimes  (1) and (2) will be referred to as the ``{weak-coupling }'' regime and we shall speak  only in terms of  $\eps$-expansions.

 Our  exact analytical calculations are  performed in the framework of resummed Mayer diagrammatics introduced in Paper I. 
 For the inverse screening length in the bulk we retrieve 
\cite{MitchellNinh68}-\cite{KjelMitchell94} that, up to order $\eps$,
\begin{equation}
\label{valuekappab}
\kappab=\kd \left[1+\eps 
\left(\frac{\Sigma_3}{\Sigma_2}\right)^2 \frac{\ln 3}{4}
+o(\eps)\right]
\end{equation}
where
\begin{equation}
\label{defSigma}
\Sigma_m\equiv \sum_{\alpha=1}^{n_s}  \roab Z_\as^m.
\end{equation}
The leading corrections involved in the notation $o(\eps)$ are given in \eqref{orders}. As announced above, only screening effects of 
the non-integrable long-range Coulomb
interaction  are involved up to order $\eps$; the diameters $\sigma_{\as}$'s of
charges appear only in higher-order terms. This property holds in the bulk
as well as in the vicinity of a wall (see e.g. Ref.\cite{AquaCor01II}). More generally, the specific form of the short-distance steric repulsion between charges does not appear in  the leading correction of order $\eps$.
The correction of order $\eps$ in the bulk screening length $\kappab^{-1}$
vanishes in a charge symmetric electrolyte, where 
species with charge $-Z_{\as}e$ has the same density as species with charge $Z_{\as}e$ ($\Sigma_3=0$). 
If the fluid is not charge symmetric,
the screening length $\kappab^{-1}$ decreases when the coupling strength increases.

Our main results are the following. First, we find
  that
\begin{equation}
 \label{valueZeff}
\Zeffba=Z_{\as} \left\{ 1  
\rule{0mm}{6mm}
+ \eps \left[ Z_\as \frac{\Sigma_3}{\Sigma_2}\frac{\ln 3}{2}
+\left(\frac{\Sigma_3}{\Sigma_2}\right)^2 
\left(\frac{1}{6}-\frac{\ln 3}{8}\right)
\right]
+o(\eps)\right\}
\end{equation} 
where $(1/6)-(1/8)\ln3>0$. In the case of a one-component plasma, formula \eqref{valueZeff} is reduced to that found in Ref. \cite{MitchellNinh68} by diagrammatic techniques. (The expression given for a multi-component electrolyte in Ref.\cite{KjelMitchell94} corresponds to another  definition of the effective charge and does not coincide with our expression \eqref{valueZeff}.)  As in the case of the screening length, there is no correction at order $\eps$ if the composition of the electolyte is charge symmetric. According to the diagrammatic origin of this correction, the contribution to $\Zeffba$ from a screened interaction via one intermediate charge has the  sign of $Z_{\as}\Sigma_3$ whereas the contribution to $\Zeffba$ from a screened interaction via two intermediate charges always increases the effective charge. We notice that the existence of the non-linear term $Z_{\as}^2$ in $\Zeffba$ implies that $w_{\as\asp}^{\rm as}$ cannot be written as $Z_{\as}\psi_{\asp}$ where $\psi_{\asp}$ would be the total electrostatic potential created at $\vecr$ by the charge $Z_{\asp}e$ at $\vecr'$ and its screening cloud in the electrolyte. $\psi_{\asp}$ does not exist beyond the framework of linear-response theory.

Second, as expected,
the screening length in the direction perpendicular to the wall proves to be  the same as in the bulk, 
 at least up to first order in $\eps$. 
Besides, the renormalized charge $\Zeffwa$ defined in \eqref{comparaison} and the renormalized charge $\Zeffba$ in the bulk do not coincide.  However, up to order $\eps$, their ratio is independent of the species $\alpha$, 
\begin{equation}
\label{rapZ}
\Zeffwa= \left[1+\gamma^{(1)}+ o(\eps)\right] \Zeffba
\end{equation} 
with
\begin{equation}
\label{valueCeff1}
\gamma^{(1)}=C_{\phi}^{ (1)}\left(\frac{\beta e^2}{\epsilon_{\rm solv}b}, \ln(\kd b),\dw\right)
-\eps\left(\frac{\Sigma_3}{\Sigma_2}\right)^2 
\left[\frac{a_c(\dw)}{4}-\frac{\ln 3}{8}\right].
\end{equation}
(We notice that the notation  $o(\eps)$ in \eqref{rapZ} contains both contributions such as those in \eqref{orders} and terms of order $\eps \times \kd b$ .)  As exhibited by their diagrammatic origins, the various
terms in $\gamma^{(1)}$ arise both from the nonuniformity of the
density profiles 
and from  screened interactions via  two intermediate charges.
These profiles, which have been calculated explicitly in the limit of vanishing $\kd b$ in Ref.\cite{AquaCor01I},  result from the competition between, on one hand, the screened self-energy arising both from the electrostatic response of the wall and its  steric deformation of screening clouds, and, on the other hand,  the profile of the electrostatic potential drop which  these two effects  induce in the electrolyte. 
 More precisely,
 $C^{(1)}_{\phi}$ written in \eqref{valueda} is the first-order
renormalization of the amplitude of  ${\overline D}_{\phi}(x)$  (see \eqref{structDphi}), which  originates
from the nonuniformity of the
density profiles. (The screened potential $\phi$  appears as an auxiliary object in the resummed Mayer diagrammatics, and the expressions of $C_{\phi}^{(1)}$ and
  ${\overline G}^{\rm exp\, (1)}_{\phi}(x,x')$ are calculated in Section \ref{section5}.)
$C^{(1)}_{\phi}$ and its sign depend on the composition of the electrolyte, on the closest approach distance $b$ to the wall (through the parameters $\kd b$ and $\beta e^2/(\epsilon_{\rm solv} b)$),  and on 
the parameter $\dw$, which charaterizes the difference between the dielectric permittivity of the wall and that of the solvent,
\begin{equation}
\label{defdw}
\dw \equiv \frac{\epsilon_{\rm \scriptscriptstyle W}-\epsilon_{\rm solv}}
{\epsilon_{\rm \scriptscriptstyle W}+\epsilon_{\rm solv}}.
\end{equation}
The second term in the r.h.s. of \eqref{valueCeff1} originates
 from the renormalization of the screening length and from the difference in the  contributions  from four-body effective interactions in the bulk and along the wall. 
The three-body effective interactions do not contribute to $\gamma^{ (1)}$ -- so that $\gamma^{(1)}$ is independent of the species $\alpha$ --, because they give  the same corrections to
the amplitudes of $w_{\as\asp}^{\rm {\scriptscriptstyle B}}$ and 
$w_{\as\asp}$.
 The  constant $a_c(\dw)$ is written in \eqref{valueac}.
If $\epsilon_{\rm solv}> \ew$, as it is the case when the solvent is water and the wall is made of glass,  $a_c(\dw)>(1/2)\ln 3$ and the second term decreases the ratio $\Zeffwa/\Zeffba$.
 We notice that
 $G_{\as}^{\rm exp\, (1)}(x)$ at first
order in $\eps$ is  given
in Section \ref{section65}.  
Contrarily to $D_{\as}^{(0)}(x)$, the sign of 
$D_{\as}^{(1)}(x)$ may vary  with the distance $x$, and  it depends on the composition of
the electrolyte, on the closest approach distance to the wall $b$, and on the ratio between the dielectric constant of the wall and  that of the solvent.

\subsection{Contents}
\label{section13}

The paper is organized as follows. The large-distance  behaviors of the effective pairwise interactions  $w_{\as\asp}$, in the bulk or along the wall, are
investigated through the large-distance decay of the Ursell function 
$h_{\as\asp}$  according
to relation \eqref{defw}. The latter decay is conveniently studied  from Mayer diagrammatics  generalized to inhomogeous situations. In Section \ref{section2}
we recall the resummed Mayer diagrammatics introduced  in Paper I in order to systematically handle  with the large-distance nonintegrability of  the bare Coulomb potential (far away or near the wall). There appears a screened potential $\phi$, which coincides with the interaction defined from the immersion free energy between two infinitesimal external point charges (see Section \ref{section22}).
  In the bulk,
$\phi$ is a solution of the usual Debye equation. Near the wall $\phi$  obeys an
inhomogeneous
Debye equation, where the inverse screening length depends on $x$. 
In Section \ref{section25} a decomposition of $h_{\as\asp}$  into four contributions enables one to show how the basic internal- and external-screening sum rules arise in resummed Mayer diagrammatics, and how they are preserved if only some subclass of diagrams is retained. It also allows one to show that, for a symmetric electrolyte, 
$\soma \roa(\vecr)h_{\as\asp}(\vecr,\vecr')$ decays faster than
$h_{\as\asp}(\vecr,\vecr')$ (see Sections \ref{densitdensit} and \ref{section53bis}). 
We also recall the graphic reorganization of diagrams devised in Paper I for the study of the general structure of large-distance tails in dilute regimes.

Systematic double-expansions in
the  dimensionless parameters $\sigma/a$ and $\eps$    can be performed from resummed diagrams.
In Section \ref{section4}  we exhibit the nature of the  first various contributions. This leads us to introduce the scaling regimes \eqref{case1bis} and \eqref{subcase2bis} 
where the correction proportional to the coupling parameter $\eps$ is the leading contribution. (We also recall the expression of the  pair
correlation at any distance  at leading order in $\eps$.)

Section \ref{section3} is devoted to bulk correlations.
We take advantage of the full
translational invariance in the bulk in order to resum the four geometric 
series which appear in the Fourier transform of the graphic decomposition of
$h_{\as\asp}^{\scriptscriptstyle B}$ recalled in Section  \ref{section25}.
Thus, we obtain a compact formula for the large-distance behavior of 
$h_{\as\asp}^{\scriptscriptstyle B}$,  where the contributions of both charges are factorized. This formula is   appropriate to obtain  systematic
 $\eps$-expansions of the screening length and of the renormalized charge from the  $\eps$-expansions of resummed diagrams.

 In Section \ref{section45}, we also show how to retrieve
the corresponding corrections of order $\eps$ by a more cumbersome method which
 will be useful for the calculations in the vicinity of a wall, where the
 translational invariance is lost in the direction perpendicular to the wall.
In position space every convolution in the graphic representation  of
$h_{\as\asp}^{\scriptscriptstyle B}$ decays exponentially over the Debye screening length $\kd^{-1}$ at large relative distances $r$,  with an amplitude which is proportional to 
$1/r$ times a polynomial in $r$. The resummation of  the  series of the leading tails in $r$ at every order in $\eps$ must be performed in order to get the exponential decay over the  screening length   $\kappab^{-1}$ calculated up to  order $\eps$ (see Appendix  \ref{AppA}). On the contrary, the correction of order $\eps$ in the renormalized charge can be retrieved from only a finite number of resummed Mayer diagrams.

In Section \ref{section5}  we recall how  the screened potential $\phi(x,x',y)$  and  the effective dipole ${\overline D}_{\phi}(x)$ in its large-$y$ tail are formally expressed in terms of the density profiles in the vicinity of the wall
 \cite{AquaCor03I}.  Then, the $\eps$-expansion
of   ${\overline D}_{\phi}(x)$ can be performed from the $\eps$-expansion
 of the density profiles, by applying the method devised in Ref.\cite{AquaCor01II}.
The density profiles, which vary
rapidly over the Bjerrum length $\beta e^2/\epsilon_{\rm solv}$ in the vicinity of a dielectric wall, have been explicitly determined up to order $\eps$ 
 in the limit where $\kd b$ vanishes in Ref.\cite{AquaCor01I}, and we  explicitly calculate  ${\overline D}_{\phi}(x)$ up to order $\eps$  in the same limit.

In Section \ref{section6} we recall how the structure of the effective dipole $D_{\as}(x)$ in the $1/y^3$ tail of $h_{\as\asp}(x,x',y)$ is given  in terms of the graphic representation written in Section \ref{section25}. We also derive a sum rule for $\soma e_{\as}\roa(x)D_{\as}(x)$.
By using the resummation method checked for
the bulk situation in Section \ref{section45}, we determine
the renormalized value of the screening length in the direction perpendicular 
to the wall at first order in $\eps$.  For that purpose, in Appendix \ref{AppB} we show that the leading term in $x$ at every order $\eps^q$ is  proportional to
$\left(x-b\right)^q\exp\left[-\kd (x-b)\right]$, and we resum the series of these leading terms. Thus, we check that
the correction of order $\eps$ in the screening length is indeed the same in the
bulk and in the direction perpendicular to the wall. Then, $D_{\as}(x)$ is  determined up to order $\eps$ by only two
 resummed Mayer diagrams. Explicit calculations are performed in the limit where $\kd b$ vanishes and the expressions of $\Zeffba$ and $\Zeffwa$ are compared.
Their physical interpretation is given thanks to the diagrammatic origins of the various contributions.

\section{General formalism}
\label{section2}

\subsection{Model}
\label{section21}

In the primitive model defined above, the hard-core effect between two species $\alpha$ and $\alpha'$ can be taken 
into account in the pair energy by an 
interaction $\vsr$ which is infinitely repulsive at  distances shorter 
than the
sum $(\sigma_{\as}+\sigma_{\asp})/2$ of the sphere radii of both species.
Its Boltzmann factor reads
\begin{equation}
\label{defvsr}
\exp\left[- \beta\vsr(|\vecr - \vecr'|; \alpha,\alpha')\right ]= \left\{ \begin{array}{ll}
                     0  & \mbox{if $|\vecr - \vecr'|
                     <(\sigma_{\as}+\sigma_{\asp})/2$},\\ 
                     1 & \mbox{if $|\vecr - \vecr'|>
                     (\sigma_{\as}+\sigma_{\asp})/2$}.
                     \end{array}
                     \right.       
\end{equation}
Since charges are reduced to points at the centers of  excluded-volume spheres with the same dielectric constant as the solvent, the Coulomb interaction between two charges can be  written in the whole space (even for $x<0$ or $x'<0$) as 
$(Z_\as Z_\asp e^2 /\epsilon_{\rm solv})v(\vecr, \vecr')$, where
$v(\vecr, \vecr')$ is the solution of Poisson equation  for unit point charges with the
adequate electrostatic boundary conditions. 
Since the  half-space $x<0$ is occupied by a   material with a dielectric constant $\epsilon_{\rm \scriptscriptstyle W}$, 
$v(\vecr,\vecr')$ in Gauss units reads, for $x>0$ and $x'>0$ and  for any  $|\vecr - \vecr'|>0$ \cite{Jackson},
\begin{equation}
  \label{valuev}
v(\vecr,\vecr')  = \frac{1}{|\vecr - \vecr'|}-\dw \frac{1}{|\vecr -
 {\vecr'}^{\star}|}.
\end{equation}
$\dw$,  defined in \eqref{defdw},  lies between $-1$ and $1$,
and  ${\vecr'}^{\star}$ is the {\it image} of the position $\vecr'$
 with respect to the plane interface between the solution and 
 the dielectric material.   
In the bulk the  Coulomb potential reads
\begin{equation}
  \label{valuevB}
  v_{\scriptscriptstyle B}(|\vecr - \vecr'|)= \frac{1}{|\vecr - \vecr'|}.
\end{equation} 
The total pair
energy $ U_{\rm pair}$ is
 \begin{equation}
\label{valueUpair}
 U_{\rm pair}= 
\frac{1}{2} \sum_{i\not= j} \vsr(\vert \vecr_i-\vecr_j\vert;\alpha_i,\alpha_j) 
 + \frac{1}{2}\sum_{i\not= j} \frac{e^2}{\epsilon_{\rm solv}}
 Z_{{\scriptscriptstyle \alpha}_i} Z_{{\scriptscriptstyle \alpha}_j}
   v( \vecr_i,\vecr_j),  
\end{equation}
 where $i$ is 
the index
of a particle.

In the vicinity of the wall, one-body potentials appear in the total energy of the system.
For every charge  a self-energy 
$Z_\as^2(e^2/\epsilon_{\rm solv})\Vself$ arises from 
the work necessary to bring a charge
$Z_{\as} e$ 
from $x=+\infty$ (in the solvent) to a point $\vecr$ in the vicinity of 
the wall. According to \eqref{valuev}, the wall electrostatic response is
equivalent to the presence of an image charge 
$-\dw Z_{\as}e$ at point $\vecr^{\star}$ inside  a wall that would have the same dielectric constant $\epsilon_{\rm solv}$ as the solvent, and
\begin{equation}
\label{defVself}
\Vself(x) =- \dw {1\over 4x}.
\end{equation}
In the case of a glass wall in contact with water, the
relative dielectric constant $\ew/\epsilon_{\rm solv}$ of the wall with respect to the solvent is
of order $(1/80)<1$, $\dw$ defined in \eqref{defdw} is negative, and $\Vself$ is a repulsive potential.
The impenetrability of the wall  corresponds to a short-ranged potential $\Vsr(x)$, the Boltzmann factor of which is
\begin{equation}
\label{}
 \exp\left[-\beta\Vsr(x)\right] = \left\{ \begin{array}{ll}
                     0 & \mbox{if $x<b$}\\ 
                     1 & \mbox{if $x>b$},
                     \end{array}
                     \right.       
\end{equation}
where $b$ is  the closest approach distance to the wall for the  centers
 of    spherical particles, which is the same  for all species. 
The confinement of all particles to the positive-$x$
region and the electrostatic self-energy 
may be gathered   
 in a one-body potential $V_{\rm wall}$,
\begin{equation}
V_{\rm wall}=  \sum_{i}  \Vsr(x_i;\alpha_i) +\sum_{i} 
\frac{ e^2}{\epsilon_{\rm solv}}
Z_{{\scriptscriptstyle \alpha}_i}^2\Vself(x_i). 
\end{equation}

\subsection{Generalized resummed Mayer diagrams}
\label{section22}

By virtue of definition  \eqref{defw},
the leading large-distance behavior $w_{\as\asp}^{\rm as}$ of $w_{\as\asp}$ is proportional to the  large-distance behavior $h_{\as\asp}^{\rm as}$ of $h_{\as\asp}$, 
\begin{equation}
\label{relhaswas}
h_{\as\asp}^{\rm as}
=-\beta w_{\as\asp}^{\rm as},
\end{equation}
because any power $[w_{\as\asp}^{\rm as}]^n$, with $n\geq 2$, has a faster decay than $w_{\as\asp}^{\rm as}$. In an inhomogeneous situation $h_{\as\asp}^{\rm as}$ is conveniently studied by means of the Mayer diagrammatic representation of 
$ h_{\as\asp}$. However, the large-distance behavior of the Coulomb pair interaction $v(\vecr,\vecr')$ is not integrable, and every integral corresponding to a standard Mayer diagram that is not sufficiently connected diverges   when the volume of the region occupied by the fluid becomes infinite.

 As shown in Paper I, thanks to  a generalization of the procedure introduced by
Meeron \cite{Meeron} in order to calculate $h_{\as\asp}$ in the bulk,
the  density-expansion of $h_{\as\asp}$ in the vicinity of the wall can  be expressed in terms of resummed Mayer
diagrams with integrable  bonds $F$. Since the procedure for the systematic resummation of Coulomb divergences relies on topological considerations, the definitions of Mayer diagrams with resummed bonds are formally the same ones in the bulk or near the wall.
The two 
differences between resummed diagrams in the  bulk and near the wall are the following. First, near the wall  the point weights are not
 constant densities but $x$-dependent density profiles. Second, the screened potential $\phi$ arising from collective effects described by the systematic resummation of Coulomb divergences is no longer the Debye potential, but it obeys an ``{inhomogeneous}'' Debye equation,
\begin{equation}
\label{Debyeeq}
   \Delta_{\vecr} \phi (\vecr, \vecr') - \kbar^2 (x) \phi(\vecr, \vecr')  
   = - 4  \pi
  \delta( \vecr - \vecr').
\end{equation}
In \eqref{Debyeeq}  $\kbar^2 (x)$ is defined as
\begin{equation}
\label{defkbar}
  \kbar ^2 (x) \equiv 
 4 \pi \beta \frac{  e^2}{\epsilon_{\rm solv}}\sum_{\alpha} Z_\as^2  \roa(x),
\end{equation}
where  all densities $\roa(x) $'s vanish for $x<b$. $\phi$ obeys the same
boundary conditions as the Coulomb potential $v$: $\phi ( \vecr, \vecr')$ is continuous everywhere and tends to $0$ when
$|\vecr-\vecr'|$ goes to $+\infty$, while its gradient times the dielectric
constant is continuous at the interface with dielectric walls.
We recall that particles are supposed to be made of a material with the same dielectric constant as the solvent.

 The two resummed bonds $F$,  called $\Fcc$ and $\Fr$ respectively, are written in terms of the screened potential $\phi$ as
\begin{equation}
\label{defFcc}
\Fcc(n,m)=- \frac{ \beta e^2}{\epsilon_{\rm solv}} Z_{\as_n}Z_{\as_m} \phi(\vecr_n,\vecr_m) 
\end{equation}
and
\begin{equation}
\label{defFr}
 \Fr(n,m) = 
\exp\left[-\beta \vsr(\vert\vecr_n-\vecr_m\vert) 
 -\frac{ \beta e^2}{\epsilon_{\rm solv}} Z_{\as_n}Z_{\as_m}
 \phi(\vecr_n,\vecr_m) \right] 
-1 + \frac{ \beta e^2}{\epsilon_{\rm solv}} Z_{\as_n}Z_{\as_m}
\phi(\vecr_n,\vecr_m),
\end{equation}
where $n$ and $m$ are point indices in the Mayer diagrams. (In the bond notations, the superscript ``{cc}'' stands for ``{charge-charge}'' and ``{R}'' means ``{resummed}''. Indeed,  $\Fcc$ is proportional to the resummed interaction $\phi(\vecr,\vecr')$ between point charges; $\Fr+\Fcc$ is equal to the original Mayer bond where the Coulomb pair interaction $v(\vecr,\vecr')$ is replaced by its resummed expression $\phi(\vecr,\vecr')$, while the short-range repulsion is left unchanged.) The resummed Mayer diagrammatics of $h_{\as\asp}$ 
is
\begin{equation}
\label{gresumed}
h_{\as\asp}(x,x',\vy) = \\
\sum_{\Pi} \frac{1}{S_{\Pi}} \int_{\Lambda}  \left[\prod_{n=1}^N
 d{\vecr}_n  \sum_{\as_n=1}^{n_s}  \rho_{\as_n}(x_n) \right] 
\left[ \prod F \right]_{\Pi} .
\end{equation}
In (\ref{gresumed}) the sum runs over all the unlabeled topologically different
connected diagrams $\Pi$ with two root points $(\vecr,\alpha)$ and 
$(\vecr',\alpha')$ (which are
not integrated over) and N internal points (which are integrated over) with
$N=0,\ldots,\infty$, and which are built according to the following rules.  Each pair
of points in $\Pi$ is linked by at most one  bond $F$, and  there is no articulation
point. (An articulation point is defined  by the fact that, if it
is taken out of the diagram, then the latter is split into two pieces, one of
which at least is no longer linked to any root point.)
Moreover, in order to avoid double counting in the resummation process, diagrams
$\Pi$ must be built with an ``{excluded-composition}'' rule : there is no
 point attached by only two  bonds $\Fcc$ to the rest of the diagram. 
$\left[ \prod F \right]_{\Pi}$ is the product of the
 bonds $F$ in the $\Pi$ diagram and $S_{\Pi}$ is its symmetry factor, i.e.,
the number of permutations of the internal points $\vecr_n$ that do not change this
product.  Every point has a
weight equal to $\roa(x) $ that is summed over all species.  We have used the convention that, if $N$ is equal to $0$, no 
$\int_{\Lambda} d{\vecr}_n \rho_{\as_n}(x_n)$ appears, and 
$(1/S_{\Pi})\left[ \prod F \right]_{\Pi}$ is reduced to $F(\vecr,\vecr')$. Near the wall, $\Lambda$ denotes a finite-size region bounded by the wall on the left, whereas, in the bulk, $\Lambda$ stands for a finite-size region far away from the wall.
The screened potential $\phi$ is integrable at large distances. (In the bulk $\phi$ decays exponentially fast in all directions; near the wall in the large-distance behavior given in \eqref{structDphi}, ${\overline D}_{\phi}(x)$ has an  exponential decay and the $1/y^3$ tail is integrable.)
As a consequence, $\Pi$ diagrams
 correspond to convergent integrals in the limit where the volume $\Lambda$
 extends to infinity inside the bulk or on the right of the wall.

\subsection{ Graphic reorganization of resummed diagrammatics}
\label{section25}

In $h_{\as\asp}(\vecr_a,\vecr'_{a'})$  we can distinguish four classes of diagrams by considering
whether a single bond $\Fcc$ is attached to root point $a$ or to root point 
$a'$. ($a$ and $a'$ are  short  notations for the
couple of variables  $(\vecr_a,\alpha)$ 
and $(\vecr_{a'},\alpha')$, respectively, which are  associated with the  root points in a Mayer diagram.)
$h_{\as\asp}$ can be rewritten as the sum 
\begin{equation}
\label{decomph}
h_{\as\asp} \equiv \hcc_{\as\asp} + \hct_{\as\asp}+ \htc_{\as\asp}+
\htt_{\as\asp},
\end{equation} 
where in $\hcc_{\as\asp}$ both $a$ and $a'$ carry a single  bond $\Fcc$, in  $\hct_{\as\asp}$ ($\htc_{\as\asp}$ ) only $a$ ($a'$) is linked to the rest of the diagram by a single  bond $\Fcc$, and in $\htt_{\as\asp}$ neither $a$ nor $a'$ are linked to the rest of the diagram by only one  bond $\Fcc$.

\subsubsection{Screening rules}
\label{genscreenrule}

A first  interest of decomposition \eqref{decomph} is that it enables one to derive the basic screening rules (recalled hereafter) from the fact that they are already fulfilled by the diagram made of a single bond $\Fcc$ (because of the corresponding sum rules obeyed by the screened potential $\phi(\vecr,\vecr')$). Moreover, since the sum rules are linked to the large-distance behavior of the charge-charge correlation function, decomposition 
 \eqref{decomph} also enables one  to show that if some diagrams are to be kept for their contributions to $h_{\as\asp}^{--\rm as}$ in some dilute regime, then the corresponding diagrams ``{dressed}'' with $\Fcc$ bonds in $\hcc_{\as\asp}$, $\hct_{\as\asp}$ and $\htc_{\as\asp}$ are also to be retained, together with the bond $\Fcc$, in order to ensure that the screening rules are still satisfied.

The basic screening rules are the following. In a charge fluid with Coulomb interactions, an internal charge of the system, as well as an infinitesimal external charge, are perfectly screened by the fluid: each charge is surrounded by a cloud which carries exactly the opposite charge. 
These properties can be written in a compact form in terms of the charge-charge correlation defined as 
\begin{equation}
\label{defCgen}
C(\vecr,\vecr')\equiv  e^2
\left\{ \sum_{\alpha}Z_\as^2 \roa(\vecr) \delta(\vecr-\vecr')+
\sum_{\alpha,\alpha'}Z_\as Z_\asp \roa(\vecr) \roap(\vecr')  h_{\as\asp}(\vecr,\vecr')\right\}.
\end{equation}
The internal-screening rule reads
 \begin{equation}
\label{sumruleint}
\int d\vecr \sum_{\alpha}Z_{\as}{\roa}(\vecr)h_{\as\asp}(\vecr,\vecr')=-Z_{\asp},
\end{equation} 
and, by performing the summation $\sum_{\alpha'}Z_{\asp}\roap(\vecr')\times\eqref{sumruleint}$,  the internal-screening sum rule implies that
\begin{equation}
\label{sum1}
\int d\vecr\, C(\vecr,\vecr')=0.
\end{equation}
By virtue of  the linear response theory, the external-screening sum rule  reads
\begin{equation}
\label{sum2}
\frac{\beta}{\epsilon_{\rm solv}} \int d\vecr\int d\vecr' v(\vecr_0,\vecr')
C(\vecr',\vecr)=1.
\end{equation}
The latter equation, derived  for inhomogeneous systems by Carnie and Chan \cite{CarnieChan81},  is the generalization of the  sum-rule first settled by Stillinger and Lovett  \cite{StilLovet} for the second moment of $C(\vecr,\vecr')$  in the homogeneous case
 (see next paragraph). As a consequence of the internal screening sum rule, \eqref{sum2} holds whatever short-distance regularization may be added to the pure Coulomb interaction $v(\vecr_0,\vecr')$ \cite{revueMartin}. 
 
In the bulk, the translational invariance in all directions implies that  sum rules \eqref{sum1} and \eqref{sum2}  are relative respectively to
 the $k=0$
value and to the coefficient of
the $k^2$-term  in the $k$-expansion of $C^{\rm{ \scriptscriptstyle B}}(k)$.
Both sum rules are summarized in the following small-$k$ behavior,
\begin{equation}
\label{StilLovrule}
C^{\rm{ \scriptscriptstyle B}}(k)
\underset{k\rightarrow 0}{\sim}\,\, \frac{\epsilon_{\rm solv}}{4\pi\beta} \,k^2.
\end{equation}
In the vicinity of a wall, there is translational invariance only in directions parallel to the plane interface, and the Carnie and Chan sum rule \eqref{sum2} takes the form of a dipole sum rule \cite{Carnie83,revueMartin},
\begin{equation}
\label{dipolerule}
\int_0^{+\infty}dx\int_0^{+\infty}dx' \int d\vy\, x' C(x,x',\vy)=
-\frac{\epsilon_{\rm solv}}{4\pi\beta}.
\end{equation} 
As shown in Ref. \cite{JancoSamaj01}, the first moment of $C(x,x',\vy)$ is linked to the amplitude 
 $f_{\scriptscriptstyle C}(x,x')$ of the 
$1/y^3$ tail of $C(x,x',\vy)$,
 \begin{equation}
\int_0^{+\infty}dx' \int d\vy\, x' C(x,x',\vy)=
 \frac{\epsilon_{\rm solv}}{\ew} 2\pi \int_0^{+\infty}dx' f_{\scriptscriptstyle C}(x,x').
\end{equation} 
$f_{\scriptscriptstyle C}(x,x')$  coincides with 
$-\beta \sum_{\alpha\alpha'}
e^2Z_{\as}Z_{\asp} \roa(x)\roap(x')f_{\as\asp}(x,x')$, where  $-\beta f_{\as\asp}(x,x')/y^3$ is the large-distance behavior of $h_{\as\asp}(x,x',\vy)$. Therefore, the   moment rule \eqref{dipolerule} can be rewritten as a sum rule for the amplitude $f_{\as\asp}(x,x')$, first derived in Ref.\cite{Janco82II},
\begin{equation}
\label{sumrule2wall}
\int_0^{+\infty}dx\int_0^{+\infty}dx' \sum_{\alpha\alpha'}
e^2Z_{\as}Z_{\asp} \roa(x)\roap(x')f_{\as\asp}(x,x')=
\frac{\ew}{8\pi^2\beta^2}.
\end{equation} 
(We notice that there is a misprint in Paper I, where the above sum rule is written in Eq.(4) with a an extra spurious coefficient $1/\epsilon_{\rm solv}$ on the r.h.s.)

Now, we show how the combination of  decomposition \eqref{decomph} with 
sum rules obeyed by $\phi$ enables one to derive the two basic screening rules.
A key ingredient of the derivation is the relations between $\hcc$ and $\htc$ on one hand, and $\hct$ and  $\htt$  on the other hand,  which arise from their definitions.

In the bulk, because of the full translational invariance, the latter relations take simple forms in Fourier space.  They  read 
\begin{equation}
\label{relhcchtcbulk}
\hcc_{\as\asp}(k)=\Fcc_{\as\asp}(k)+
\sum_{\gamma_1} 
\rho_{\scriptscriptstyle\gamma_1}^{\rm\scriptscriptstyle B}
\Fcc_{\as\gs_1}(k)\htc_{\gs_1\asp}(k),
\end{equation} 
and
\begin{equation}
\label{relhcthttbulk}
\hct_{\as\asp}(k)= \sum_{\gamma_1} 
\rho_{\scriptscriptstyle\gamma_1}^{\rm\scriptscriptstyle B}
 \Fcc_{\as\gs_1}(k)\htt_{\gs_1\asp}(k).
\end{equation}
(For the sake of clarity, in the present paragraph, we forget superscripts ${\rm B}$, except in the densities, in $h_{\as\asp}$ and in $C$.) 
On the other hand, by virtue of the explicit expression \eqref{defphib} of  ${\phi}_{\rm\scriptscriptstyle B}$,
\begin{equation}
\label{saturbulk}
 \sum_{\alpha}Z_{\as}{\roab}\Fcc_{\as\asp}(k=0)=-Z_{\asp}.
\end{equation} 
In other words, the part $\Fcc$ in $h_{\as\asp}^{\rm\scriptscriptstyle B}$ already fulfills the internal-screening sum rule.  When relations \eqref{relhcchtcbulk} and \eqref{relhcthttbulk} are inserted in decomposition   \eqref{decomph} of $h_{\as\asp}^{\rm\scriptscriptstyle B}$, 
\begin{equation}
\label{decompbis}
h_{\as\asp}^{\rm{\scriptscriptstyle B}}(k)=\Fcc_{\as\asp}(k)
+\left[\sum_{\gamma_1} \rho_{\scriptscriptstyle\gamma_1}^{\rm\scriptscriptstyle B}
\Fcc_{\as\gs_1}(k)\htc_{\gs_1\asp}(k)
+\htc_{\as\asp}(k)\right]
+\left[\sum_{\gamma_1} \rho_{\scriptscriptstyle\gamma_1}^{\rm\scriptscriptstyle B}
\Fcc_{\as\gs_1}(k)\htt_{\gs_1\asp}(k)
+\htt_{\as\asp}(k)\right]
\end{equation} 
Then property
\eqref{saturbulk} implies that, in  
$ \sum_{\alpha}Z_{\as}{\roab}h_{\as\asp}^{\rm{\scriptscriptstyle B}}(k=0)$, the contribution  from $\hcc-\Fcc$, given in  \eqref{relhcchtcbulk}, cancels that from $\htc$ , and the contribution from $\hct$, given in \eqref{relhcthttbulk},
 compensates that from $\htt$, so that the internal-screening rule 
 is  indeed satisfied. 
 
In the case of  the bulk external-screening rule  \eqref{StilLovrule}, the same mechanism operates  when the $k^2$-term in the small-$k$ expansion of $C^{\rm{ \scriptscriptstyle B}}(k)$ is considered. The charge-charge correlation  $C^{\rm{ \scriptscriptstyle B}}_{\Fcc}$, where $h_{\as\asp}^{\rm{\scriptscriptstyle B}}$ is replaced by $\Fcc_{\as\asp}$, fulfills the second-moment sum-rule,
\begin{equation}
C^{\rm{ \scriptscriptstyle B}}_{\Fcc}(k)
\underset{k\rightarrow 0}{\sim}\,\, \frac{\epsilon_{\rm solv}}{4\pi\beta}\, k^2.
\end{equation}
 Again, by virtue of \eqref{saturbulk}, decomposition \eqref{decompbis}
 implies that the $k^2$-term in $\htc$  is canceled by  the part of the $k^2$-term in  $\hcc-\Fcc=\sum_{\gamma_1} 
\rho_{\scriptscriptstyle\gamma_1}^{\rm\scriptscriptstyle B}
\Fcc_{\as\gs_1}(k)\htc_{\gs_1\asp}(k)$  that arises from the $k^2$-term in  $\htc$. 
Similarly, the $k^2$-term in $\htt$  is canceled by  the part of the $k^2$-term in  $\hct
= \sum_{\gamma_1} 
\rho_{\scriptscriptstyle\gamma_1}^{\rm\scriptscriptstyle B}
 \Fcc_{\as\gs_1}(k)\htt_{\gs_1\asp}(k)$  that arises from the $k^2$-term in  $\htt$. Moreover, \eqref{relhcthttbulk} and \eqref{saturbulk} imply that 
$ \sum_{\alpha'}Z_{\asp}{\roapb}\htc_{\gs\asp}(k=0) =-\sum_{\gamma'}Z_{\gsp}\rogpb \htt_{\gs\gsp}(k=0)$, so that the part of the $k^2$-term in  $\hcc-\Fcc$ that comes from the $k^2$ term in $\Fcc$ is opposite to the   part of the $k^2$-term in  $\hct$ that is generated by the $k^2$-term in $\Fcc$.
 We notice that the present argument is analogous to that found  in Ref.\cite{Cornu96II} for an analogous decomposition in a quantum charge fluid.

 In the vicinity of the wall, the derivation of  screening rules \eqref{sumruleint} and \eqref{sumrule2wall} also relies on the analog of decomposition \eqref{decompbis} and on two sum rules derived for  
$\phi$ in Paper I, namely, if $x'>b$,
\begin{equation}
\label{sumrulephi}
\int_0^{+\infty} dx\,  \kbar ^2 (x) \int  d\vy\, \phi(x,x',\vy)=4\pi,
\end{equation} 
and
\begin{equation}
\label{sumrulefphi}
\int_0^{+\infty} dx\int_0^{+\infty} dx'\,  \kbar ^2 (x) \kbar ^2 (x') f_{\phi}(x,x')=2\frac{\ew}{\epsilon_{\rm solv}}.
\end{equation} 
The Fourier transform of a function $f(\vy)$ at wave vector $\vl$ is defined as 
$f(\vl)\equiv \int d\vy \exp(i\vl\cdot\vy)f(\vy)$.
Thanks to the translational invariance in the direction $\vy$ parallel to the plane interface, the relations, which arise from their definitions, between $\hcc$ and $\htc$ on one hand, and $\hct$ and  $\htt$  on the other hand, take the simple form,
\begin{eqnarray}
\label{relhcchtc}
\hcc_{\as\asp}(x,x',\vl)&=&\Fcc_{\as\asp}(x,x',\vl)\\
&&+
\int_0^{+\infty} dx_1\sum_{\gamma_1} 
\rho_{\gs_1}(x_1)
  \Fcc_{\as\gs_1}(x,x_1,\vl)\htc_{\gs_1\asp}(x_1,x',\vl),
 \nonumber
\end{eqnarray} 
and
\begin{equation}
\label{relhcthtt}
\hct_{\as\asp}(x,x',\vl)=\int_0^{+\infty} dx_1 \sum_{\gamma_1} 
 \rho_{\gs_1}(x_1)
 \Fcc_{\as\gs_1}(x,x_1,\vl)\htt_{\gs_1\asp}(x_1,x',\vl).
\end{equation} 
\eqref{sumrulephi} and \eqref{sumrulefphi} imply  that $\Fcc$ saturates the internal sum rule \eqref{sumruleint} and the external sum rule  \eqref{sumrule2wall}, respectively.

The external-screening sum rule \eqref{sumrule2wall} in the vicinity of the wall is studied  again in Section \ref{section53bis}. We show that, in the case where all species have the same closest approach distance to the wall,   decomposition  \eqref{decomph} enables 
one to derive a sum rule fulfilled by the effective dipole amplitude ${\overline D}_{\alpha}(x)$.

\subsubsection{Large-distance tails}

Another interest of decomposition \eqref{decomph} is that  the large-distance behavior of the Ursell function $h_{\as\asp}$ 
 can be conveniently analyzed from this decomposition,  after a suitable reorganization of resummed Mayer diagrams, which has been introduced in Paper I. The resummed Mayer diagrammatics \eqref{gresumed} for 
$h_{\as\asp}$ is reexpressed 
in terms of  ``{graphs}'' made of two kinds of bonds: the bond  $\Fcc$ and  the bond $\Ir$ that is defined as
the sum of all subdiagrams that either contain no $\Fcc$ bond or remain connected in a single piece when a bond $\Fcc$ is cut. 
$\Fr$ falls off faster than $\Fcc$ at large distances (namely, as $[\Fcc]^2/2$) and the topology of
subdiagrams involved in $\Ir$ implies that $\Ir$ decays faster than $\Fcc$ at large
distances  in a sufficiently dilute regime. 
Since the  reorganization is purely topological, it is valid
for correlations in the bulk as well as in the vicinity of the wall.
According to the excluded-composition rule obeyed by resummed 
$\Pi$ diagrams, the functions in the r.h.s. of \eqref{decomph} are equal to the series  
represented in
Figs.\ref{hcc}-\ref{htt} respectively,
\begin{eqnarray}
\label{defhcc}
 h^{\rm cc}_{\as\asp}(\vecr,\vecr')=
 \Fcc(a,a') &+&
 \int d\vecr_1d\vecr'_1 \sum_{\gamma_1,\gamma'_1} 
 \rho_{\gs_1}(\vecr_1)\rho_{\gspu}(\vecr'_1)
 \Fcc(a,1)\Ir(1,1')\Fcc(1',a')\nonumber\\
 &+ &\int d\vecr_1d\vecr'_1 \sum_{\gamma_1,\gamma'_1}
 \rho_{\gs_1}(\vecr_1)\rho_{\gspu}(\vecr'_1) 
 \int d\vecr_2d\vecr'_2 \sum_{\gamma_2,\gamma'_2} 
 \rho_{\gs_2}(\vecr_2)\rho_{\gspd}(\vecr'_2)\nonumber\\  
 &&\qquad \Fcc(a,1)\Ir(1,1')\Fcc(1',2)\Ir(2,2')\Fcc(2',a')+\cdots\, ,
\end{eqnarray}
\begin{eqnarray}
\label{defhct}
 h^{\rm c-}_{\as\asp}(\vecr,\vecr')&\equiv&
 \int d\vecr_{c'} \sum_{\gamma'}  \rho_{\gsp}(\vecr_{c'})
 \Fcc(a,c')\Ir(c',a')\nonumber\\
&+& \int d\vecr_{c'} \sum_{\gamma'}  \rho_{\gsp}(\vecr_{c'})
\int d\vecr_1d\vecr'_1 \sum_{\gamma_1,\gamma'_1}
 \rho_{\gs_1}(\vecr_1)\rho_{\gspu}(\vecr'_1)\nonumber\\
&&\qquad \Fcc(a,1)\Ir(1,1')\Fcc(1',c')\Ir(c',a')
 + \cdots\, ,
\end{eqnarray}
while $h^{\rm -c}$ is defined in a symmetric way, and
\begin{eqnarray}
\label{defhtt}
 h^{\rm --}_{\as\asp}(\vecr,\vecr')\equiv
 \Ir(a,a') &+&
 \int d\vecr_{c}\int d\vecr_{c'} \sum_{\gamma,\gamma'}  
 \rho_{\gs}(\vecr_{c})\rho_{\gsp}(\vecr_{c'})
  \Ir(a,c)\Fcc(c,c')\Ir(c',a')\nonumber\\
 &+& \int d\vecr_{c}\int d\vecr_{c'} \sum_{\gamma,\gamma'}  
 \rho_{\gs}(\vecr_{c})\rho_{\gsp}(\vecr_{c'})
 \int d\vecr_1 d\vecr'_1 \sum_{\gamma_1,\gamma'_1}
 \rho_{\gs_1}(\vecr_1)\rho_{\gspu}(\vecr'_1) \nonumber\\
 &&\qquad \Ir(a,c)\Fcc(c,1)\Ir(1,1')\Fcc(1',c')\Ir(c',a')+\cdots\, .
\end{eqnarray}
In the previous definitions $c$ is a short  
notation for  $(\vecr_c,\gamma)$, and $i$  stands for $(\vecr_{i},\alpha_i)$.

 We notice that, according to previous section, any contribution to $\Ir$ automatically generates a change in  $h_{\as\asp}$ that preserves the two basic  screening sum rules.  In the bulk, the external-screening rule \eqref{StilLovrule} is also retrieved from the compact formulae obtained by resummations in the graphic expansion \eqref{defhcc}--\eqref{defhtt}, as shown in Section \ref{densitdensit}.

\section{Weak-coupling regime}
\label{section4}

\subsection{Small parameters}
\label{section41}

Now we take into account the fact that in an electrolyte all species have charges and diameters   of 
the same magnitude orders $e$ and $\sigma$, respectively. Moreover,  all bulk densities are comparable, and the typical interparticle distance does not depend on species: it  is denoted by $a$. 
First,
we assume that the densities are so low that
 the   volume fraction $(\sigma/a)^3$ of particles 
is small,
\begin{equation}
\label{hypdilute}
\left(\frac{\sigma}{a}\right)^3\ll 1.
\end{equation}
Our second assumption is that the  temperature is  high enough
for
the mean closest approach distance between charges of the same sign at 
temperature $T$, of order $\beta e^2/{\epsilon_{\rm solv}}$,
to be small compared with the mean interparticle distance $a$.
In other words, the
coulombic coupling parameter $\Gamma$
   between charges of the fluid is 
negligible,
\begin{equation}
\label{defGamma}
\Gamma\equiv\frac{\beta e^2}{\epsilon_{\rm solv}a}
\propto \left(\frac{a}{\xid}\right)^2
\ll 1.
\end{equation}
(The proportionality relation in \eqref{defGamma} arises from definition \eqref{defkd}.)
The high-dilution condition \eqref{hypdilute} implies the weak-coupling condition $\Gamma^3\ll1$, if $\beta e^2/({\epsilon_{\rm solv}}\sigma)$ is of order unity or smaller than 1.

In fact, conditions \eqref{hypdilute} and  \eqref{defGamma} can be realized in  two different kinds of expansions in the density and temperature parameters. In the first situation, the  density vanishes at fixed temperature; then the ratio between the pair energy at contact and the mean kinetic energy, $\beta e^2/(\epsilon_{\rm solv}\sigma)$, is also fixed, namely
\begin{equation}
\label{case1}
\left(\frac{\sigma}{a}\right)^3\propto \Gamma^3\quad\mbox{\rm case (1)}.
\end{equation}
In the second situation the density vanishes  while the temperature goes to infinity, so that
$\beta e^2/(\epsilon_{\rm solv}\sigma)$ also vanishes, namely
\begin{equation}
\label{case2}
\Gamma^3\ll \left(\frac{\sigma}{a}\right)^3\quad\mbox{\rm case (2)}.
\end{equation}

\subsection{Expansions of resummed diagrams}
\label{section42}

The discussion of the $\Gamma$- and $(\sigma/a)$-expansions of the integrals associated with 
resummed $\Pi$  diagrams is easier if
we split the bond $\Fr$ into two pieces
\begin{equation}
\label{split}
\Fr=\frac{1}{2}[\Fcc]^2+\Frt.
\end{equation}
(The notation $\Frt$ refers to the truncation with respect to $\Fr$.)
Diagrams built with bonds $\Fcc$, $[\Fcc]^2/2$ and $\Frt$ -- and with the same
exclusion
rule for  bonds $\Fcc$ as in $\Pi$ diagrams  -- will be called ${\widetilde{\Pi}}$.
 The Ursell
function $h_{\as\asp}$ is represented in terms of ${\widetilde{\Pi}}$ diagrams by the same
formula \eqref{gresumed} as in the case of $\Pi$ diagrams. The splitting \eqref{split} has
already been used for a classical plasma in the vicinity of a dielectric wall in
Ref.\cite{AquaCor01II}, and its use was detailed for quantum plasmas 
in the bulk in Refs.\cite{AlCorPerez, Cor98II}.

For the sake of simplicity, the scaling analysis of diagrams is now discussed in the case of the bulk. The  bulk screened
potential $\phi_{\scriptscriptstyle B}$ obeys  \eqref{Debyeeq} 
 far away from any boundary,  where $\kbar (x)$ no longer depends  on $x$ and coincides with the inverse Debye screening length $\kd$. Then, \eqref{Debyeeq} is reduced to the usual
Debye equation, and, since $\phi_{\scriptscriptstyle B}$ is a
function of $|\vecr-\vecr'|$ that vanishes when $|\vecr-\vecr'|$ goes to
infinity, it is equal to the well-known Debye potential $\phid$,
\begin{equation}
\label{defphib}
{\phi}_{\rm\scriptscriptstyle B}(|\vecr-\vecr'|)=\phid(|\vecr-\vecr'|)\equiv 
\frac{e^{-\kd |\vecr-\vecr'|}}{|\vecr-\vecr'|}.
\end{equation}
 The integrals of the diagrams with a single bond can be calculated explicitly, and their orders in $\Gamma$ and $\sigma/a$ are the following:
  \begin{equation}
\int d\vecr'\,\roab \Fcc(\vecr,\vecr')=\cO(\Gamma^0)  
\end{equation} 
and
\begin{equation}
\int d\vecr'\, \roab\frac{1}{2}[\Fcc]^2(\vecr,\vecr')= \cO(\Gamma^{3/2}),
\end{equation}
where $\cO(\Gamma^0)$ and $\cO(\Gamma^{3/2})$  denote  terms of orders unity and
$\Gamma^{3/2}$, respectively.
According to \eqref{defFr} 
\begin{equation}
\label{valuebisFrt}
\Frt(\vecr, \vecr';\alpha,\alpha') = \left\{ \begin{array}{ll}
                     -1-\Fcc(\vecr, \vecr')
                     -\frac{1}{2}\left[\Fcc\right]^2(\vecr, \vecr')  
                     & \mbox{if $|\vecr - \vecr'|
                     <(\sigma_{\as}+\sigma_{\asp})/2$}\\ 
                     \sum_{n=3}^{+\infty}\frac{1}{n!}
                     \left[\Fcc\right]^n(\vecr, \vecr')  
                       & \mbox{if $|\vecr - \vecr'|>
                     (\sigma_{\as}+\sigma_{\asp})/2$}.
                     \end{array}
                     \right.       
\end{equation}
If we assume, for the sake of simplicity, that all particles have the
 same diameter $\sigma$, the expression $\Frt^{\scriptscriptstyle B}$ of $\Frt$ in the bulk leads to
 \cite{Haga} 
 \begin{eqnarray}
\label{Haga}
\int d\vecr'\, \Frt^{\scriptscriptstyle B}(\vecr, \vecr';\alpha,\alpha') 
=&&-\frac{4\pi}{3}\sigma^3
+2\pi \frac{ \beta  e^2 Z_\as Z_{\asp}}{\epsilon_{\rm solv}}\sigma^2 
-2\pi\left(\frac{  \beta  e^2 Z_\as Z_{\asp}}{\epsilon_{\rm solv}} \right)^2 \sigma\nonumber\\
&& +\frac{2\pi}{3} 
\left(\frac{  \beta  e^2 Z_\as Z_{\asp}}{\epsilon_{\rm solv}}\right)^3
\left[C+\ln (3\kd \sigma)\right]\nonumber\\
&&-4\pi 
\left(\frac{  \beta  e^2 Z_\as Z_{\asp}}{\epsilon_{\rm solv}}\right)^3
\sum_{n=1}^{+\infty} \frac{(-1)^n}{(n+3)!n} 
\left(\frac{  \beta  e^2 Z_\as Z_{\asp}}{\epsilon_{\rm solv}\sigma}\right)^n +R_{\Gamma^{3/2}}, 
\end{eqnarray}
where  $R_{\Gamma^{3/2}}$ denotes terms which are of relative order $\Gamma^{3/2}$ with respect to those written on the r.h.s. of \eqref{Haga}.
Therefore, since $\roab$ is of order $1/a^3$, at leading order 
$\int d\vecr'\,\roab  \Frt(\vecr, \vecr')$ is a sum of terms with respective orders 
\begin{equation}
\label{orders}
\left(\frac{\sigma}{a}\right)^3,\quad
\left(\frac{\sigma}{a}\right)^2 \Gamma,\quad \frac{\sigma}{a}\Gamma^2,\quad\Gamma^3,\quad
\Gamma^3\ln\left[\left(\frac{\sigma}{a}\right)^2 \Gamma\right]
,\quad \mbox{and} \quad
\Gamma^3 
f\left(\frac{\beta e^2}{\epsilon_{\rm solv}\sigma}\right),
\end{equation} 
 where $\beta e^2/(\epsilon_{\rm solv}\sigma)=\Gamma/(\sigma/a)$ and the function  $f(u)\equiv
\sum_{n=1}^{+\infty}  (-1)^n u^n/[(n+3)!n] $ vanishes for $u=0$. The last term in \eqref{orders} arises from the short-distance behavior of the Boltzmann factor, the explosion of which for oppositely charged species is prevented by the  cut-off distance $\sigma$ provided by the hard-core repulsion.

As already noticed in Paper I,
the contributions from excluded-volume effects in  the primitive model are not involved in $\Fcc$ but they are contained in $\Fr$. Indeed,
the potential $\phi$ solution of \eqref{Debyeeq}
 describes resummed interactions between point charges at the centers of  penetrable spheres, because it corresponds to the integral equation
 \begin{equation}
 \label{eqintphi}
\phi(\vecr,\vecr')=v (\vecr,\vecr')-\frac{\beta e^2}{\epsilon_{\rm solv}}\int d\vecr''
\soma Z_{\as}^2\roa(x'') v(\vecr,\vecr'')\phi(\vecr'',\vecr').
 \end{equation}
 We notice that, in the bulk, for the primitive model again, in a linearized mean-field Poisson-Boltzmann theory  where   excluded-volume spheres are taken into account  \cite{McQuarrie}, an extra Heaviside function $\theta
 [\vert\vecr''-\vecr'\vert-(\sigma_{\aspp}+\sigma_{\asp})/2)]$ appears in an equation analogous to \eqref{eqintphi}, and the effective interaction between two charges $e_{\as}$ and $e_{\asp}$ behaves as \linebreak
 $e_{\as}e_{\asp} \exp\{-\kd[r-(\sigma_{\as}+\sigma_{\asp})/2)]\}/\left\{[1+\kd (\sigma_{\as}+\sigma_{\asp})/2]r\right\}$ at large relative distances $r$. The latter interaction  is equal to  $e_{\as}e_{\asp}\phi_{\scriptscriptstyle B}(r)$  up to a steric correction  of order $(\kd \sigma)^2\propto \Gamma(\sigma/a)^2$.  This is also the case in the so-called DLVO  theory
\cite{VerweyOver, MedinaMcQuarrie80} for another model where every charge is spread over the surface of the excluded-volume sphere instead of being concentrated at its  center. In the corresponding effective interaction at large distances, the denominator of the steric factor which multiplies $\exp(-\kd r)/r$  takes the slightly different form
$[1+\kd (\sigma_{\as}+\sigma_{\asp})/4]^2$. The  order $\Gamma(\sigma/a)^2$ of this steric correction is  one among the contributions  listed in \eqref{orders}.

By using the variable change 
$\vecr\equiv {\widetilde \vecr}/\kd $, it can be shown that, when the number of
internal points in a ${\widetilde{\Pi}}$ diagram increases, then the lowest order in
$\Gamma$ at which it contributes to various integrals also increases. (See e.g. Refs. \cite{AlCorPerez} or \cite{Cor98II}.)

\subsection{Scaling regimes}
\label{sectionregime}

As shown in previous section, in the bulk the leading coupling correction is of order $\Gamma^{3/2}$, and the next correction without any steric contribution is of order $\Gamma^3$. The orders of the first corrections induced by  steric  effects involve $\sigma/a$ and $\Gamma$ through the combinations written in \eqref{orders}.

In the first  scaling regime \eqref{case1}, all terms in \eqref{orders} are of order $\Gamma^3$, and the leading correction is indeed provided by the correction of order $\Gamma^{3/2}$ arising only from Coulomb interactions for point charges in the Debye approximation. Moreover, we notice that in this  regime,
where the temperature $T$ is fixed,   $\Gamma^{3/2}$ is proportional to $\sqrt{\soma \roab Z_{\as}^2}$: the density-expansions prove to involve  powers of  the square root  of a linear combination of  densities. (The appearance of such  square roots instead  of integer powers in density expansions is an effect  of the long
range of Coulomb interactions, which makes the  infinite-dilution and vanishing-coupling limit singular.)

In the second case \eqref{case2}, $\beta e^2/(\epsilon_{\rm solv}\sigma)$ vanishes,  and  terms in \eqref{orders} are of orders $(\sigma/a)^3$ and $(\sigma/a)^3$ times a  function of $\beta e^2/(\epsilon_{\rm solv}\sigma)$ which tends to zero when $\beta e^2/(\epsilon_{\rm solv}\sigma)$ goes to zero.
The  explicit calculations will be performed in a subcase where the leading coupling correction of order $\Gamma^{3/2}$ is large compared with all corrections involving steric effects. This property is fulfilled if $(\sigma/a)^3/\Gamma^{3/2}$ goes to zero, and the corresponding subregime reads
\begin{eqnarray}
\label{subcase2}
\Gamma^3\ll \left(\frac{\sigma}{a}\right)^3\ll\Gamma^{3/2}
 \quad{\rm subcase \, (2)}.
\end{eqnarray}

In place of $\Gamma$, we shall use the parameter
 $\eps$  defined in \eqref{defeps}, because the first coupling correction is of order
\begin{equation}
\Gamma^{3/2}\propto \eps.
\end{equation} 
(See \eqref{defGamma} and  the definition \eqref{defkd} of $\kd$.)
In the first scaling regime, relation \eqref{case1}  can be written as $(\sigma/a)^3\propto \eps^2$.
Then all terms in \eqref{orders} are of order $\eps^2$, $\eps^2\ln \eps$  and $\eps^2f(\beta e^2/(\epsilon_{\rm solv}\sigma)$, and the whole double expansion in powers of $\eps$ and $\sigma/a$ proves to be a series in integer powers of $\eps$ times some possible powers of $\ln \eps$, at fixed 
$\beta e^2/(\epsilon_{\rm solv}\sigma)$. 
In the second regime,  condition \eqref{case2} reads
$\eps^2\ll (\sigma/a)^3$
and the extra condition in \eqref{subcase2} is $(\sigma/a)^3\ll\eps$.

In the following, the so-called 
``{weak-coupling}'' regime refers to the scaling limits \eqref{case1bis} or \eqref{subcase2bis}. Moreover the term ``{$\eps$-expansions}'' refers to $\eps$- and $\sigma/a$- expansions, as if they were always performed in the scaling regime
\eqref{case1bis}.

\subsection{Pair correlation  at any distance in the weak-coupling limit}
\label{section43}

The scaling analysis of $\eps$-expansions for resummed 
${\widetilde{\Pi}}$ diagrams (see Section \ref{section42}) shows that 
the $\eps$-expansions of integrals involving  $\Ir$ start at least at relative order 
$\eps$.  As a
consequence, at any relative distance, the pair correlation 
$h^{(0)}_{\as\asp}$ in the infinite-dilution and vanishing-coupling limit arises only from the sum of ${\Pi}$  diagrams with a single bond and  where the screened potential $\phi$ is replaced by its leading value
 $\phi^{(0)}$: 
$h^{(0)}_{\as\asp} =F^{\rm c  c\, {(0)}}+\Fr^{(0)}$.
(In  other words, only the graph with one bond $\Fcc$ in $\hcc$ and the graph $\Ir$ in $\htt_{\as\asp}$ where $\Ir$ is replaced by $\Fr$ do contribute at finite distances: $\hcczero_{\as\asp}= F^{\rm c  c\, {(0)}}$, 
$\hctzero_{\as\asp}=\htczero_{\as\asp}=0$, and 
$\httzero_{\as\asp}= (1/2)\left[F^{\rm c  c\, {(0)}}\right]^2+\Frt^{(0)}=\Fr^{(0)}$.)
At any finite distance $|\vecr-\vecr'|$, $h^{ (0)}_{\as\asp}$ reads
\begin{equation}
\label{valuehzerobulk}
h^{ (0)}_{\as\asp}(\vecr,\vecr')=
\theta\left(|\vecr-\vecr'|-\frac{\sigma_\as+\sigma_\asp}{2}\right)
\exp\left[- \frac{\beta e^2 }{\epsilon_{\rm solv}} 
\, Z_\as Z_\asp
\phi^{(0)}(\vecr,\vecr')\right]-1.
 \end{equation}
In the bulk the inverse screening length $\kbar$ in \eqref{Debyeeq} does not depend on $x$, $\kbar=\kd$ and $\phi^{(0)}_{\rm\scriptscriptstyle B}=\phi_{\rm\scriptscriptstyle B}$ given in \eqref{defphib}, where $\phi_{\rm\scriptscriptstyle B}$  obeys the Debye equation with the same boundary conditions as the  bare Coulomb potential $v_{\scriptscriptstyle B}$ far away from any vessel surface. Near the wall, since  the density profiles  created by interactions depend  on the coupling strength, $\kbar^2(x)$ has  an $\eps$-expansion, and so has $ \phi(x,x',y)$. In the infinite-dilution and vanishing-coupling limit, $\kbar^2(x)$ tends to $\kd^2$ and $\phi^{(0)}$ obeys Debye equation with the same boundary conditions as the bare Coulomb potential $v$, which take into account the dielectric response of the wall.

The large-distance 
 behavior 
 of $h^{}_{\as\asp}$
 at leading order, $h^{{\rm  as (0)}}_{\as\asp}$,  is equal to the large-distance behavior of
 $h^{ (0)}_{\as\asp}$, namely
 \begin{equation}
 \label{valuehbslowzero}
 h^{{\rm  as}\,(0)}_{\as\asp}(\vecr,\vecr')
 =-\frac{\beta  e^2 }{ \epsilon_{\rm solv}} 
Z_\as Z_\asp\phi^{ (0)\, {\rm as}}(\vecr,\vecr').
 \end{equation}
In other words, since $\Fr$ decays only as the square of $\Fcc$, in the diagrammatic representation $h^{{\rm  as}\,(0)}_{\as\asp}$ arises
    only from the diagram with one bond $\Fcc$, where $\phi$ is replaced by $\phi^{(0)}$. (The diagram with one bond $\Fcc$ is called ${\widetilde{\Pi}}_a$ in the following and is shown in Fig.\ref{diaga}.)
 Subsequently, the first term in the $\eps$-expansion  of 
 $\kappa$ is
 \begin{equation}
 \label{valuekappabulk0}
\kappa^{(0)}=\kd.
\end{equation}

\section{Bulk correlations}
\label{section3}

In the bulk, the Ursell function $h_{\as\asp}$ decays exponentially fast in all
directions \cite{BrydgesFederbush80}. In the high-dilution and weak-coupling regime, the leading tail at large distances is 
a monotonic exponential
decay over the  screening length $1/\kappa^{\rm {\scriptscriptstyle B}}$
(see Ref.\cite{BrydgesMartin99} for a review), while  damping might 
become oscillatory in regimes with higher densities, as
expected from various approximate theories (see e.g.  Refs.\cite{KjelMitchell94} and
\cite{LeeFisher97}). 

The resummed Meeron diagrammatic expansions used in the present
paper enable one  to retrieve the existence of an
exponential decay in the dilute regime. Indeed, all resummed Mayer diagrams $\Pi$ are built with bonds $\Fcc$
\eqref{defFcc} and
$\Fr$ \eqref{defFr}, the large-distance decays of which are ruled by the  screened
potential $\phi_{\scriptscriptstyle B}$ that is the solution of \eqref{Debyeeq} in the bulk. By virtue of \eqref{defphib}, $\phi_{\scriptscriptstyle B}$
falls off exponentially over the length scale $1/\kd$ defined in \eqref{defkd}. 
 The
monotonic exponential decay  of  $h_{\as\asp}$  over the length scale $1/\kappab$ in the dilute regime
is expected to be given by partially
resumming the tails of $\Pi$ diagrams, which decrease exponentially over the scale $1/\kd$, though the convergence of the corresponding series is
not controlled. 

Before going into details, we introduce the following definitions. Let $f(r)$ be a rotationally invariant function that decays exponentially fast at large distances $r$. Let $\kd$ be the smallest inverse screening length in the exponential tails of $f$. $f$ may contain several tails $\exp[-\kd r]/r^{\gamma}$ with various exponents $\gamma$'s, which may be negative. We define 
the slowest of the exponential tails of $f$, denoted by
$f^{\rm slow}(r)$ hereafter, as the sum of all tails $\exp(-\kd r)/r^{\gamma}$, with any exponent $\gamma$. In other words, $f^{\rm slow}(r)$
is the large-distance behavior with the largest
screening length in the exponential and all possible powers of $r$. The notation $f^{\rm as}(r)$ will be restricted to the leading tail in the large-distance behavior of $f(r)$: 
$f^{\rm as}(r)$ is the leading term in $f^{\rm slow}(r)$, namely the contribution in $f^{\rm slow}(r)$ with the smallest exponent $\gamma$. For instance, as argued in Appendix \ref{App0}, if $f=\phid\ast[\phid]^2$, 
$f^{\rm slow}(r)=a \exp(-\kd r)/r$ and $f^{\rm as}=f^{\rm slow}$, whereas, if $f=\phid\ast[\phid]^2\ast\phid$, 
$f^{\rm slow}(r)=[b+cr] \exp(-\kd r)/r$ and $f^{\rm as}=c\exp(-\kd r)$.

\subsection{Resummations of geometric series in Fourier space}
\label{section32}

The
 translational invariance in the bulk 
implies that the graph series in the decomposition \eqref{decomph}--\eqref{defhtt} of
$h_{\as\asp}$
are sums of convolutions. In Fourier space, they  become geometric series which are resummed into compact formulae.
$\hccb_{\as\asp}(k)$ merely reads
\begin{equation}
\label{resumhcc}
\hccb_{\as\asp}(k)=-
\frac{\beta  e^2}{\epsilon_{\rm solv}} Z_\as Z_\asp
\frac{\phid(k)}
{1+\phid(k)\Irbar(k)},
\end{equation}
where
\begin{equation}
\label{valuephibk}
\phid(k)=\frac{4\pi}{k^2+\kd^2}
\end{equation}
and
\begin{equation}
\label{defIrbar}
\Irbar(k)\equiv \frac{\beta  e^2}{\epsilon_{\rm solv}} 
\sum_{\gamma,\gamma'}\rogb\rogpb 
Z_\gs Z_\gsp
\Ir(k;\gamma,\gamma').
\end{equation}
$\hccb_{\as\asp}(k)$  is reduced to a fraction
\begin{equation}
\label{resumhccbis}
\hccb_{\as\asp}(k)=- \frac{\beta  e^2}{\epsilon_{\rm solv}} 
\frac{4\pi}
{k^2+\kd^2+4\pi\Irbar(k)}Z_\as Z_\asp.
\end{equation}
The same   geometric series appears in the case of $\hctb_{\as\asp}$ and 
$\httb_{\as\asp}$ with the results
\begin{equation}
\label{resumhctbis}
\hctb_{\as\asp}(k)= 
-\frac{\beta  e^2}{\epsilon_{\rm solv}} 
\frac{4\pi}
{k^2+\kd^2+4\pi\Irbar(\vk)} Z_\as\sum_{\gamma'} \rogpb Z_{\gsp}
\Ir(k;\gamma',\alpha')
\end{equation} 
and
\begin{equation}
\label{resumhttbis}
\httb_{\as\asp}(k)=\Ir(k;\alpha,\alpha')
-\frac{\beta  e^2}{\epsilon_{\rm solv}}
\frac{4\pi}{k^2+\kd^2+4\pi\Irbar(k)} 
\sum_{\gamma} \rogb Z_\gs \Ir(k;\gamma,\alpha)
\sum_{\gamma'} \rogpb Z_\gsp\Ir(k;\gamma',\alpha').
\end{equation}

Finally, $h^{\rm{\scriptscriptstyle B}}_{\as\asp}(k)$ takes the half factorized form
\begin{equation}
\label{resumh}
h^{\rm{\scriptscriptstyle B}}_{\as\asp}(k)=\Ir(k;\alpha,\alpha')
-\frac{\beta  e^2}{\epsilon_{\rm solv}}\,
\frac{4\pi}{k^2+\kd^2+4\pi\Irbar(k)} 
\left[ Z_{\as}+\sum_{\gamma} \rogb Z_\gs \Ir(k;\gamma,\alpha)\right]
\left[ Z_\asp +\sum_{\gamma'} \rogpb Z_\gsp\Ir(k;\gamma',\alpha')\right].
\end{equation}
From this expression we readily get that $\soma Z_{\as}\roab
h^{\rm{\scriptscriptstyle B}}_{\as\asp}(k=0)=-Z_{\asp}$, which is another writing of the internal-screening sum rule \eqref{sumruleint} in the bulk case. (We also notice that the writing of $h^{\rm{\scriptscriptstyle B}}_{\as\asp}(k)$ in \eqref{resumh} is analogous to Eq.(2.110) of Ref.\cite{BrownYaffe01}, where the authors consider the
Feynman
diagrammatics for a field theory, with some short-distance regularization, which modelizes a charge fluid.)

\subsection{Large-distance behavior of bulk correlations}
\label{section33}

When the bulk Ursell function  $h_{\as\asp}(\vecr,\vecr')$ is only a 
function of
$|\vecr-\vecr'|$ and decays faster than any inverse power of $|\vecr-\vecr'|$
when $|\vecr-\vecr'|$ becomes infinite, its large-distance behavior
$h_{\as\asp}^{\rm as}(\vecr,\vecr')$ is determined by the  general formulas recalled in Appendix \ref{App0}.
As checked at first order in $\eps$ in
next Section, 
 the singular points of $\Ir(k;\gamma,\gamma')$ in the weak-coupling regime
 are more distant from 
the real axis 
in the upper complex 
 half-plane $k=k'+ik"$ than the pole $k_0$ of the fraction  
 $1/[k^2+\kd^2+\Irbar(k)]$ that has the smallest positive imaginary part. Moreover,
 $k_0$ is purely imaginary,  
 $k_0=i\kappab$, and $k_0$ is a pole of rank $1$.  

Therefore, in the weak-coupling regime,
 the slowest exponential tail of $h^{\rm{ \scriptscriptstyle B}}_{\as\asp}$
is a purely $\exp(-\kappab r)/r$ function, as well as the slowest exponential tails of
$\hccb_{\as\asp}$, $\htcb_{\as\asp}$, $\hctb_{\as\asp}$ and 
$\httb_{\as\asp}$. 
By inserting    the property 
\begin{equation}
\label{residufraction}
\left.{\rm Res}\left[\frac{1 }{k^2+\kd^2+4\pi\Irbar(k)}\right]
\right\vert_{k=k_{0}} =
\left[2k_{0}+4\pi\left. \frac{\partial
\Irbar(k)}{\partial k}\right\vert_{k_{0}}\right]^{-1}
\end{equation} 
into the general formula \eqref{fslow} applied to
$f=h_{\as\asp}$ given in \eqref{resumh} with
$k_{0}=i\kappab$, we find that the
large-distance behavior $h^{\rm{ \scriptscriptstyle B}\, as}_{\as\asp}(r)$ 
of $h^{\rm{ \scriptscriptstyle B}}_{\as\asp}(r)$ 
 takes the form  
 \begin{equation}
\label{hasbulk}
h_{\as\asp}^{\rm {\scriptscriptstyle B} \,as}(r)
=-\frac{\beta e^2}{\epsilon_{\rm solv}}
\Zeffba
\Zeffbap
\frac{ e^{- \kappab r}}{r},
\end{equation}
where $i\kappab$ is the pole of $1/[k^2+\kd^2+\Irbar(k)]$ with the positive imaginary part, and
\begin{equation}
\label{slowfund}
\Zeffba=
\frac{Z_\as+\sum_{\gamma} \rogb Z_\gs \Ir(i\kappab;\gamma,\alpha)}
{\sqrt{1-i(2\pi/\kappab)\partial\Irbar(k)/\partial k\vert_{i\kappab}}}.
\end{equation}

\subsection{Large-distance tail at order $\eps$}
\label{section44}

According to the scaling analysis of  Section \ref{section42},
the first term  in the $\eps$-expansion of $\Ir(k) $
is $\Ir^{(1)}(k)$ with $\Ir^{(1)}=[\Fcc]^2/2$. 
$\Irbar^{(1)}$ is calculated from the  definition
\eqref{defIrbar} of $\Irbar(k)$ with $\Ir^{(1)}$ in place of $\Ir$.
By using 
\begin{equation}
\label{valuephib2k}
\left[\phidd\right](\vk)=\frac{4\pi}{k}\arctan\left(\frac{k}{2\kd}\right)
\end{equation}
and
\begin{equation}
\frac{1}{2}\left(\frac{\beta e^2}{\epsilon_{\rm solv}}\right)^3
\left(4\pi\sum_{\gamma} \rogb Z_{\gs}^3\right)^2= \kd^3 \eps
 \left(\frac{\Sigma_3}{\Sigma_2}\right)^2,
\end{equation} 
where $\Sigma_m$ is defined in \eqref{defSigma}, we get
\begin{equation}
\label{valueIrbar1}
4\pi \Irbar^{(1)}(\vk)= \eps\, \kd^2 \left(\frac{\Sigma_3}{\Sigma_2}\right)^2
\frac{\kd}{k}\arctan\left( \frac{k}{2\kd}\right).
\end{equation}
$\Ir^{(1)}$ and $\Irbar^{(1)}(k)$  have a branch point at $k=2i\kd$, while  
 $\left[k^2+\kd^2+4\pi\Irbar^{(1)}(k)\right]^{-1}$ has a pole at the  value of $k$ equal to
\begin{equation}
i \kd\left[ 1+ \frac{2\pi}{\kd^2}\Irbar^{(1)}(i\kd)+o(\eps) \right].
\end{equation}
The leading corrections involved in the notation $o(\eps)$ are given in \eqref{orders}. 
The latter pole is closer to the real axis than the branch point at $k=i2\kd$.
 Therefore, at first order in
 $\eps$, the singular 
 point in   $h^{ {\scriptscriptstyle B}}_{\as\asp}(k)$ 
  that is the
 closest one to the real axis  in the upper complex half-plane of $k$ is 
 the pole of $1/[k^2+\kd^2+4\pi\Irbar^{(1)}(k)]$.

The scenario of Section \ref{section33}
 does happen at leading order in 
$\eps$ and the
large-distance behavior  $h_{\as\asp}^{ \rm {\scriptscriptstyle B}\, as}(r)$ 
up to order $\eps$  takes the form  \eqref{hasbulk} where $k_0=i[\kd+\kbun]$ with
 \begin{equation}
\label{renormkappa1}
\kbun=\frac{2\pi}{\kd}\Irbar^{(1)}(i\kd). 
\end{equation}
According to \eqref{valueIrbar1} we find
\begin{equation}
\label{valuecorkappabulk}
\frac{\kbun}{\kd}=\eps 
\left(\frac{\Sigma_3}{\Sigma_2}\right)^2\frac{\ln 3}{4}.
\end{equation}
We retrieve the formula of 
Ref.\cite{KjelMitchell94} obtained from integral
equations. It  is reduced to the results obtained by Mitchell and Ninham
through  diagrammatic techniques for the one-component plasma 
\cite{MitchellNinh68} or for a two-component 
electrolyte
\cite{MitchellNinh78}. (The formulae for the one-component plasma  can be
derived from those calculated for a two-component plasma with charges $e_+$ and
$e_-$ and densities $\rho_+$ and $\rho_-$ by taking the limit where  $e_-$
vanishes while $\rho_-$ diverges under the constraint $e_-\rho_-=-e_+\rho_+$.)
The correction $\kbun$ vanishes in the case of a 1:1 electrolyte. If the
electrolyte is not charge symmetric, 
the expression \eqref{valuecorkappabulk} shows that the screening length
$1/\kappa^{\rm {\scriptscriptstyle B}}$ is  a decreasing function of the coupling parameter
$\eps$ at first order in $\eps$.

The bulk effective charge $\Zeffba$ up to order $\eps$ is
 calculated 
by formula \eqref{slowfund}. The explicit result is written in 
\eqref{valueZeff}. In view of the discussion of next Section, we rewrite $\Zeffba$ as
\begin{equation}
\label{valuehbslow}
 \Zeffba=Z_\as[1+A_\as^{(1)}+o(\eps)].
 \end{equation} 
 The amplitude \eqref{hasbulk} of $h_{\as\asp}^{ \rm {\scriptscriptstyle B}\, as}(r)$ up to
 order $\eps$ can be rewritten as
\begin{equation}
\label{reecrit}
h_{\as\asp}^{ \rm {\scriptscriptstyle B} \, as}(r)
=- \frac{\beta  e^2}{\epsilon_{\rm solv}}
\frac{Z_\as Z_\asp}{r}\left\{ \left[1+  A_\as^{(1)}+ A_\asp^{(1)}\right]
e^{-(\kd+\kbun) r} +o(\eps) \right\}.
\end{equation}
To our knowledge,
the amplitude of $h_{\as\asp}^{ \rm {\scriptscriptstyle B}\, as}(r)$ for a
multicomponent plasma has not been calculated in the litterature previously.
In the limit of the one-component plasma, it is  the same as that
found by Mitchell and Ninham in Ref.\cite{MitchellNinh68} by a  diagrammatic method.

We notice that the use of \eqref{slowfund} with $\Irbar^{(1)}$ in place of
$\Irbar$ is equivalent to replacing the diagrammatic series of 
$\hccb_{\as\asp}$, $\hctb_{\as\asp}$  and
 $\httb_{\as\asp}$
shown in Figs.\ref{hcc}-\ref{htt} by the corresponding series represented in 
Figs.\ref{hccun}-\ref{httun}. As shown in Appendix \ref{AppA},
the correction $\kbun$ to the screening length in $\hccb$, $\hctb$,  and $\httb$ arises from the whole series  in Figs.\ref{hccun}, \ref{hctun}, and \ref{httun}, respectively. 

On the contrary, the first corrections to the effective bulk charges can be seen as arising from only a finite number of diagrams in Figs.\ref{hccun} and \ref{hctun}. This will be shown in next section. The property relies on the following rewriting of
$h_{\as\asp}^{ \rm {\scriptscriptstyle B} \, as}(r)$,
  \begin{equation}
 \label{diageps1}
h^{\rm {\scriptscriptstyle B} \,as}_{\as\asp}(r)
=-\frac{\beta  e^2 }{ \epsilon_{\rm solv}} \frac{Z_\as Z_\asp}{r}
\left\{
 \left[1+  A^{(1)}_{\as} + A^{(1)}_{\asp} -\kbun \, r\right]e^{-\kd r}
  +o(\eps)
 \right\}.
 \end{equation}
 We point out that \eqref{diageps1} is
 valid for any distance $r$. Indeed
 $\exp[-(\kd+\kbun)r]= \left(1-\kbun \, r\right)\exp(-\kd r)+o(\eps)$ for any
 distance, whereas
 $\exp[-(\kd+\kbun)r]= \left[1-\kbun \, r+o(\eps)\right]\exp(-\kd r)$ only for distances
 $r<L_{\rm max}\equiv \xid/\eps^\nu$ with $\nu<1/2$.
 The condition $\nu<1/2$ ensures that for $r<L_{\rm max}$ 
every $n^{\rm th}$ term with $n\geq 2$ in
the expansion of the exponential
 $\exp[-\kbun r]$ is indeed a correction, of order 
 $\eps^{n(1-\nu)}=o(\eps)$,  with respect to  the $\cO(\eps^{1-\nu})$-term $\kbun r$.

\subsection{Alternative  derivation of the bulk large-distance tail at order $\eps$}
\label{section45} 

In view of calculations in the vicinity of a wall, where the  infinite series 
in $f=\hccb$, $\hctb$ or $\httb$ can no longer be
resummed in Fourier space, because of the lost of translational invariance in the direction perpendicular to the wall,
we show how to retrieve the expression \eqref{reecrit} for 
$h^{\rm {\scriptscriptstyle B} \,as}_{\as\asp}$ in a less systematic
way than the method involving the resummed formulae \eqref{hasbulk} and \eqref{slowfund}. (For the sake of simplicity we omit the indices $\alpha$ and $\alpha'$ for
charge species  in the notation $f$.)

In the general method of Sections \ref{section32} and \ref{section33},
we performed  
Fourier transforms, then  we resummed the four infinite 
series $f=\hccb$, $\hctb$,  
$\htcb$,   or $\httb$
that define $h^{\rm {\scriptscriptstyle B}}_{\as\asp}$ through the graphs  shown in
Figs.\ref{hcc}-\ref{htt}, and  we got a compact formula for $h^{\rm {\scriptscriptstyle B}}_{\as\asp}(k)$ from which we calculated the  
large-distance  behavior $h^{\rm {\scriptscriptstyle B} \,as}_{\as\asp}(r)$ of 
$h^{\rm {\scriptscriptstyle B} }_{\as\asp}(r)$.  Here, on the contrary,  in each
series $f$  we formally calculate the  slowest exponential tail 
$f_m^{{\rm slow}}(r)$ of every
graph $f_m$, with $m$ bonds $\Fcc$, directly in position space 
by using the residue theorem, and  
the large-distance behavior $f^{\rm as}$ of
$f=\hcc$, $\hct$, $\htc$, or $\htt$ is given by  the sum (over $m$) of the slowest 
tails $f_m^{{\rm slow}}(r)$'s in each case.
(The slowest exponential tail is defined in the introduction of
Section \ref{section3}.)

The second procedure is more  cumbersome, because the series sums $f(k)$'s  have a pole of rank $1$ at $k=i\kappab$ and their inverse Fourier transform decay as $\exp(-\kappab r)/r$, whereas each term $f_m(k)$ in the series has a multiple pole of rank $m$ at $k=i\kd$ and its inverse Fourier transform behaves as $\exp(-\kd r)/r$ times a polynomial in $r$ of degree $m-1$.
We point out that, in the present  $\eps$-expansion 
of $f^{\rm as}(r)$ around its $\exp[-\kd r]/r$ limit 
 behavior  in the infinite-dilution and vanishing-coupling
limit (where only the bond $\Fcc$ contributes),
 for every graph $f_m$ we must retain the entire slowest tail 
$f_m^{\rm slow}$ -- namely the entire polynomial in $r$ --, and 
we only disregard 
tails $\exp[-l \kd r]/r$ with $l\geq 2$. (See example in \eqref{exemple2}.) The procedure  is legitimate as long as dilution is sufficiently high. 
 Details are given in Appendix \ref{AppA} and we 
give only a summary in  the present section.

\subsubsection{General structure of the $\eps$-expansion of $h^{\rm as}_{\as\asp}$}

As long as densities are low enough, the graph $I$
decays faster than the bond $\Fcc$, and, as shown in Appendix \ref{AppA}, the slowest tail
$f_m^{{\rm slow}}(r)$ of $f_m(r)$ is equal to $\exp[-\kd r]/r$ times a polynomial
in $r$ of rank $m-1$, $\sum_{p=0}^{m-1} F_{m,p} r^p$.
As a consequence, the
$\eps$-expansion of the large-distance behavior of
 $h_{\as\asp}^{\scriptscriptstyle B}$ reads
\begin{equation}
\label{devtail}
h_{\as\asp}^{\rm {\scriptscriptstyle B}\, as}(r)=\frac{e^{-\kd r}}{r}\sum_{p=0}^{+\infty}
 r^p\, H_p (\alpha,\alpha';\eps),
\end{equation}
 where the coefficient $H_p(\alpha,\alpha';\eps)$ -- which is the 
 the sum of the
contributions from $\hccb$, $\hctb$, $\htcb$, and $\httb$ -- arises only from the
graphs $f_m$ with
 $m\geq p+1$ in
 the series representations  shown in
Figs.\ref{hcc}-\ref{htt}. (We recall that $\eps$ is a short notation for parameters $\eps$ and $\sigma/a$ in the scaling regimes \eqref{case1} and \eqref{subcase2}, as explained at the end of Section \ref{sectionregime}.)

 Moreover, according to the
scaling analysis  for $\eps$-expansions in Section \ref{section42},   all coefficients 
$F_{m,p}$ in the polynomial in the slowest tail 
of every graph $f_m$ with $m\geq 1$ start at order
$\eps^{n_0+m-1}$, where $n_0=0$ if $f=\hccb$, $n_0=1$ 
 if $f=\hctb$ or $\htcb$, and $n_0=2$ if 
 $f=\httb$. 
 Therefore, the $\eps$-expansion of $H_p(\alpha,\alpha';\eps)$ starts 
 at order
 $\eps^{p}$, and, after reversing the summation orders,
 \begin{equation}
\label{devepshas}
h_{\as\asp}^{\rm {\scriptscriptstyle B}\, as}(r)=\frac{e^{-\kd r}}{r}\sum_{q=0}^{+\infty} \eps^q
\sum_{p=0}^{q}  r^p\,H_p^{(q)} (\alpha,\alpha')
\end{equation}
with
 \begin{equation}
\label{devepshasbis}
H_q^{(q)} (\alpha,\alpha')=F^{ (q)}_{q+1,q}\left( f=\hccb_{\as\asp}\right).
\end{equation}
\eqref{devepshas} displays that, in the $\eps$-expansion of 
$h_{\as\asp}^{{\scriptscriptstyle B}\,{\rm as}}(r)$ around 
its $\exp[-\kd r]/r$ behavior in the limit where
$\eps$ vanishes, 
the leading tail of $h_{\as\asp}^{\rm {\scriptscriptstyle B}\,as}(r)$ at order $\eps^q$ behaves as $ r^q$ times 
$\exp[-\kd r]/r$. Moreover, by virtue of \eqref{devepshasbis},  it coincides with
 the first term  in the $\eps$-expansion  of  the 
leading tail in the slowest exponential decay $f_{q+1}^{{\rm slow}}(r)$ of the graph $f_{q+1}$ with $(q+1)$  bonds $\Fcc$ in $f=\hccb_{\as\asp}$.

\subsubsection{Renormalization of the screening length}
\label{section451}

  As shown in Appendix \ref{AppA}, for each $f=\hccb$, $\hctb$, $\htcb$, or $\httb$,
 the sum  $f^{{\rm as}\, \star}(r)$ of the 
  leading  tails 
at every order  $\eps^{n_0+q}$
 in the $\eps$-expansion of $f^{\rm as}(r)$ around its lowest-order 
$\eps^{n_0}\exp[-\kd r]/r$ limit can be performed explicitly. 
(Indeed,  the  coefficient $a_q^{(n_0+q)}\eps^{n_0+q}$ of the   leading $r^{q}\exp[-\kd r]/r$ tail
 at order $\eps^{n_0+q}$ in the $\eps$-expansion of $f^{\rm as}$ coincides with the first term  in the $\eps$-expansion
of the coefficient $F_{q+1,q}$
 of the  leading $r^{q}\exp[-\kd r]/r$ term in the slowest exponential decay $f_{q+1}^{{\rm slow}}(r)$ of the graph $f_{q+1}$ with $(q+1)$  bonds $\Fcc$, and the formal expression  of  $F_{q+1,q}$ in terms of $\Ir$ is given in \eqref{valueA}.)
$f^{{\rm as}\, \star}(r)$
proves to be
equal to $\exp[-\delta \kappa^{\star}_{\rm{\scriptscriptstyle B}}\,r]$ times 
$f^{{\rm as}(n_0)}_1$, the value at the first order
$\eps^{n_0}$ of
the large-distance behavior of the graph $f_1$ with only one bond $\Fcc$. As a consequence,
the sum $h^{\rm{ {\scriptscriptstyle B}\,as}\, \star}_{\as\asp}(r)$ of the 
  leading $ r^q \exp[-\kd r]/r$ tails 
at every order  $\eps^q$
 in the $\eps$-expansion of $h^{\rm{ {\scriptscriptstyle B}\,as}}_{\as\asp}(r)$ around its lowest-order 
$\exp[-\kd r]/r$ limit, namely
\begin{equation}
h^{\rm{ {\scriptscriptstyle B}\,as}\, \star}_{\as\asp}(r)\equiv\sum_{q=0}^{+\infty}
H_q^{(q)} (\alpha,\alpha')\eps^{q}r^q \frac{e^{-\kd r}}{ r}, 
\end{equation} 
reads
\begin{equation}
\label{relhstarh0}
h^{\rm{ {\scriptscriptstyle B}\,as}\, \star}_{\as\asp}(r)= 
 F^{\rm c  c\, {(0)}}(r)\,
e^{-\delta\kappa^{\star}_{\rm{\scriptscriptstyle B}}\,r}.
\end{equation}
It arises from the    leading $\eps^q r^{q}\exp[-\kd r]/r$ tails of $\hccb_{\as\asp}$ only.

Moreover,
  $\delta\kappa^{\star}_{\rm{\scriptscriptstyle B}}$ coincides with the first-order correction 
$\kbun$  to the bulk screening length
(see \eqref{renormkappa1}) calculated from the exact procedure  of
Section \ref{section33},
 \begin{equation}
\delta\kappa^{\star}_{\rm{\scriptscriptstyle B}}=\kbun.
\end{equation} 
 Eventually, the resummation of the  series of leading 
 tails at every order in $\eps$  for
$\hccb$
proves to be   a way to retrieve the value of $\kbun$.

\subsubsection{ Diagrams with slowest exponential tails of order $\eps$}
\label{section452}
 
As already seen in Section \ref{section43}, diagram ${\widetilde \Pi}_a$ 
in Fig.\ref{diaga} is the only diagram
whose slowest exponential tail (proportional to $\exp(-\kd r)/r$)  has an amplitude of order $\eps^0\times \beta e^2 /\epsilon_{\rm
solv}$.
Diagrams whose slowest exponential tails have  amplitudes  of order $\eps$ are diagrams ${\widetilde \Pi}_b$,
 ${\widetilde \Pi}_{b^{\star}}$ and ${\widetilde \Pi}_c$ shown in Figs.\ref{diagbbe} and \ref{diagc} (which come from the series $\hct$, $\htc$ and $\hcc$, respectively, where $\Ir$ is replaced by $\Ir^{(1)}$, as shown in Figs.\ref{hccun} and \ref{hctun}). 
 The contribution of ${\widetilde \Pi}_b$ to
 $h_{\as\asp}^{ \rm {\scriptscriptstyle B} \,as}(r)$ reads
 \begin{equation}
 \label{tailPib}
 - \frac{\beta  e^2}{\epsilon_{\rm solv}}Z_\as Z_\asp^2
 \eps \frac{\Sigma_3}{\Sigma_2}\frac{\ln 3}{2}\frac{e^{-\kd r}}{r},
 \end{equation}
 while the contribution of  ${\widetilde \Pi}_c$ is
 \begin{equation}
 \label{tailPic}
 - \frac{\beta  e^2}{\epsilon_{\rm solv}}Z_\as Z_\asp
 \left[ -\eps
 \left(\frac{\Sigma_3}{\Sigma_2}\right)^2 \frac{\ln 3}{4} \, \kd r
 +2\eps 
  \left(\frac{\Sigma_3}{\Sigma_2}\right)^2
  \left(\frac{1}{6} - \frac{\ln 3}{8}\right)
 \right]
 \frac{e^{-\kd r}}{r}.
 \end{equation}
 Indeed, the contribution of diagram  ${\widetilde \Pi}_{b}$ 
 is proportional to the  convolution $\phid\ast[\phid]^2$ calculated in
 \eqref{exemple1}, while the  
  diagram ${\widetilde{\Pi}}_c$ involves the convolution
 $\phid\ast[\phid]^2\ast\phid$, whose expression at any distance is given by 
  \eqref{exemple2}. We notice that the
  diagrams ${\widetilde \Pi}_b$,
 ${\widetilde \Pi}_{b^{\star}}$ and ${\widetilde \Pi}_c$
 have already been  calculated in the case of the electron gas 
 \cite{MitchellNinh68}.

By virtue of \eqref{tailPib} and \eqref{tailPic},
 the sum of the slowest  exponential tails of  diagrams ${\widetilde \Pi}_b$,
 ${\widetilde \Pi}_{b^{\star}}$ and ${\widetilde \Pi}_c$
coincides with the expression \eqref{diageps1} of 
$h^{\rm {\scriptscriptstyle B} \,as }_{\as\asp}$ up to order $\eps$,
 where  $\kbun$ is given 
 in \eqref{valuecorkappabulk} and
\begin{equation}
 \label{valueAun}
A^{(1)}_{\as} =Z_\as {\overline A}^{(1)}_{[b]} + A^{(1)}_{[c]}
 \end{equation} 
with
\begin{equation}
 \label{defA1b}
{\overline A}^{(1)}_{[b]}= \eps \frac{\Sigma_3}{\Sigma_2}\frac{\ln 3}{2}
 \end{equation}
and 
 \begin{equation}
 \label{defA1c}
A^{(1)}_{[c]}=\eps 
  \left(\frac{\Sigma_3}{\Sigma_2}\right)^2
  \left(\frac{1}{6} - \frac{\ln 3}{8}\right).
 \end{equation}
The $-\kbun \, r$ term in \eqref{diageps1} comes from diagram ${\widetilde{\Pi}}_c$. 
 The term $Z_\as {\overline A}^{(1)}_{[b]}$ 
($Z_\asp {\overline A}^{(1)}_{[b]}$) proportional to $Z_{\as}$ ($Z_{\asp}$) in $A^{(1)}_{\as}$ comes 
from ${\widetilde{\Pi}}_b$ (${\widetilde{\Pi}}_{b^{\star}}$), whereas  the
 other term, $A^{(1)}_{[c]}$, arises from ${\widetilde{\Pi}}_c$.

 Therefore, the constants $A^{(1)}_{\as}$ and $A^{(1)}_{\asp}$ are
 determined by the exponential tails of only three diagrams 
  ${\widetilde \Pi}_b$,
 ${\widetilde \Pi}_{b^{\star}}$, and ${\widetilde \Pi}_c$. (See the comment after
 \eqref{diageps1}  for a comparison with the exact method of
 Section \ref{section44}.) 
 Moreover, as a
 consequence of the analysis summarized in Section \ref{section451},
  the coefficient of the $r$-term in
 the amplitude of the slowest tail of
  diagram ${\widetilde{\Pi}}_c$ with
two 
$\Fcc$ bonds (See Fig.\ref{diagc}) must coincide with the opposite of the first-order correction $\kbun$ in the inverse screening length.

\subsection{Density-density and charge-charge correlation}
\label{densitdensit}

By virtue of the bulk local charge neutrality, 
\begin{equation}
\label{localneut}
\sum_{\alpha}e_\as\roab=0,
\end{equation}
the Fourier transform of the density-density 
correlation function
 takes the form 
\begin{equation}
\label{defC}
\sum_{\alpha,\alpha'}\roab \roapb 
h^{\rm{ \scriptscriptstyle B}}_{\as\asp}(k)=
\sum_{\alpha,\alpha'}\roab \roapb\Ir(k;\alpha,\alpha')
\\ -\frac{\beta  e^2}{\epsilon_{\rm solv}}
\frac{4\pi}{k^2+\kd^2+4\pi\Irbar(k)} 
\left[\sum_{\alpha} \roab\sum_{\gamma} \rogb Z_\gs
\Ir(k;\gamma,\alpha)\right]^2,
\end{equation}
while the charge-charge structure factor, defined from \eqref{defCgen}, 
\begin{equation}
\label{valueCB}
C^{\rm{ \scriptscriptstyle B}}(k)\equiv  e^2
\left\{ \sum_{\alpha} \roab Z_\as^2+
\sum_{\alpha,\alpha'}\roab \roapb Z_\as Z_\asp h_{\as\asp}(k)\right\},
\end{equation}
reads
\begin{equation}
\label{expCB}
C^{\rm{ \scriptscriptstyle B}}(k)=\frac{\epsilon_{\rm solv}}{4\pi\beta}
\left\{\kd^2+4 \pi \Irbar(k)
-\frac{\left[\kd^2+4\pi\Irbar(k)\right]^2}
{k^2+\kd^2+4\pi\Irbar(k)}
\right\}.
\end{equation}

As announced in Subsection \ref{genscreenrule}, the expression
\eqref{expCB}
of the charge-charge
structure factor $C^{\rm{ \scriptscriptstyle B}}(k)$ indeed obeys the sum rule \eqref{StilLovrule}, which summarizes both the internal-screening sum rule \eqref{sum1} and the external-screening sum rule \eqref{sum2}. (If a phase transition gave rise to nonintegrable  algebraic tails in $\Irbar(r)$ and subsequent non-analytic terms of order $k^{\eta}$ with $\eta\leq 0$ in the Fourier transform $\Irbar(k)$, then $C^{\rm{ \scriptscriptstyle B}}(k)$ would still vanish at $k=0$, but the coefficient of the $k^2$-term would be different from the universal value in \eqref{StilLovrule}, as exhibited in the exactly-soluble spherical model of Ref.\cite{AquaFisher04}.) 

We stress that the $k^2$-term in the Fourier expansion of $C^{\rm{ \scriptscriptstyle B}}(k)$  is independent of the short-range potential \hfill\break
$\vsr(\vert \vecr-\vecr\vert;\alpha,\alpha') $, which must be introduced in three dimensions in order to avoid the collapse  under the attraction between charges with opposite signs. This property is a consequence of the internal screening rule \cite{revueMartin}, and it is retrieved from the structure of expression \eqref{expCB}. On the contrary, the
$k^2$-term in $\sum_{\alpha,\alpha'}\roab \roapb 
h^{\rm{ \scriptscriptstyle B}}_{\as\asp}(k)$ given in \eqref{defC} has not any universal value: it depends on the short-distance repulsion in the generic case. However, this is not the case for a symmetric 1:1 electrolyte in two-dimensions
\cite{Janco00,Jancoetal00}, where the pure logarithmic Coulomb interaction needs not be regularized at short distances and is scale-invariant. Then, for point charges,  
scale-invariance arguments lead to a value of the dimensionless second-moment of $\sum_{\alpha,\alpha'}\roab \roapb 
h^{\rm{ \scriptscriptstyle B}}_{\as\asp}(r)$ that depends only on the coupling parameter $\beta e^2$.
We also notice that formulae \eqref{defC}
 and  \eqref{expCB} enable one to retrieve the leading low-density values of the  coefficients of the $k^2$ and $k^4$ terms in the Fourier transforms of the density-density correlation and of the charge-charge structure factor derived for a symmetric 1:1  electrolyte  in Ref.\cite{BekiranovFisher99}.

 When there is no charge symmetry in the composition of the electrolyte,  
the same argument as that used  in Section \ref{section33} implies that, according to
\eqref{defC} and \eqref{expCB}, 
 the large-distance behaviors of the density-density and charge-charge
correlations in the high-dilution and  weak-coupling 
regime are determined by the zero 
$k_0=i\kappab$ of $k^2+\kd^2+\Irbar(k)$ : they decay over the same screening
length as the  correlation $h^{\rm{ \scriptscriptstyle B}}_{\as\asp}$.

In the case of a symmetric electrolyte  made of two species  with
opposite charges $+Ze$ and $-Ze$ and with the same radii, 
$\sum_{\alpha} \roab\sum_{\gamma} \rogb Z_\gs \Ir(\vk;\gamma,\alpha)$ vanishes by
virtue of the local neutrality \eqref{localneut} and of the symmetries ($\Ir(\vk;++)=\Ir(\vk;--)$ and 
$\Ir(\vk;+-)=\Ir(\vk;-+)$). As a consequence,
$\sum_{\alpha}\roab  h_{\as\asp}(r)$ and $\sum_{\alpha,\alpha'}\roab \roapb h_{\as\asp}(r)$ decay as
$\Ir(r;\alpha,\alpha')$ by virtue of \eqref{defC}.  $\Ir(r;\alpha,\alpha')$ is expected to decay over 
the length $1/(2\kappab)$, 
by analogy with the infinite-dilution and  vanishing-coupling  limit where it behaves  as the diagram 
$[\Fcc]^2/2$, which  falls off
over the scale $1/(2\kd)$. Therefore, in this peculiar case,  the ``{screening}'' length of the 
density-density correlation is expected to be $1/(2 \kappab)$ at low density, in agreement with
the result of Ref.\cite{BekiranovFisher99}.

\section{Screened potential along the wall}
\label{section5}

\subsection{ Formal expression of the screened potential}
\label{section51}

Near the plane dielectric wall  located at $x=0$, interactions create density profiles and  \eqref{Debyeeq} is
an inhomogeneous Debye equation, where the  inverse squared screening length
$\kbar^{2}$ depends on the distance $x$ to the wall. Moreover,
$\phi ( \vecr, \vecr')$
obeys the same boundary conditions as $v(\vecr,\vecr')$ (defined after \eqref{defvsr}), 
\begin{equation}
\label{boundcond1}
\lim_{x\rightarrow 0^{-}} 
\frac{\epsilon_{\rm \scriptscriptstyle W}}
{\epsilon_{\rm solv}}
\frac{\partial \phi}{\partial
  x} (\vecr, \vecr')
 =\lim_{x\rightarrow 0^{+}} 
\frac{\partial \phi}{\partial x} (\vecr, \vecr') 
\end{equation}
and
\begin{equation}
\label{boundcond2}
\lim_{x\rightarrow b^{-}} 
\frac{\partial \phi}{\partial x} (\vecr, \vecr')
 =\lim_{x\rightarrow b^{+}} 
\frac{\partial \phi}{\partial x} (\vecr, \vecr'),
\end{equation}
since particles are  made of a material with the same dielectric constant as the solvent.
In order to take advantage of the invariance along directions parallel to the
wall, we introduce the Fourier transform with respect to the $\vy$ variable,
and we write
\begin{equation} 
  \label{}
  \phi (x,x',\vy) = \kd \int \frac{d^2\vq}{(2 \pi)^2} \, e^{- i \vq
  . (\kd \vy)} \phit (\kd x, \kd x'; \vq).
\end{equation}
In the following, the tilde index denotes dimensionless quantities, such as the
Fourier transform $\phit (\kd x, \kd x'; \vq)$ and  the dimensionless
 coordinate 
$\xt\equiv \kd x$.

 The solution of the
inhomogeneous Debye equation \eqref{Debyeeq} requires one to distinguish only 
three
regions: region I for $x<0$, region II for $0<x<b$ and region III for $b<x$.
In  regions I and II, $\kbar(x)$  vanishes by virtue of \eqref{defkbar}. According to  \eqref{Debyeeq}, the dimensionless Fourier transform 
$\phit(\xt, \xt'; \vq)$ obeys a  one-dimensional
differential equation. When $x'>b$ it reads
\begin{equation}
\label{equadiffuu}
\left\{ \frac{\partial^2}{\partial \xt^2} - \vq^2
  \right\} \phit (\xt, \xt'; \vq)= 0 \quad \mbox{if $\xt  <\bt$}.
\end{equation}
The solution with boundary conditions \eqref{boundcond1} and \eqref{boundcond2}
is
\begin{equation}
\label{phijtild}
\phit (\xt,\xt',\modq )= \left\{ 
\begin{array}{ll}
B (\xt',\modq) \left( 1 - \dw \right) e^{\modq \xt} 
& \mbox{ if}\quad \mbox{ $ \xt <0$},\\
B (\xt',\modq) \left[ e^{\modq \xt} - \dw e^{-\modq \xt} \right]
& \mbox{ if $ 0< \xt < \bt$}.
\end{array} 
\right.
\end{equation}
(A similar equation is solved in Ref.\cite{AquaCor01I} with a misprint in
Eq.(4.20).)

In region III, when $x$ goes to $+\infty$,
 $\kbar(x)$ tends to the inverse Debye length
$\kd^{ -1}$ \eqref{defkd},  and we
rewrite the Fourier transform of \eqref{Debyeeq} as 
\begin{equation}
  \label{equadiffuq}
  \left\{ \frac{\partial^2}{\partial \xt^2} - \left( 1 + \vq^2 \right) -U(\xt)
  \right\} \phit (\xt, \xt'; \vq)= - 4
  \pi \delta ( \xt - \xt') \quad \mbox{if $\xt  > \bt$}
\end{equation}
with
\begin{equation}
  \label{defU}
 U(\xt)\equiv  \frac{4\pi\beta e^2}{\epsilon_{\rm solv}\kd^2 }\soma 
 Z_\as^2 [\roa(x)-\roab].
\end{equation}
The solution of the one-dimensional equation \eqref{equadiffuq} can be written
in terms of the solutions $h$ of the associated ``{homogeneous}'' equation (with a zero in place of 
the Dirac distribution)  which is valid for $-\infty<x<+\infty$. Indeed, 
 the general solution of \eqref{equadiffuq} for $x>b$ and $x'>b$  is the following sum: a
linear combination of two 
independent solutions $h^+$ and $h^-$  plus a particular solution $\phi_{{\rm sing} }$ of
\eqref{equadiffuq}, which is singular when $x=x'$ and which is  
calculated in terms of  $h^+$ and $h^-$ by the so-called Wronskian method \cite{Zwillinger}.
In the following, $h^+$ ($h^-$) is chosen to be  a solution which  vanishes (diverges) when $x$
tends to $+\infty$.
In the bulk, $\kbar(x)$ is a constant equal to the inverse Debye length
$\kd$: $h^+$ and $h^-$ can be chosen to be equal to
$\exp[\mp x\rac]$.
When $\kbar$ depends on $x$, we look for the solutions $h^+$ and $h^-$
in terms of the bulk solutions as
\begin{equation}
\label{defH}
e^{\mp\xt\rac} [1+H^{\pm} (\xt, \vq^2)].
\end{equation}
Moreover, the particular solutions $H^+$ and $H^-$ can be chosen to vanish at $\xt=\bt$.
As shown in Ref.\cite{AquaCor01II}, when $\xt>b$ and $\xt'>b$
\begin{equation}
\label{structphi}
\phit(\xt,\xt',\vq)= \phit_{\rm sing}(\xt,\xt',\vq^2)
  + Z(\modq) 
e^{-(\xt+\xt')\rac}[1+H^+(\xt,\vq^2)][1+H^+(\xt',\vq^2)]
\end{equation}
and
\begin{equation}
\phit_{\rm sing}(\xt,\xt',q)=-\frac{4\pi}{W(\vq^2)} 
e^{- \vert\xt-\xt'\vert \rac}
 [1+H^-(\inf(\xt,\xt'),\vq^2)][1+H^+(\sup(\xt,\xt'),\vq^2)],
\end{equation}
where $\inf(\xt,\xt')$ ($\sup(\xt,\xt')$) is the infimum (supremum) of $\xt$ and $\xt'$.
Since $H^+$ and $H^-$  vanish at $\xt=\bt$,
 $\partial H^-/\partial \xt\vert_{\xt=\bt}$ is also equal to zero,
as  can be checked from the formal solutions given in next
paragraph. Therefore the Wronskian  $W(\vq^2)$ takes the simple
 form
\begin{equation}
\label{valueW}
W(\vq^2)= -2\rac+\left.\frac{\partial H^+(\xt,\vq^2)}{\partial \xt}\right\vert_{\xt=\bt}.
\end{equation}
For the same reasons, the value of $Z(\modq)$  depends only on
$\partial H^+/\partial \xt\vert_{\xt=\bt}$. Indeed, $Z(\modq)$
is entirely determined by
the ratio of 
the continuity equations 
\eqref{boundcond1} and \eqref{boundcond2} obeyed by 
 $\phit$ and $\partial\phit/\partial\xt$ at $\xt=\bt$, 
and the amplitude $B (\xt',\modq)$ in region $0<\xt<\bt$ (see \eqref{phijtild}) disappears 
 in the latter ratio.

As shown in Ref.\cite{AquaCor01II},
 $H^+$  can be represented by a formal alternating series, which will be used in the following,
\begin{equation}
\label{formalseries}
  H^+(\xt,\vq^2) =-{\cal T}^+[1] (\xt;\vq^2) + {\cal T}^+ 
  \left[{\cal T}^+ [1] \right] (\xt;\vq^2)
- \cdots, 
\end{equation}
where the operator ${\cal T}^+$ acting on a function $f(\xt)$ reads
\begin{equation}
\label{defintT}
  {\cal T}^+ [f] (\xt;\vq^2) \equiv  \int_{\bt}^{\xt} dv \, e^{2 \rac \, v} 
 \int_v^{+\infty} dt \, e^{-2 \rac \, t}  \, U(t) f(t).
\end{equation}
Similarly $H^-(\xt,\vq^2)$ is equal to the series
\begin{equation}
\label{formalseriesbis}
 H^-(\xt,\vq^2) ={\cal T}^-[1] (\xt;\vq^2) + {\cal T}^- 
  \left[{\cal T}^- [1] \right] (\xt;\vq^2)
+ \cdots 
\end{equation}
with
\begin{equation}
\label{defT}
  {\cal T}^- [f] (\xt;\vq^2) \equiv  \int_{\bt}^{\xt} dv \, e^{-2\rac \, v} 
 \int_\bt^{v} dt \, e^{2 \rac \, t}  \, U(t) f(t).
\end{equation}

In the infinite-dilution and vanishing-coupling limit, density profiles become uniform and $\kbar^2(x)$ tends to $\kd^2$. The corresponding  screened potential  $\phi^{(0)}$ obeys the Debye equation and satisfies the same boundary conditions as the bare Coulomb potential $v$.
In other words, $H^+$ and $H^-$ in expressions
 \eqref{structphi} vanish for $\phitzero(\xt,\xt',\vq)$ according to their definitions \eqref{defH}, while the expression \eqref{valueW}   is reduced to $-2\rac$ for 
$W^{(0)}(\vq^2)$. $Z^{(0)}(\modq)$ is then determined by the
 continuity equations 
\eqref{boundcond1} and \eqref{boundcond2}. The result reads
\begin{equation}
\label{valuephi0}
\phi^{(0)}(x,x',\vy)=
\phi_{\rm sing}^{(0)}(\vecr-\vecr')
+\kd \int \frac{d^2\vq}{(2 \pi)^2} \, e^{- i \vq . (\kd \vy)}
Z^{(0)}(\modq) e^{-(\xt+\xt')\rac},
\end{equation}
where
\begin{equation}
\label{valueZ0}
Z^{(0)}(\modq)=\frac{2\pi}{\rac}e^{2\bt\rac}
\frac{1-\dw e^{-2 \bt \modq}(\rac+\modq)^2}
{(\rac+\modq)^2-\dw e^{-2\bt\modq}}.
\end{equation}
The particular solution  $\phi^{(0)}_{\rm sing}(\vecr-\vecr')$ that is singular when $\vecr=\vecr'$ coincides
with the bulk screened potential in Debye theory,
\begin{eqnarray}
\label{phizerobulk}
\phi_{\rm sing}^{(0)}(\vecr-\vecr')
=&&\kd \int \frac{d^2\vq}{(2 \pi)^2} \, e^{- i \vq . (\kd \vy)}
\frac{2\pi}{\rac} e^{-\vert\xt-\xt'\vert\rac}\nonumber\\
=&& \phid(\vecr-\vecr')
 \end{eqnarray}
where $\phid$ is written in \eqref{defphib}.

\subsection{Large-distance tail of the screened potential }

\label{section52}

When $x>b$ and $x'>b$, $\phi(x,x',\vy)$  falls off as $1/y^3$, because of the boundary conditions at the interface $x=b$. 
The reason is the following. The  appearance of an $1/y^3$ tail in the large-$y$ behavior of a function $f(\vy)$ corresponds to the existence of a term proportional to $\modq$, which  is not analytical in the Cartesian components of $\vq$, in the small-$\vq$ expansion of $f(\vq)$ \cite{Gelfand}. 
Functions different from $Z(\vq)$ in $\phit(\xt,\xt',\vq)$ [see  \eqref{structphi}]  prove to be functions of $\vq^2$, but the boundary conditions at $\xt=\bt$ imply that, as well as the  small-$\vq$ expansion of $\phit(\xt,\xt',\vq)$ when $\xt<\bt$ (and $\xt'>\bt$) [see \eqref{phijtild}], the small-$\vq$ expansion of $Z(\vq)$ contains a term proportional to $\modq$.

As shown in Paper I, the $1/y^3$ tail of $\phi$ takes the product structure \eqref{tailphi} where
\begin{equation}
\label{valueDphi}
{\overline D}_{\phi}(x) = -
\sqrt{\frac{(-B_{Z})}{2\pi}}\,
 \frac{e^{-\xt}}{\kd}[1+H^{+}(\xt,\vq^2={\bf 0})].
\end{equation} 
In \eqref{valueDphi} $B_{Z}$ is the coefficient of the $\modq$-term in the small-$\vq$ expansion of $Z(\modq)$,
\begin{equation}
\label{Zexp}
Z(\modq)=Z(\vq={\bf 0})+B_Z\modq+ O(\modq^2).
\end{equation}  
We notice that, as shown in Paper I,  sum rules obeyed by $\phi(x,x',y)$ imply that 
${\overline D}_{\phi}(x)$ has the same sign for all $x$'s, and the $1/y^3$ tail of $\phi(x,x',y)$ is repulsive at all distances $x$ and $x'$ from the wall.
In \eqref{valueDphi} the minus sign in front of the square root is {\it a priori} arbitrary. It has been introduced, because in the infinite-dilution and vanishing-coupling  limit and
in the case of a plain wall ($\ew=\epsilon_{\rm solv}$), ${\overline D}_{\phi}(x) $ is expected to have the same sign as the dipole $d(x)$ carried by the set made of a positive unit charge and its screening cloud repelled from the wall, and  $H^+(\xt,\vq^2={\bf 0})$ vanishes in this limit.

The large-distance behavior of $\phi(x,x',y)$ at leading  order, $\phi^{\rm as\,(0)}$, is equal to the
  leading tail $\phi^{\rm (0)\, as}$ of 
$\phi^{(0)}$: 
$\phi^{\rm as\,(0)}
={\overline D}^{(0)}_{\phi}(x){\overline D}^{(0)}_{\phi}(x')/y^3$
with ${\overline D}^{(0)}_{\phi}=
{\overline D}_{\phi^{(0)}}$.  ${\overline D}_{\phi^{(0)}}(x)$    is given by \eqref{valueDphi}, where $H^{+}$ vanishes and 
 $B_Z$ is calculated for $\phi^{(0)}$, namely, $B^{(0)}_Z$ is equal to the coefficient of the $\modq$-term in the small-$\vq$ expansion of $Z_{\phi^{(0)}}(\vq)\equiv Z^{(0)}(\vq)$.  According to \eqref{valueZ0}
\begin{equation}
\label{expvalueZ0}
Z^{(0)}(\modq)=2\pi e^{2\bt}\left[1-2\frac{\ew}{\epsilon_{\rm solv}}\modq\right] +O(\modq^2)
\end{equation}
and the resulting expression for ${\overline D}_{\phi^{(0)}}$ is written in   \eqref{valueDzero}.
The expression of the distance $y_{\star}^{(0)}(x)$ at which the $1/y^3$ tail in $\phi^{(0)}$ overcomes the exponential tails  in $\phi^{(0)}$ has been estimated in Paper I. In the case where the solvent is water and where the dielectric wall is made of glass, 
$\ew/\epsilon_{\rm solv}\sim 1/80$ and $y_{\star}^{(0)}(x=b)= 7\xid$,
$y_{\star}^{(0)}(x=b+\xid)= 10\xid$, 
  $y_{\star}^{(0)}(x=b+3\xid)= 15\xid$ and $y_{\star}^{(0)}(x=b+5\xid)= 20\xid$.

\subsection{Large-distance tail of the screened potential up to order $\eps$}
\label{section63}

\subsubsection{Formal $\eps$-expansion of the tail}
\label{section62}

Because of the nonuniformity of the density profiles in the vicinity of the wall,
$\phi$ has an $\eps$-expansion. More precisely,
the $\eps$-expansion of the screened potential $\phi$ originating from the  $\eps$-expansion of density profiles can be determined by 
\eqref{structphi} from the $\eps$-expansion of the functions $H^+$  and $H^-$,
which themselves are derived from the formal series
\eqref{formalseries} and \eqref{formalseriesbis} respectively.

According to  \eqref{valueDphi} and with the notations of \eqref{structDphi}, the first correction ${\overline D}^{(1)}_{\phi}(x)$ in the $\eps$-expansion of ${\overline D}_{\phi}(x)$ is obtained  from 
$H^{+(1)}(\xt,\vq^2=0)$ and from the $\eps$-expansion of the  coefficient of $\modq$ in the small-$\vq$ expansion 
\eqref{Zexp} of $Z(\modq)$,
$B_Z=B^{(0)}_Z +B^{(1)}_Z+o(\eps)$. It reads
\begin{equation}
\label{formDunphi}
{\overline D}^{(1)}_{\phi}(x)={\overline D}^{(0)}_{\phi}(x)
\left[ C^{(1)}_{\phi}+
{\overline G}_{\phi}^{\rm exp\, (1)}(\xt)\right],
\end{equation}
where the constant $C_{\phi}^{(1)}$ is equal to
\begin{equation}
 \label{formvalueC1}
C^{(1)}_{\phi}=\frac{B^{(1)}_Z}{2B^{(0)}_Z}
+\lim_{\xt\rightarrow +\infty}H^{+\,(1)}(\xt,\vq^2=0),
\end{equation}
and  the function 
 ${\overline G}_{\phi}^{\rm exp\, (1)}(\xt)$, which vanishes exponentially fast when $\xt$
 goes to infinity, is
\begin{equation}
\label{formvalueGphi1}
{\overline G}_{\phi}^{\rm exp\, (1)}(\xt)=
H^{+\,(1)}(\xt,\vq^2=0)
-\lim_{\xt\rightarrow +\infty}H^{+\,(1)}(\xt,\vq^2=0).
\end{equation}

If we write the
$\eps$-expansion of  $Z(\modq)$ up to order $\eps$ as
$Z(\modq)=Z^{(0)}(\modq)+Z^{(1)}(\modq)+o(\eps)$,
then $B^{(0)}_Z=B_{Z^{(0)}}$ and $B^{(1)}_Z=B_{Z^{(1)}}$.
By virtue of \eqref{expvalueZ0}, $B_{Z^{(0)}}=-4\pi(\ew /\epsilon_{\rm solv}) \exp(2\bt)$.
As already
mentioned in Section \ref{section51},  $Z(\modq)$, and subsequently $B_Z$,  is entirely determined from the expression
of $\partial H^{+}/\partial \xt$ at
$\xt=\bt$ by the ratio of the continuity equations
\eqref{boundcond1} and \eqref{boundcond2}. 
When the $\eps$-expansions of $Z(\modq)$ and 
$\partial H^{+}/\partial \xt\vert_{\bt}$ up to order $\eps$ 
are introduced in the expression \eqref{structphi} of $\phit(\xt,\xt',\vy)$, the continuity equations at $\xt=\bt$ lead to
 \begin{eqnarray}
Z^{(1)}(\modq)=&&
\left.Z^{(0)}(\modq)
\frac{\partial H^{+}(\xt,\vq^2)}{\partial \xt}\right\vert^{(1)}_{\xt=\bt}\times
 \frac{1}{2\rac}\nonumber\\
&&\times \qquad
\frac{3\rac+\modq -\dw e^{-2 \bt\modq}(3\rac-\modq) }
{\rac+\modq -\dw e^{-2\bt\modq}(\rac-\modq)}.
\end{eqnarray}
 Then the coefficient $B_{Z^{(1)}}$
of the $\modq$-term in the $\vq$-expansion of
 $Z^{(1)}(\modq)$ is determined by using \eqref{expvalueZ0}, and the expression of $C^{(1)}_{\phi}$ is given by \eqref{formvalueC1} where
 \begin{equation}
\label{rapBZ}
 \frac{B^{(1)}_Z}{2B^{(0)}_Z}=
 \left.\frac{\partial H^+(\xt,\vq^2=0)}{\partial \xt}
\right\vert_{\xt=\bt}^{(1)}.
 \end{equation}

The expression of $H^{+(1)}(\kd x,\vq^2=0)$ is calculated from $\soma Z_{\as}^2 \roa(x)$ as the term of order $\eps$ in the $\eps$-expansion of the formal series \eqref{formalseries}.
The first term in the latter series  reads
\begin{equation}
  \label{defTp}
  {\mathcal T}^+ [1] (\xt;\vq^2) =
  \int_{\bt}^{\xt} dv \, e^{2  \, v\rac} 
 \int_v^{+\infty} dt \, e^{-2  \, t\rac}  \, 
 \left[\frac{\kbar^2(t/\kd)}{\kd^2}-1\right].
\end{equation}
As shown in next section, the contribution to $\kbar^2(x)$ from each
species varies over two length scales, $\beta e^2/\epsilon_{\rm solv}$ (times $Z_\as^2$) and $1/\kd$. Therefore, 
${\cal T}^+ [1] (\xt;\vq)$, as well as all other terms in the series
\eqref{formalseries}, can be expanded in powers of the ratio $2\eps=\kd \beta
e^2/\epsilon_{\rm solv}$ of these two lengths.  As shown in Ref.\cite{AquaCor01II}, for an 
operator ${\cal T}^+$ associated with a function similar to $\kbar^2(x)$, the
$\eps$-expansion of ${\cal T}^+ [1] (\xt;\vq)$ starts at  order $\eps$ (for any
value of $\bt$), and the $\eps$-expansions of next terms in the formal series \eqref{formalseries} are of larger order in $\eps$. Therefore $H^{+(1)}$ is reduced to the contribution from ${\cal T}^+ [1]$
\begin{equation}
\label{HTun}
H^{+(1)}(\xt,\vq^2)=-\left. {\cal T}^+ [1] (\xt;\vq)
\right\vert^{(1)}.
\end{equation}
Similarly, $\partial H^{+}/\partial\xt\vert_{\xt=\bt}^{(1)}=-\left. \partial{\cal T}^+ [1] (\xt;\vq)
/\partial x\right\vert^{(1)}$. We notice that  the latter derivative originates both from  the derivative of 
$\left. {\cal T}^+ [1] (\xt;\vq)\right\vert^{(1)}$ and of 
$\left. {\cal T}^+ [1] (\xt;\vq)\right\vert^{(2)}$, because the latter term can be written as $\eps$ times a function of the two arguments $\xt$ and $\xt/\eps$
(see Appendix  \ref{AppC}). Similar results hold for $H^{-(1)}$, with 
$H^{-(1)}(\xt,\vq^2)={\cal T}^- [1] (\xt;\vq)\vert^{(1)}$.

\subsubsection{$\eps$-expansion of density profiles}
\label{section61}

The density profiles in the vicinity of a dielectric wall have been calculated 
in the high-dilution and weak-coupling regime in Refs.\cite{AquaCor01II} and \cite{AquaCor01I}.
(The
systematic approach in \cite{AquaCor01II} is based on the Mayer diagrammatics for 
the fugacity
expansions of density profiles. Resummations of Coulomb divergences are performed along a
scheme which is similar to -- but more complicated than --  the procedure used in 
Section \ref{section22}, because of differences in the topological definition of
Mayer diagrams in the two cases.) Up to corrections of  first order in the coupling 
parameter $\eps$, for $\kd b$ and $\beta e^2/(\epsilon_{\rm solv} b)$ fixed, the density profile  reads
\begin{equation}
\label{profileform}
\roa(x)=\roab 
\exp\left[-Z_{\as}^2\frac{ \beta e^2}{\epsilon_{\rm solv}}\Vimsc(x;\kd)\right]
\times \left[1-Z_{\as}^2\eps  \Lb(\kd x;\kd b) - Z_\as \beta e\Phi^{(1)}\left(x;\kd,\kd b,\frac{\beta e^2}{\epsilon_{\rm solv} b}\right)+\cO(\eps^2)\right].
\end{equation}
In \eqref{profileform} $\cO(\eps^2)$ is a short notation for terms of orders written in \eqref{orders} with $\Gamma\propto\eps^{2/3}$.

More precisely, in \eqref{profileform} $(Z_\as^2e^2/\epsilon_{\rm solv})\Vimsc(x;\kd)$, called the bulk-screened self-image interaction in the following,  is the part of the  screened self-energy that is reduced  to a mere {\it bulk} Debye exponential screening of
 the bare
self-image interaction \eqref{defVself}  due to the dielectric response of the wall.  
For two charges separated by a distance $2x$, the bulk screening factor at leading order
is  $\exp(-2\kd x)$. After multiplication by $\beta$,
\begin{equation}
\label{valueVselfscreened}
\beta\frac{\left(Z_{\as}  e\right)^2}{\epsilon_{\rm solv}} \Vimsc(x;\kd)=
- Z_{\as}^2\, \dw\frac{ \beta e^2}{\epsilon_{\rm solv}}  {e^{-2\kd x}\over 4x}
= Z_{\as}^2\, f_{\rm im}\left(\kd x, \frac{\beta e^2}{\epsilon_{\rm solv} x}\right).
\end{equation}
  The other part of  the screened self-energy comes from the  deformation of the set made by a charge, its  screening cloud inside the electrolyte, and their images inside the wall, with respect to the spherical symmetry of a charge and its  screening cloud in the bulk. The deformation stems both from  the impenetrability of the wall (steric effect) and from the contribution of  its electrostatic response if  $\dw\not=0$ (polarization effect). When it is multiplied by  $\beta$, one gets
\begin{equation}
\label{valueVselfsteric}
\beta  \frac{\left( Z_{\as}e\right)^2}{\epsilon_{\rm solv}} \,\frac{1}{2}\,\kd \, \Lb(\kd x;\kd b)= Z_{\as}^2 \,\eps\, \Lb(\kd x;\kd b)
\end{equation}
 $\Phi^{(1)}$ is the electrostatic potential created by the
charge-density profile at first order in $\eps$. It is given by 
\begin{multline}
\label{relPhiphi0}
\Phi^{(1)}\left(x;\kd,\kd b,\frac{\beta e^2}{\epsilon_{\rm solv} b}\right)\\
=\left.\frac{e}{\epsilon_{\rm solv}}\int_{b}^{+\infty} dx'\int d\vy
\,\phi^{(0)}(x,x',\vy) \somg Z_\gs \rogb 
\exp\left[-Z_{\gs}^2\frac{ \beta e^2}{\epsilon_{\rm solv}} \Vimsc(x';\kd)\right]
\times\left\{1-Z_{\gs}^2\eps  \Lb(\kd x';\kd b)\right\}\right\vert^{(1)}
\end{multline}
where $\phi^{(0)}(x,x',\vy)$ is written in \eqref{valuephi0} and $\vert^{(1)}$ means that the integral must be calculated at first order in $\eps$ with $\kd b$ and $\beta e^2/(\epsilon_{\rm solv}b)$ kept fixed.  As a consequence, 
\begin{equation}
Z_\as\,\beta e
  \Phi^{(1)}\left(x;\kd,\kd b,\frac{\beta e^2}{\epsilon_{\rm solv} b}\right)
=Z_{\as}\,\eps\, f_{\Phi}\left(\kd x;\kd b,\frac{\beta e^2}{\epsilon_{\rm solv} b}\right) 
\end{equation}
$ \Lb$ and $\Phi^{(1)}$ are functions of $x$ which 
are bounded in the interval $0<x<+\infty$, and which decay exponentially  fast over a few $\kd^{-1}$'s
 when $x$ goes to $+\infty$. In the case of an electrolyte confined between two walls, the density profile  exhibits an analogous structure \cite{Cornu2paroi}.

\subsubsection{Explicit results in the limit $\kd b\ll 1$ at fixed $\beta e^2/(\epsilon_{\rm solv}b)$}
\label{titato}

Density profiles have been explicitly calculated at leading order in a double expansion in $\eps$ and  $\kd b$  with $\beta e^2/(\epsilon_{\rm solv} b)$ fixed in Ref.\cite{AquaCor01I}.
Indeed, in  regimes where $\kd b\ll 1$
 the  density profile written in \eqref{profileform} can be  explicitly calculated at leading order by considering the limit of $\eps \Lb(\kd x;\kd b)$ and of 
 $\eps\, f_{\Phi}\left(\kd x;\kd b,\beta e^2/(\epsilon_{\rm solv} b)\right) $ when
 $\kd b$ vanishes at fixed $\beta e^2/(\epsilon_{\rm solv}b)$, and by keeping only the terms of order $\eps\ln(\kd b)$ and $\eps$. The corresponding expressions 
 are valid in regime (1) where the temperature is fixed (see \eqref{case1bis}). In regime (2) (see \eqref{subcase2bis}), the temperature goes go infinity,  and  the density profiles are obtained from those  of regime (1) by taking the limit where $\beta e^2/(\epsilon_{\rm solv}b)$ vanishes while $\kd b$ is kept fixed.

We notice that the corresponding results enables one to calculate the surface tension of the electrolyte-wall interface at leading order in $\eps$ and $\kd b$ at fixed $\beta e^2/(\epsilon_{\rm solv}b)\propto \eps/\bt$ \cite{CornuBeng}. From the generic expression, one retrieves  results  already known in some special cases.

In regime (1) 
 $\kd b$ vanishes at fixed $\beta e^2/(\epsilon_{\rm solv}b)$, and the explicit expressions of functions in \eqref{profileform} are the following.
\begin{equation}
  \label{valueLb}
  \Lb (\xt;\bt) = (1-\dw^2) \int_1^{\infty} dt \,
\frac{ e^{-2 t \xt} }{\left(t+\ract \right)^2-\dw}+O(\bt)
\end{equation}
and 
\begin{multline}
  \label{valuePhi}
 -\beta e \Phi^{(1)} \left(x;\kd, \bt ,\frac{\beta e^2}{\epsilon_{\rm solv}b}\right)\\
=  \eps \, \left\{\frac{\somt}{\somd}\Mb (\xt) +\frac{\dw}{2} 
\left[ \frac{\somt}{\somd}\left(C+\frac{\ln 3}{2}+\ln\bt\right)
+\frac{ \somg Z_\gs^3 \rogb g_{\gamma}}{\somd}\right]e^{-\xt}
+ \frac{\dw}{4} \frac{\somt}{\somd}
e^{-2\xt}S_-(\xt)
\right\}+\cO(\eps\bt),
\end{multline}
where $\cO(\eps\bt)$  stands for a term of order $\eps\bt$. By virtue of \eqref{relPhiphi0}, the electrostatic potential profile $\Phi^{(1)}(x)$ at first order in $\eps$  arises from the screened self-energy: the term with $\Mb $ comes from the deformation of screening clouds with respect to the bulk spherical symmetry, which is  described by  $\Lb$ \eqref{valueVselfsteric}, and the other terms originate from the bulk-screened self-image interaction $\Vimsc(x)$  \eqref{valueVselfscreened}. If $\dw=0$, $\Phi^{(1)}(x)$ is reduced to $ \eps \, (\somt/\somd)\Mb (\xt)$.  In \eqref{valuePhi}  
\begin{equation}
  \label{valueMb}
\Mb (\xt) = \int_1^{\infty} dt \, \frac{e^{-2 t \xt} -2t e^{-\xt}}
{1-(2t)^2} \, \frac{1-\dw^2}{\left(t+\ract \right)^2-\dw}\,\,,  
\end{equation}
$C$ is the Euler constant,
\begin{equation}
\label{defggamma}
g_{\gamma}\equiv g\left( \frac{\dw}{4} Z_\gs^2\frac{\beta e^2 }
{ \epsilon_{\rm solv} b}\right),
\end{equation}
where
\begin{equation}
\label{defg}
g(u)\equiv-1+\frac{e^u-1}{u}-\int_0^udt \frac{e^t-1}{t}.
\end{equation}
$S_-$ is defined by
\begin{equation}
  \label{defSpm}
  S_{\pm}(u) \equiv
 e^{3u} \Ei (-3u) \pm e^u \Ei (-u),  
\end{equation}
 where ${\rm Ei}(-x)$ is the
Exponential-Integral function:  for $x>0$ 
\begin{equation}
\label{defEi}
{\rm Ei}(-x)\equiv-\int_x^{+\infty}dt\, \frac{e^{-t}}{t}\,=\,C+\ln x+\int_0^x dt\frac{e^{-t}-1}{t}.
\end{equation}
$S_-(u)$ decays proportionally to $1/u$   when $u$ goes to $\infty$, since $\Ei (-u)$  behaves as $\exp(-u)/u$ for large $u$.

We notice that, in  the calculation of the part in $\Phi^{(1)}(x)$ that comes from the bulk- screened image contribution $\Vimsc$, a key ingredient  is the decomposition \eqref{decompint} combined with  the expression  of the Exponential-Integral function \eqref{defEi}. 
$g(u)$ arises because
\begin{equation}
\label{defgbis}
\int_{\bt}^{+\infty} dv \left[e^{(\eta/v)}-1-\frac{\eta}{v}\right]=
-\eta \,g\left(\frac{\eta}{\bt}\right).
\end{equation}
We point out that $g(u\!=\!0)=0$.

In regime (2) (see \eqref{subcase2bis}), $ \beta e^2/(\epsilon_{\rm solv}b)$ vanishes,  whatever the sign of $\dw$ is,  because $(\beta e^2/\epsilon_{\rm solv})\ll b\ll \xid$. In this regime,  $\eps$ vanishes faster than $\bt$, and  $\eps/\bt$ must be set to zero  while $\bt$  is kept fixed in $\ln\bt+ g_{\gamma}(\eps/\bt)$; then the latter sum is reduced to $\ln \bt$.  The result is the same as if the exponential involving the bulk-screened self-image interaction $\Vimsc(x')$ in the expression \eqref{valuePhi} of $\Phi^{(1)}(x)$ had been linearized at all distances  $x'$, as it is the case in the second integral in decomposition \eqref{decompint}.
In the following, we write expressions only for  the more general regime where $\eps$ and $\bt$  vanish with their ratio  kept fixed.

In regime (1) (see \eqref{case1bis}),  $ \beta e^2/(\epsilon_{\rm solv}b)$ is finite. 
 For  an electrostatically attractive wall ($\dw>0$), we cannot consider the limit  $b\ll (\beta e^2/\epsilon_{\rm solv})\ll \xid$, where $\dw \beta e^2/(\epsilon_{\rm solv} b)$ tends to $+\infty$:  there is an irreducible dependence on $b$. On the contrary, for an electrostatically repulsive wall ($\dw<0$), we can take the previous limit, where $\dw \beta e^2/(\epsilon_{\rm solv} b)$ goes to $-\infty$. In this limit,  $\bt$ vanishes faster than $\eps$, and we must set   $\kd b=0$  at fixed   $\varepsilon$  in the term $\ln \bt+g_{\gamma}(\eps/\bt)$ in \eqref{valuePhi}; then this term becomes equal to $\ln\left(|\dw|  Z_\gs^2\eps/2\right) +C-1$.

From the expression of the density profiles up to order $\eps$ in the vanishing-$\bt$ limit, we explicitly calculate the vanishing-$\bt$ limit of the term ${\overline D}^{(1)}_{\phi}(x)$ of order $\eps$ in the coefficient  ${\overline D}_{\phi}(x)$.  The  formal expression of ${\overline D}_{\phi}^{(1)}(x)$ has been derived in Section \ref{section62}.  The calculations of the zero-$\bt$ limits of
$H^{+(1)}(\xt,\vq^2=0)$ and $\partial H^{+}/\partial \xt\vert^{(1)}(\bt,\vq^2=0)$ are performed  in Appendix \ref{AppC}.   
$C_{\phi}^{(1)}$ and ${\overline G}_{\phi}^{\rm exp\, (1)}(\xt)$ in \eqref{formDunphi} are given by
 \begin{subequations}
   \label{valueda}
\begin{eqnarray}
  \frac{C_{\phi}^{(1)}}{\eps} & = &  
\frac{\dw}{2}
\left[\frac{\somq}{\somd} \left(C+\frac{3}{2}\ln 2+\ln \bt\right)+
\frac{\somg  \rogb Z_\gs^4 g_{\gamma}}{\somd}\right]\label{a}\\
& & + \frac{\somq}{\somd} \int_1^{\infty} dt \, \frac{1-\dw^2}
{\left( t + \ract \right)^2 - \dw} \left[ \frac{t+1/2}{2t(t+1)} \right] \label{b}\\
& & + \frac{1}{4}\left(\frac{\somt}{ \somd}\right)^2 \left\{ \int_1^{\infty} dt \, \frac{1-\dw^2}
{\left( t + \ract \right)^2 - \dw} \, \frac{1}{t^2 - 1/4} \left[ 
\frac{t+1/2}{2t(t+1)} - \frac{4}{3} t \right] \right\}\label{c}\\
& & - \frac{\dw}{3} \frac{\somt}{ \somd}\left[
\frac{\somg  \rogb Z_\gs^3 g_{\gamma}}{\somd} +
\frac{\somt}{ \somd}\left(C+\frac{\ln 3}{2}+\ln \bt\right)\right]
\label{e}\\
& & + \frac{\dw}{4} \left(\frac{\somt}{ \somd}\right)^2
\left[\frac{2}{3} \ln 2 - \ln 3 \right]\label{d}
\end{eqnarray}
\end{subequations}
and
\begin{subequations}
 \label{valuedabis}
\begin{eqnarray}
  \frac{{\overline G}_{\phi}^{\rm exp\, (1)} (\xt)}{\eps} & = & 
  \frac{\dw}{4} \frac{\somq}{\somd}
\left\{ e^{2\xt} \Ei(-4\xt) - \Ei(-2\xt) \right\} \label{abis}\\
& & -  \frac{\somq}{\somd} \int_1^{\infty} dt \, \frac{1-\dw^2}
{\left( t + \ract \right)^2 - \dw}
 \frac{e^{-2 t \xt}}{4 t (t+1)}   \label{bbis}\\
& & + \frac{1}{4}\left(\frac{\somt}{ \somd}\right)^2 \left\{ \int_1^{\infty} dt \, \frac{1-\dw^2}
{\left( t + \ract \right)^2 - \dw} \, \frac{1}{t^2 - 1/4} \left[ \frac{2}{3} t
e^{-\xt} - \frac{e^{-2t\xt}}{4t(t+1)} \right] \right\}\label{cbis}\\
& & + \frac{\dw}{6} 
\frac{\somt}{ \somd}\left[
\frac{\somg  \rogb Z_\gs^3 g_{\gamma}}{\somd} +
\frac{\somt}{ \somd}\left(C+\frac{\ln 3}{2}+\ln \bt\right)\right] e^{-\xt}\label{ebis}\\
& & + \frac{\dw}{4} \left(\frac{\somt}{ \somd}\right)^2
\left[ \frac{1}{3} e^{2\xt}\Ei(-4\xt) +\Ei(-2\xt) 
-\frac{1}{3} e^{-\xt} \Ei(-\xt) - e^{\xt}\Ei(-3\xt)\right].\label{dbis}
\end{eqnarray}
\end{subequations}
The expressions \eqref{a} and \eqref{abis} arise from the contribution of the screened 
self-image interaction $\Vimsc$ \eqref{valueVselfscreened} to the density profile 
\eqref{profileform}.
The terms \eqref{b} and \eqref{bbis} originate from the
deformation of screening clouds described by the function $\Lb$ given in \eqref{valueLb}, which does not vanish even when $\ew=\epsilon_{\rm solv}$. 
The three last ligns in $C_{\phi}^{(1)}/\eps$ and 
${\overline G}_{\phi}^{\rm exp\, (1)} (\xt)/\eps$ come from the  contribution of the electrostatic 
potential $\Phi(x)$ to the density profile. More precisely,
 \eqref{c}-\eqref{cbis},  \eqref{e}-\eqref{ebis} and
\eqref{d}-\eqref{dbis} 
originate from the functions $\Mb(\xt)$, $\exp(-\xt)$ and $\exp(-2\xt)S_-(\xt)$, in \eqref{valuePhi}, respectively.

\section{Correlations along  the wall }
\label{section6}

\subsection{Tails at large distances along the wall}
\label{section53}

The Ursell function $h_{\as\asp}$ cannot decay faster than $1/y^3$. Indeed,
by an argument based on linear response theory and 
screening in  macroscopic electrostatics, the 
 correlation between global 
 surface-charge densities 
at points separated by a distance $y$ is shown to decay as  $1/y^3$ with a universal
negative amplitude  \cite{Janco82II}: $f_{\as\asp}$ in the amplitude $-\beta f_{\as\asp}$ of the $1/y^3$ tail of $h_{\as\asp}$ obeys sum rule \eqref{sumrule2wall}.
The latter sum rule holds whether all species have the same closest approach distance $b_{\as}$ to the wall or not. We recall that it is a consequence of external screening, as sum rule \eqref{StilLovrule}.

On the other hand, 
as a consequence of the $1/y^3$ decay of the screened potential $\phi$, according to \eqref{defFcc} and \eqref{defFr}, the bonds $\Fcc$ and $\Fr$ in resummed Mayer diagrams  behave as $1/y^3$ and 
$1/y^6$, respectively, at large distances $y$. Since $h_{\as\asp}$ does not fall off faster than $1/y^3$, no compensation mechanism kills the $1/y^3$ tail arising from the slowest one among the algebraic bonds in the Mayer diagrammatics. Thereore, in a regime where only a finite number of Mayer diagrams -- or only some infinite class of diagrams --  contribute to the large-distance behavior of $h_{\as\asp}$, $h_{\as\asp}$ indeed decays as $1/y^3$. This is the case in the dilute regime studied hereafter. (We notice that if, in some regime, the summation of some infinite series of subdiagrams led to an infinite contribution to $f_{\as\asp}(x,x')$, then $h_{\as\asp}$ would fall off more slowly than $1/y^3$. However, since 
$h_{\as\asp}$ is integrable by definition, it cannot decay more slowly than $1/y^2$.)

\begin{figure*}[H]
\includegraphics[width=0.96 \textwidth]{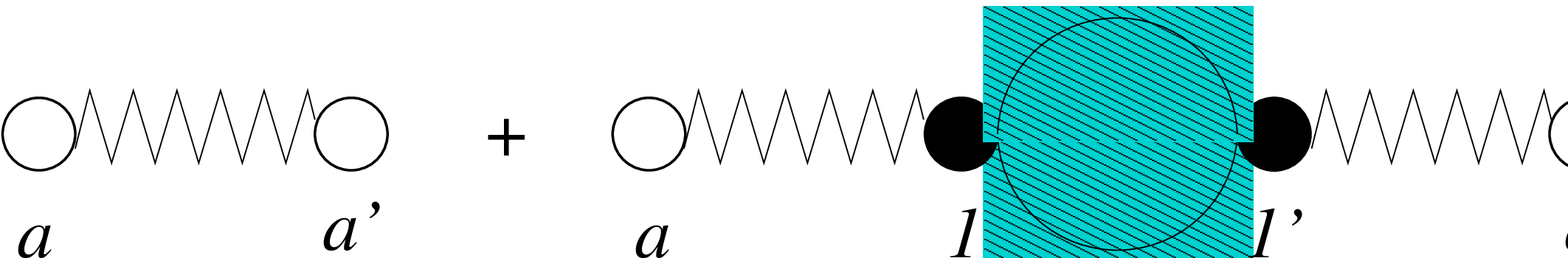}
\caption{\label{hcc}}
\end{figure*}

As shown in Paper I, the large-$y$ behavior of $h_{\as\asp}$ along the wall is conveniently studied from the decomposition described by 
 \eqref{decomph}--\eqref{defhtt}, as in the case of  bulk correlations.
In the latter graphic representation of $h_{\as\asp}$,
the topology of
diagrams involved in $\Ir$ implies that the bond $\Ir$ decays algebraically faster than $\Fcc$ at large
distances $y$ (see Section \ref{section25}). Moreover, as exhibited in 
Figs.\ref{hcc}-\ref{htt}, all graphs in $h^{\rm cc}$,
 $h^{\rm -c}$, $h^{\rm c-}$, and $h^{\rm --}$ are chain graphs, and, because of the translation invariance in the
 direction parallel to the wall, the chain graphs can be seen as multiple
 convolutions with respect to the variable ${\bf y}$.
Therefore, 
every term, except $\Ir$,  in the graphic representation of  $h^{\rm cc}$,
 $h^{\rm -c}$, $h^{\rm c-}$ and $h^{\rm --}$ has  $1/y^3$ tails arising from
 all its $\Fcc$ bonds. 
 The $1/y^3$ tail of every graph
  in $\hcc$, $\hct$,
 $\htc$ and $\htt$ (see Figs.\ref{hcc}-\ref{htt}) is a sum of contributions, each of which is determined by 
 replacing one of the bonds $\Fcc$ by its $1/y^3$ behavior at large $y$, while
 the other part of the graph is replaced by its Fourier transform at the value
 $\vq={\bf 0}$.

Eventually, as shown in Paper I, when  all species have the same closest approach distance to the wall, 
\begin{equation}
\label{hlargey}
h_{\as\asp}(x,x',y)\underset{y\rightarrow +\infty}{\sim}-\beta
\frac{ D_{\as}(x)D_{\asp}(x')}{y^3}
\end{equation}
and
\begin{equation}
\label{valueD}
D_\as(x)=
\frac{e}{\sqrt{\epsilon_{\rm solv}}}
\Bigl\{ Z_{\as} 
\left[{\overline D}_{\phi}(x) +\overline{C}^{{\rm c}-}(x)\right]
 + C^{--}_{\as}(x)\Bigr\},
 \end{equation}
where  $C^{--}_{\as}(x)$ and
$ \overline{C}^{{\rm c}-}(x)$ are related to $\htt_{\as\asp}$ and $\hct_{\as\asp}$, respectively, by 
\begin{equation}
\label{valueCtt}
C^{--}_{\as}(x)\equiv
\int dx''
\sum_{\gamma''} \rho_{\gs''}(x'')Z_{\gs''} 
{\overline D}_{\phi}(x'')   h_{\as\gs''}^{--}(x,x'', \kd \vq={\bf 0})\end{equation} 
and
\begin{equation}
\label{valueCct}
Z_\as {\overline C}^{{\rm c}-}(x)\equiv
\int dx''
\sum_{\gamma''} \rho_{\gs''}(x'')Z_{\gs''} 
{\overline D}_{\phi}(x'')   \hct_{\as\gs''}(x,x'', \kd \vq={\bf 0}).
\end{equation}

An advantage of the resummed Mayer diagrammatic representation is that the contribution from every diagram ${\widetilde \Pi}$ can be associated with some physical effect. For instance, diagram 
${\widetilde{\Pi}}_a$ made of the single bond $\Fcc$  describes Coulomb screening at leading order, the sum of the two diagrams made of bonds $[\Fcc]^2/2$ and $\Frt$ respectively contains the short-distance repulsion, while   diagrams ${\widetilde{\Pi}}_b$ ${\widetilde{\Pi}}_{b^{\star}}$ and 
${\widetilde{\Pi}}_c$ shown in Figs.\ref{diagbbe}-\ref{diagc} involve many-body corrections to the mean-field contribution from ${\widetilde{\Pi}}_a$.

In order to trace back the physical effects, we have to identify the contributions to $D_{\as}(x)D_{\asp}(x')$ from the various diagrams ${\widetilde \Pi}$ defined in Section \ref{section42}. In other words, we have to recognize in \eqref{hlargey}-\eqref{valueD} the tails of $\hcc_{\as\asp}$, $\hct_{\as\asp}$, $\htc_{\as\asp}$ and $\htt_{\as\asp}$, the sum of which is equal to $h_{\as\asp}$. As shown in
 the Appendix of Paper I, the latter tails read
\begin{equation}
\label{tailhcc}
h_{\as\asp}^{\rm cc}(x,x',\vy)
\underset{y\rightarrow +\infty}{\sim}
 -\frac{\beta e^2}{\epsilon_{\rm solv}}
 Z_\as Z_\asp\left[{\overline D}_{\phi}(x) + \overline{C}^{c-}(x)\right]
 \left[{\overline D}_{\phi}(x') + \overline{C}^{c-}(x')\right] \frac{1}{y^3},
 \end{equation}
\begin{equation}
\label{tailhct}
h_{\as\asp}^{\rm c-}(x,x',\vy)
\underset{y\rightarrow +\infty}{\sim}
 -\frac{\beta e^2}{\epsilon_{\rm solv}} 
  Z_\as\left[{\overline D}_{\phi}(x) + \overline {C}^{c-}(x)\right]C_{\asp}^{--}(x')\frac{1}{y^3},
 \end{equation}
\begin{equation}
\label{tailhtt}
h_{\as\asp}^{\rm -- }(x,x',\vy)
\underset{y\rightarrow +\infty}{\sim}
 -\frac{\beta e^2}{\epsilon_{\rm solv}}
  C_{\as}^{--}(x)C_{\asp}^{--}(x') \frac{1}{y^3}.
 \end{equation}
  
\subsection{Sum rule for the effective dipole $D_{\as}(x)$}
\label{section53bis}

According to \eqref{hlargey},  $f_{\as\asp}(x,x')=D_{\as}(x)D_{\asp}(x')$ so that the sum rule \eqref{sumrule2wall} for $f_{\as\asp}(x,x')$ can be rewritten as
\begin{equation}
\label{sumrule2wallbis}
\int_0^{+\infty}dx \sum_{\alpha}
e_{\as} \roa(x)D_{\as}(x)=
\sqrt{\frac{\ew}{8\pi^2\beta^2}}.
\end{equation} 
Similarly to what happens for the internal-screening rule \eqref{sumruleint}, the latter external-screening sum rule can be derived  from the decomposition \eqref{valueD}, the integral relation \eqref{relhcthtt} between $\hct$ and $\htt$, and two sum rules obeyed by $\phi$, namely \eqref{sumrulephi} and
 a sum rule for ${\overline D}_{\phi}(x)$ in the $f_{\phi}(x,x')/y^3$ tail of $\phi(x,x',y)$,
\begin{equation}
\label{sumrule2phi}
\int_0^{+\infty}dx\, \kbar^2(x){\overline D}_{\phi}(x)=
\sqrt{\frac{2\ew}{\epsilon_{\rm solv}}}.
\end{equation} 
The latter equation arises from the sum rule \eqref{sumrulefphi} obeyed by  the amplitude $f_{\phi}(x,x')$ (derived  in Paper I), and from the fact that  $f_{\phi}(x,x')$ takes the factorized form 
${\overline D}_{\phi}(x){\overline D}_{\phi}(x')$ in the case where all species have the same closest approach distance $b$ to the wall.

More precisely, the derivation of \eqref{sumrule2wallbis} is as follows.
The integral relation \eqref{relhcthtt} between $\hct$ and $\htt$ and sum rule \eqref{sumrulephi} imply that the contributions
from $C^{--}_{\as}(x)$ and $Z_{\as} \overline{C}^{{\rm c}-}(x)$ to the integral in \eqref{sumrule2wallbis} cancel each other. On the other hand, sum rule \eqref{sumrule2phi} ensures that the contribution from $Z_{\as}{\overline D}_{\phi}(x)$ to the integral in  \eqref{sumrule2wallbis} is already equal to  the constant in the r.h.s. of the equation. 
In other words, the bond $\Fcc$, namely diagram ${\widetilde{\Pi}}_a$, already fulfulls sum rule \eqref{sumrule2wallbis}.

As a consequence, if some diagrams are to be kept for their contributions to $C^{--}_{\as}(x)$ in some dilute regime, then the corresponding diagrams ``{dressed}'' with a  bond $\Fcc$ must also be retained  in $Z_{\as} \overline{C}^{{\rm c}-}(x)$  in order to ensure that  screening rule  \eqref{sumrule2wallbis} is still obeyed.

In the case of a symmetric electrolyte  made of two species  with
opposite charges $+Ze$ and $-Ze$ and with the same radii, 
$\sum_{\as} \roa(x) h_{\as\asp}(\vecr,\vecr')$ decays faster than  
$h_{\as\asp}(\vecr,\vecr')$ in the $y$-direction, similarly to what happens in the bulk (see Section \ref{densitdensit}). Indeed, symmetries enforce that
the local neutrality is satisfied not only in the bulk, where $\rho^{\rm\scriptscriptstyle B}_+=
\rho^{\rm\scriptscriptstyle B}_-$, but also in the vicinity of the wall, where $\rho_+(x)=\rho_-(x)$. As a consequence, by virtue of \eqref{valueD},
$\rho_+(x)D_+(x)+\rho_-(x)D_-(x)=(e/\sqrt{\epsilon_{\rm solv}}) \soma\roa(x) C^{--}_{\as}(x)$. Symmetries also imply that $h^{\rm --}_{++}=h^{\rm --}_{--}$
and $h^{\rm --}_{+-}=h^{\rm --}_{-+}$, and  the definition \eqref{valueCtt} of 
$C^{--}_{\as}(x)$ yields $\rho_+(x) D_+(x)+\rho_- (x)D_-(x)=0$. Subsequently, $\rho_+ h_{+\asp}+\rho_- h_{-\asp}$ decays faster than $1/y^3$. The latter property has been exhibited in Eq.(3.4) of 
Ref.\cite{SamajJanco02}, where the density-density correlation 
$\sum_{\as,\asp} \roa(x)\roap(x') h_{\as\asp}(\vecr,\vecr')$ in the infinite-dilution and vanishing-coupling  limit (where
$\roa(x)=\roab$) is shown to decay as $\exp[-2\kd(x+x')]/y^6$.

\subsection{ $\eps$-expansions}
\label{section64}
 
\subsubsection{Method}
\label{method}

In the general formula \eqref{valueD} for $D_\as(x)$, 
$Z_{\as} \overline{C}^{{\rm c}-}(x)$, as well as $C^{--}_{\as}(x)$, is a series of functions, each  of which decays as a polynomial in $x$ times $\exp(-\kd x)$,  plus functions which vanish faster (see the general structure  \eqref{structDphi} of ${\overline D}_{\phi}(x)$  in the introduction). However,
   since there is 
 no translational invariance in the direction
perpendicular to the wall, the series $C^{--}_{\as}(x)$ and 
$Z_{\as} \overline{C}^{{\rm c}-}(x)$ cannot be expressed as sums  of geometric series that could be calculated by such a simple formula as \eqref{resumh}. Therefore, the expression of $(\Zeffwa/\kappa)\exp[-\kappa(x-b)]$ in the large-distance behavior \eqref{comparaison} of $D_{\as}(x)$ 
cannot be calculated by the mere
 determination of the pole of an analytic function and the calculation of a residue.

 Though the lost of translational invariance in the direction perpendicular to
 the wall prevents one from performing systematic resummations,
$D_{\as}(x)$ can be determined up to order $\eps$ at any distance $x$ (in the sense of the comment after \eqref{diageps1}) by
the alternative procedure derived for bulk correlations in Section \ref{section45}. In a first step, 
 the  correction of order $\eps$ in the screening length $\kappa^{-1}$
of the leading exponential decay of $D_{\as}(x)$ 
 has to be  calculated by  the partial resummation mechanism whose validity has been checked  
in the case of 
  bulk correlations  (see Section \ref{section451}).
In a second  step, the amplitude factor in $D_{\as}(x)$ up to order $\eps$ is determined as follows. First,  we calculate $D_{\as}^{(1)}(x)$ in a form analogous to \eqref{diageps1}, which arises from the  contributions of only a few diagrams whose amplitude is of order $\eps$ and  which  decay at large $x$ as $\exp[-\kd (x-b)]$ times a possible linear term in $x$; in a second step we  check 
that the coefficient of the $(x-b)\exp[-\kd (x-b)]$ term, which  arises from the second diagram ${\widetilde{\Pi}}_c$ in $\hcc$,  indeed coincides with the 
 opposite of the first correction
to the screening length in the direction perpendicular to the wall, which has already been 
calculated independently.

  The $\eps$-expansions of ${\widetilde {\Pi}}$ diagrams are more complicated than in the case of the
bulk, because 
the screened potential $\phi$ also has an $\eps$-expansion when 
the vicinity of the wall  is studied. The first correction to $\phi^{(0)}$ yields ${\overline D}_{\phi}^{(1)}(x)$ in the expression \eqref{valueD} of $D_{\as}^{(1)}(x)$.
The leading term in the $\eps$-expansion of 
$ C^{--}_{\as}(x)$ or
$ Z_{\as} \overline{C}^{{\rm c}-}(x)$ is obtained as follows:
 densities $\rho_{\gs}$'s are replaced by their bulk values $\rogb$'s and both
   functions $\phi(x,x',\vq={\bf 0})$ and 
$I(x,x',\vq={\bf 0})$ are replaced by their leading values $\phi^{(0)}(x,x',\vq={\bf 0})$ and $I^{(1)}(x,x',\vq={\bf 0})=(1/2)[F^{\rm c\,c\, (0)}]^2(x,x',\vq={\bf 0})$, respectively. As in the bulk case, only the subseries shown in Figs.\ref{hccun}-\ref{httun} do contribute to $D^{(1)}_{\as}(x)$.

\subsubsection{Renormalization of the screening length}
\label{sectionrenormkappa}

We recall that, in the bulk case, the leading tail at order $\eps^q$ in the $\eps$-expansion of the large-distance behavior $h_{\as\asp}^{\rm cc\,  {\scriptscriptstyle B}\,as}$  of $h_{\as\asp}^{\rm cc\,  {\scriptscriptstyle B}}$  
around its $\exp[-\kd r]/r$ limit decays as $r^q$ times $\exp[-\kd r]/r$, and the sum $h_{\as\asp}^{\rm {cc\,{\scriptscriptstyle B}\, as}\,\star}$ of the latter tails decays as 
  $\exp\left[-(\kd+\delta\kappa^{\star}_{\scriptscriptstyle B}) r\right]$, with
$ \delta\kappa^{\star}_{\scriptscriptstyle B}=\kbun$  (see Section \ref{section451}). In the vicinity of the wall, the contribution from $\hcc_{\as\asp}$ to $D_{\as}(x)$ is equal to 
$Z_{\as}\left[{\overline D}_{\phi}(x) + \overline{C}^{c-}(x)\right]$ according to \eqref{tailhcc}. As shown in Appendix \ref{AppB}, in the $\eps$-expansion of 
 $Z_{\as} \overline{C}^{{\rm c}-}(x)$, at order $\eps^q$ the leading term at large $x$ 
  is proportional to $(x-b)^q\exp[-\kd(x-b)]$. The sum over $q$ of the latter leading terms is proportional to  $\exp\left[-(\kd+\delta\kappa^{\star}) (x-b)\right]$ with
   \begin{equation}
   \label{riri}
  \delta\kappa^{\star}=\kbun.
  \end{equation}
According to the discussion of Section \ref{section451},  
  the latter partial resummation  determines the correction $\delta\kappa^{(1)}$ of
  order $\eps$ to $\kd$ in the $x$-direction, 
 $\delta\kappa^{\star}=  \delta\kappa^{(1)}$. According to \eqref{riri}, the correction
$ \delta\kappa^{(1)}$ to the screening length in the direction perpendicular to the wall coincides at first order in $\eps$ with the value found for bulk correlations,
  \begin{equation}
  \label{idkappa}
  \delta\kappa^{(1)}=\kbun.
  \end{equation}

\subsubsection{Renormalization of the amplitude of $D_{\as}(x)$}
\label{renormz}

According to the general method summarized above, the amplitude of
$D_{\as}^{(1)}(x)$ can be determined from only a few diagrams in 
$ C^{--}_{\as}(x)$ and $ Z_{\as} \overline{C}^{{\rm c}-}(x)$.
Before turning to the explicit calculations in the regime $\bt\ll 1$, we interpret the various contributions in $D_{\as}(x)D_{\asp}(x')/y^3$ in terms of diagrams which are representative of physical effects.
The diagrams that
are involved in the determination of the  $1/y^3$ tail of
$h_{\as\asp}(x,x',y)$ up to order in $\eps$ are the same as in the case of the bulk.
Diagram ${\widetilde{\Pi}}_a$ in Fig.\ref{diaga} describes the leading screening effect and therefore gives the 
 zeroth-order contribution
\begin{equation}
D_{\as}^{(0)}(x)D_{\asp}^{(0)}(x')=
\frac{e^2}{\epsilon_{\rm solv}}Z_{\as}Z_{\asp}
{\overline D}^{(0)}_{\phi}(x){\overline D}^{(0)}_{\phi}(x').
\end{equation} 
Contrary to the bulk case, because of the nonuniformity of the density profiles in the vicinity of the wall,
$\phi$ has an $\eps$-expansion, and
the first correction $\phi^{(1)}$ to 
$\phi^{(0)}$ gives a correction of order $\eps$ in the $1/y^3$ tail of 
${\widetilde{\Pi}}_a$.
The contribution from diagram ${\widetilde{\Pi}}_a$ to
the correction of order $\eps$
\begin{equation}
\label{prodorderun}
\left[D_{\as}(x)D_{\asp}(x')\right]^{(1)}=
D_{\as}^{(0)}(x) D_{\asp}^{(1)}(x')
+D_{\as}^{(1)} (x) D_{\asp}^{(0)}(x')
\end{equation} 
reads
\begin{equation}
\label{structdabis}
\frac{e^2}{\epsilon_{\rm solv}}Z_{\as}Z_{\asp}
\left[{\overline D}^{(0)}_{\phi}(x){\overline D}^{(1)}_{\phi}(x')
+{\overline D}^{(1)}_{\phi}(x){\overline D}^{(0)}_{\phi}(x')
\right].
\end{equation}
The other contributions to \eqref{prodorderun}
arise from  diagrams
${\widetilde{\Pi}}_b$ 
${\widetilde{\Pi}}_{b^{\star}}$ and ${\widetilde{\Pi}}_c$, shown in
Figs.\ref{diagbbe}-\ref{diagc}, where fixed charges interact through screened interactions via one or two other charges.
(The $\eps$-expansions of the contributions from the latter diagrams
  to the $1/y^3$ tail of $h_{\as\asp}(x)$ 
start at
the order $\eps$, because they all involve a bond $(1/2)\left[\Fcc\right]^2$.) 
The  contribution to \eqref{prodorderun} from diagram ${\widetilde{\Pi}}_b$  reads
\begin{equation}
\frac{e^2}{\epsilon_{\rm solv}}Z_{\as}
{\overline D}^{(0)}_{\phi}(x)
{\overline C}^{--\,(1)}_{\asp}(x'),
\end{equation}
while  ${\widetilde{\Pi}}_{b^{\star}}$ leads to a symmetric term in the variables $x$ and $x'$, and 
${\widetilde{\Pi}}_c$ yields
\begin{equation}
\label{structdcbis}
\frac{e^2}{\epsilon_{\rm solv}}Z_{\as}Z_{\asp}
\left[{\overline C}^{\rm -c\, (1)}_{\as}(x)
{\overline D}^{(0)}_{\phi}(x')
+{\overline D}^{(0)}_{\phi}(x)
{\overline C}^{\rm -c\,(1)}_{\asp}(x')\right].
\end{equation}

\subsection{Explicit results in the limit $\kd b\ll 1$ at fixed $\beta e^2/(\epsilon_{\rm solv}b)$}
\label{section65}

\subsubsection{Separate contributions}

In the limit $\bt\ll 1$, $D_{\as}^{(0)}(x)$ is given by \eqref{valueDzero} where the $\exp(\kd b)$ term disappears in 
the expression of ${\overline D}_{\phi}^{(0)}(x)$. 
$D_{\as}^{(1)}(x)$ is calculated from formula \eqref{valueD}.
 ${\overline D}_{\phi}^{(1)}(x)$ has been studied in Section \ref{section5}, and
 in the limit
$\bt\ll 1$ ${\overline D}_{\phi}^{(1)}(x)$  is given by  \eqref{formDunphi}, \eqref{valueda}, and \eqref{valuedabis}.
 The other contributions $C^{--\, (1)}_{\as}(x)$ and 
$Z_{\as} \overline{C}^{{\rm c}-\, (1)}(x)$ are obtained by replacing $\rho_{\gs''}(x'')$ by $\rho_{\gs''}^{\scriptscriptstyle B}$, $\phi$ by $\phi^{(0)}$, $\htt$ and $\hct$ by the graphs with one $\Ir$ in their series representations, and  $\Ir$  by $[F^{\rm c\,c\,(0)}]^2/2$  in the expressions \eqref{valueCtt} and \eqref{valueCct} for 
$C^{--}_{\as}(x)$ and 
$Z_{\as} \overline{C}^{{\rm c}-}(x)$, respectively.
The contribution from $C^{--}_{\as}(x)$ is
\begin{equation}
\label{structdb}
 C^{-- \, (1)}_{\as}(x)=
{\overline D}_{\phi}^{(0)}(x)
Z_{\as}^2\left[{\overline B}_{[b]}^{(1)}+{\overline G}_{[b]}^{\rm exp\, (1)}(\xt)\right],
\end{equation}
where
\begin{equation}
\label{valueBb1}
{\overline B}_{[b]}^{(1)} =\eps \frac{\ln3}{2}\frac{\somt}{ \somd}
\end{equation}
and
\begin{multline}
  \label{}
{\overline G}_{[b]}^{\rm exp\, (1)}(\xt)=   \eps\frac{\somt}{ \somd} \frac{1}{2}
\left\{   -e^{-\xt} S_-(\xt)   \rule{0mm}{6mm} \right.\\
  \left. 
+  \int_1^{\infty} dt \, \frac{e^{(1-2t)\xt}}{\left[ t +\ract - \dw (t-\ract)
\right]^2} 
\left[ 2 (1-\dw)^2  + \frac{(1-\dw)^2 (1-2 e^{-\xt}) 
- 8 \dw (t^2-1)(1-e^{-\xt})}{t+1/2} \right] \right\}
\end{multline}
with $S_-(\xt)$  defined in \eqref{defSpm}.
The contribution from $Z_{\as} \overline{C}^{{\rm c}-}(x)$ reads
 \begin{equation}
\label{structdc}
Z_{\as} \overline{C}^{\rm c-\, (1)}(x)
=Z_{\as}{\overline D}_{\phi}^{(0)}(x)
\left[- \eps \frac{\ln 3}{4}\left(\frac{\somt}{ \somd}\right)^2 \, \xt
+B_{[c]}^{(1)} +G_{[c]}^{\rm exp\, (1)}(\xt)\right].
\end{equation}
In \eqref{structdc} the coefficient of the linear term $\xt$ coincides with $-\kbun$ given by 
\eqref{valuecorkappabulk}, while
\begin{equation}
\label{valueBc1}
B_{[c]}^{(1)}=\eps\frac{1}{2}
\left[\frac{1}{3}-\frac{\ln 3}{2}-
\frac{a_c(\dw)}{2}\right]\left(\frac{\somt}{ \somd}\right)^2 
\end{equation}
with 
\begin{multline}
\label{valueac}
a_c(\dw)\equiv\\
 \int_1^{\infty} dt \, 
\frac{1}{\left[ t +\ract - \dw (t-\ract)\right]^2}
\left[ \frac{(1-\dw)^2 \left[ 16 t^4 (t+1/2)-1 \right]}{8t(t-1/2)^2(t+1/2)^2(t+1)}
-\frac{\dw(t-1)(4t^2+2t+1)}{t(t-1/2)(t+1/2)^2} \right] .
\end{multline}
If $\epsilon_{\rm solv}/\ew=80$, $a_c=1.2$, and if 
$\epsilon_{\rm solv}=\ew$, $a_c=0.84$.
$G_{[c]}^{\rm exp\, (1)}(\xt)$ is the exponentially decaying function
\begin{multline}
\label{lulu}
G_{[c]}^{\rm exp\, (1)}(\xt)= 
- \frac{\eps}{4}\left(\frac{\somt}{ \somd}\right)^2  \\ 
\left\{\frac{4}{3}e^{-\xt} + \int_1^{\infty} dt \, \frac{e^{(1-2t)\xt}}
{\left[t+\ract - \dw (t-\ract)\right]^2} \frac{4\dw(t^2-1)-(1-\dw)^2(t+1)}
{(t-1/2)(t+1/2)^2}  
\rule{0mm}{7mm}\right.\\
\left. 
+ e^{-\xt} \left[-S_-(\xt)+ \xt S_+(\xt) \right]+\int_1^{\infty} dt \, 
\frac{e^{-2t\xt} \left[(1-\dw)^2-4\dw(t^2-1)\right]}
{t(t+1)(t+1/2)\left[t+\ract - \dw (t-\ract)\right]^2} \rule{0mm}{7mm} \right\}.
\end{multline}

\subsubsection{Global results}

Eventually, the sum of the various contributions at order $\eps$ reads
\begin{equation}
\label{}
D^{ (1)}_{\as}(x)=
\left[- \eps \frac{\ln 3}{4}\left(\frac{\somt}{ \somd}\right)^2 \,
\xt +B_{\as}^{(1)}+G_{\as}^{\rm exp\, (1)}(\xt)\right]
D^{(0)}_{\as}(x),
\end{equation}
where $B_{\as}^{(1)}$ is the sum 
 of the 
various constants,
 \begin{equation}
 \label{valueB1}
B_{\as}^{(1)}\equiv C_{\phi}^{(1)}+Z_\as {\overline B}_{[b]}^{(1)}+B_{[c]}^{(1)},
\end{equation}
and
\begin{equation}
\label{valueGexp}
 G_{\as}^{\rm exp\, (1)}(\xt)\equiv G_{[a]}^{\rm exp\, (1)}(\xt)+ 
Z_\as {\overline G}_{[b]}^{\rm exp\,(1)}(\xt)+
  G_{[c]}^{\rm exp\, (1)}(\xt).
\end{equation}
 $G_{\as}^{\rm exp\, (1)}(\xt)$ is a bounded function of order $\eps$ which decays to zero at least as $\exp(-\xt)$ when $\xt$ goes to infinity.
The coefficient of the term $(-x) \exp[-\kd x]$ in 
$D^{(1)}_{\as}(x)$ indeed coincides with the first-order correction
$\delta\kappa^{\scriptscriptstyle(1)}$  to the screening length in 
the direction perpendicular to the wall calculated in Appendix \ref{AppB} with 
the result \eqref{idkappa}. Therefore
the $\eps$-expansion of 
$D_{\as}(x)$ can be rewritten in terms of the explicit expression
\eqref{valueDzero} of $D^{(0)}_{\as}(x)$  as
\begin{eqnarray}
\label{structDun}
&& D_{\as}(x)=- \sqrt{\frac{2\ew}{\epsilon_{\rm solv}}}
\frac{ e}{\sqrt{\epsilon_{\rm solv}}}\,
  \frac{Z_\as }{\kd}
\nonumber\\
&&\qquad \times
\left\{e^{-(\kappab +\kbun)x}
\left[1+ B_{\as}^{(1)}+G_{\as}^{\rm exp\, (1)}(\xt)
\right] +o(\eps)\right\}.
\end{eqnarray}

The  effective dipole associated with a charge at leading order in $\eps$, $D_{\as}^{(0)}(x)$, is proportional to the mere exponential function $\exp(-\kd x)$. \eqref{structDun} shows that, when first-order corrections are taken into account, the effective dipole varies with the distance from the wall in a more complicated way described by $G_{\as}^{\rm exp\, (1)}(\xt)$, the value of which is derived from \eqref{valueGexp}. The sign of 
$ B_{\as}^{(1)}+G_{\as}^{\rm exp\, (1)}(\xt)$ may vary with the distance $x$  from the wall and depends drastically upon the composition of the electrolyte, the value of the closest approach distance $b$, and the relative dielectric constant of the wall with respect to that of the solvent.

Since $G_{\as}^{\rm exp\, (1)}(\xt)$ tends to zero at large $x$, the effective charge near the wall $\Zeffwa$,  defined from the dipolar interaction by \eqref{comparaison}, reads
\begin{equation}
\Zeffwa=Z_\as\left[1+B_{\as}^{(1)}
+\frac{\kbun}{\kappab}+o(\eps)\right],
\end{equation}
where the term $\kbun/\kappab$ arises from the $1/\kappa$ coefficient in the definition
\eqref{comparaison}. By virtue of \eqref{valueB1} and \eqref{valuecorkappabulk}
\begin{equation}
 \label{valueZeffwa}
\Zeffwa=Z_\as\left\{1+
C_{\phi}^{(1)}+Z_\as \eps \frac{\ln3}{2}\frac{\somt}{ \somd}
-\eps
\left[\frac{a_c(\dw)}{4}-\frac{1}{6}
\right]\left(\frac{\somt}{ \somd}\right)^2+o(\eps)\right\}.
\end{equation}
As exhibited by their diagrammatic origins, the various
terms in $\Zeffwa$ arise both from the nonuniformity of the
density profiles described by  diagram
${\widetilde{\Pi}}_a$ at order $\eps$ 
and from the leading screened interactions via one or two other charges
that appear in  
diagrams ${\widetilde{\Pi}}_b$ and
${\widetilde{\Pi}}_{b^{\star}}$ (Fig.\ref{diagbbe}) and in diagram ${\widetilde{\Pi}}_c$ (Fig.\ref{diagc}). If $\epsilon_{\rm solv}>\ew$, $a_c(\dw)>(2/3)$ and the four-body effective interactions tend to decrease $\Zeffwa$ with respect to its bulk value.

The comparison of the effective charge $\Zeffwa$ near the wall  with its value $\Zeffba$ in the bulk given in \eqref{valuehbslow} leads to
\begin{equation}
\frac{\Zeffwa}{\Zeffba}=1+B_{\as}^{(1)}-A_{\as}^{(1)} 
+\frac{\kbun}{\kappab}+o(\eps).
\end{equation} 
where
\begin{equation}
B_\as^{(1)}-A_\as^{(1)}
=C_{\phi}^{(1)}+B_{[c]}^{(1)}-A_{[c]}^{(1)}.
\end{equation} 
Indeed, according to \eqref{defA1b} and \eqref{valueBb1}
${\overline B}^{(1)}_{[b]}={\overline A}^{(1)}_{[b]}$: 
 the contributions in the bulk and along the wall from diagram ${\widetilde{\Pi}}_b$ compensate each other and there is no term proportional to  
 $Z_\as^2$ in the ratio $\Zeffwa/\Zeffba$.
The final result is written in  \eqref{valueCeff1}.

\section{Conclusion}

In the present Paper we have  introduced the renormalized charge $Z_{\as}^{\rm eff\,{\scriptscriptstyle W} }$ associated with the large-distance dipolar-like  effective interaction between two charges   along an insulating wall, when all charges have the same closest approach distance to the wall. This charge has been explicitly calculated up to order $\eps$ in some limit of infinite dilution and weak coupling when the wall is neutral. 

The renormalized charge could also be calculated in the case of an insulating wall with an external surface charge on it, as it is the case for instance when the wall mimicks a cell membrane.
Indeed, the general method  presently devised for the calculation of $Z_{\as}^{\rm eff\,{\scriptscriptstyle W} }$ holds for any density profiles when the $\eps$-expansions of the latters are known. On the other hand, such $\eps$-expansions could  be obtained by a generalization of the method presented in Refs.\cite{AquaCor01I} and \cite{AquaCor01II}, where the external one-body potential created by the surface charge would be  incorporated in the fugacity.


\appendix

\section{ }
\label{App0}

The present Appendix is devoted to the determination of the large-distance behaviors of exponentially decaying functions, such as those which appear in the resummed Mayer diagrammatics for bulk correlations.
When a function $f(\vecr)$ is  rotationally invariant, its Fourier transform
$f(\vk)\equiv\int d\vecr \exp[-i\vk\cdot\vecr]f(\vecr)$ depends only on the modulus
$k$ of $\vk$. On the other hand, when
$f(\vecr)$ decays faster than any inverse power law of the modulus $r$ of
$\vecr$, then 
its Fourier transform
is an analytic function of the components of $\vk$. When both conditions are
fulfilled by $f(\vecr)$, the $\vk$-expansion of $f(\vk)$ contains only powers of
$\vk^2$. Then the analytic continuation of $f(\vk)=f(k)$ to negative values of $k$
is an even function of $k$ and its inverse Fourier 
transform can be rewritten as the following
integral
\begin{equation}
\label{invFTf}
f(r)=-\frac{i}{4\pi^2} \frac{1}{r} \int_{-\infty}^{+\infty} dk\,
e^{ikr}k f(k),
\end{equation}
where $f(k)$ is a derivable function
of $k$. The one-dimensional integral in
  \eqref{invFTf} can be performed by the method of contour integrals in the complex plane $k=k'+ik''$. (We notice
  that, when $f(\vecr)$ decays algebraically, 
 then the small-$\vk$ expansion of $f(\vk)$ contains  nonanalytic terms 
 involving either $\ln |\vk|$ or odd powers of $|\vk|$ \cite{Gelfand}, and
the present method does not hold.)

The slowest  exponential tail $f^{\rm slow}(r)$ of $f$, defined at the begining of Section \ref{section3}, is determined by 
the singular point of
$f(k)$ 
that is the closest one to the real axis $k''=0$ in the upper complex 
 half-plane. If the latter singular point is a pole $k_{0}$, 
 its contribution to $f(r)$ is given by the residue theorem
\begin{equation}
\label{fslow}
f^{\rm slow}(r)=\frac{1}{2\pi r}
\left.{\rm Res}\left[e^{ikr}k f(k)\right]
\right\vert_{k=k_{0}}. 
\end{equation} 
In the present paper we  consider functions $f(k)$ that contain no exponential term and such that $k_{0}$ is purely imaginary, $k_0=i\kd$. In that case  the inverse decay length
of  $f^{\rm slow}(r)$ is equal 
to the imaginary part of the pole $k_{0}$. (If there were two poles with the same imaginary part and
opposite real parts, then $f^{\rm slow}(r)$ would be an oscillatory exponential tail.)

Moreover, when  $k_{0}=i\kd$ is a pole of rank $1$, $f^{\rm slow}(r)$ is a pure $\exp(-\kd r)/r$ function, whereas  
if $k_{0}=i\kd$ is a pole of rank $m$,
$f^{\rm slow}(r)$ is equal to $\exp(-\kd r)/r$ times a polynomial in $r$ of rank $m-1$.
For instance, if
    $f$ is equal to the convolution $\phid\ast[\phid]^2$,
 the complete contour integral which determines the r.h.s. of \eqref{invFTf}
 gives the expression of  $\phid\ast[\phid]^2$ at any distance $r$. According to \eqref{valuephibk} and \eqref{valuephib2k}, $\phid$ has a pole at $k=i\kd$, while $[\phid]^2$ has a cut
that starts at $k=2i\kd$  and goes along
the imaginary axis up to   $+i\infty$. As a consequence,
\begin{equation}
\label{exemple1}
\left[\frac{e^{-\kd r}}{ r}\ast
\left(\frac{e^{-\kd r}}{ r}\right)^2\right](r) = 
2 \pi\ln 3\, 
\frac{e^{-\kd r}}{\kd r}
+\frac{e^{-2\kd r}}{\kd r}F_b(\kd r),
\end{equation}
where the first term comes from the residue of 
$\exp[ikr]k f(k)$ at the pole $k=i\kd$  of $\phid(\vk)$, while the second term arises from the cut in the definition of $[\phid^2](\vk)$.  At large distances the first term falls off as $\phid$ and the second term as $[\phid^2]$, because
$F_b(\kd r)$ decays
 as a constant times $1/(\kd r)$. Indeed,
\begin{eqnarray} 
\label{valueFb}
F_b(\kd r)&\equiv &-8\pi 
\int_0^{+\infty} dt \frac{e^{-2\kd r\, t}}{4(1+t)^2-1}\nonumber\\
&=&-2\pi\left[e^{3\kd r} {\rm Ei}(-3\kd r)-e^{\kd r} {\rm Ei}(-\kd r)\right],
\end{eqnarray}
 where ${\rm Ei}(-x)$ is the
Exponential-Integral function defined in \eqref{defEi}.
(The large-distance 
behavior of
 the convolution $\phid\ast[\phid]^2$ is indeed dominated by  the pole of
 $\phid(\vk)$ at $k=i\kd$ and not by the branch point of the Fourier transform of 
 $[\phid(r)]^2$ at $k=2i\kd$.) In the case of $\phid\ast[\phid]^2$, the corresponding slowest exponential 
  tail $f^{\rm slow}(r)$ is exactly
 proportional to $\phid=\exp[-\kd r]/r$, since the pole at $k_0=i\kd$ is of rank $1$.

 For the convolution
 $\phid\ast[\phid]^2\ast\phid$, $[\phid(\vk)]^2$ has a pole of rank $2$, and a calculation
 similar to the previous one  gives
\begin{equation}
\label{exemple2}
\left[\frac{e^{-\kd r}}{ r}\ast
\left(\frac{e^{-\kd r}}{ r}\right)^2
\ast\frac{e^{-\kd r}}{ r}\right](r)
= 16 \pi^2\frac{1}{\kd^2}
\left[\frac{\ln 3}{4}\, \kd r +\left(\frac{\ln
3}{4}-\frac{1}{3}\right)\right] \frac{ e^{-\kd r}}{\kd r} 
+ \frac{ e^{-2\kd r}}{\kd r} \, F_c(\kd r),
\end{equation}
where $F_c(\kd r)$ decays as a constant times $1/(\kd r)$, since
\begin{equation}
F_c(\kd r)\equiv 32 \pi^2\frac{1}{\kd^2} 
 \,\int_{0}^{+\infty} dt \frac{e^{-2\kd r t}}{\left[4(1+t)^2-1\right]^2}.
\end{equation}
 The slowest
exponential tail $f^{\rm slow}(r)$ of
$f(r)=\phid\ast[\phid]^2\ast\phid$ is given by the pole of $f(k)$
 at $k=i\kd$, which is of order 2, and $f^{\rm slow}(r)$ is equal to $\exp[-\kd r]/r$ times a
polynomial in $r$  of rank $1$: it is not merely equal to $\phid$.

\section{ }
\label{AppA}

In the present Appendix we study the $\eps$-expansion of the large-distance behavior $h^{\rm as}_{\as\asp}$. The meaning of $\eps$-expansions is detailed in Section \ref{sectionregime}.

First, we calculate the expression of the slowest exponential tail of the graph
$f_m$  with exactly $m$ bonds $\Fcc$ in the definitions 
\eqref{defhcc}-\eqref{defhtt} of 
$f=h^{\rm cc\, {\scriptscriptstyle B}}$, $h^{\rm c-\, {\scriptscriptstyle B}}$, 
$h^{\rm -c\, {\scriptscriptstyle B}}$ or 
$h^{\rm --\, {\scriptscriptstyle B}}$. (For the sake of simplicity we omit the indices for
charge species.) The slowest exponential tail has been defined at the beginning of Section \ref{section3}. 
For $h^{\rm cc\, {\scriptscriptstyle B}}$ and  $h^{\rm c-\, {\scriptscriptstyle B}}$, 
 $f_1$ corresponds
to the first graph in Figs.\ref{hcc} and \ref{hct} respectively, whereas, in the case of 
$h^{\rm --\, {\scriptscriptstyle B}}$, $f_1$
corresponds to the second graph in Fig.\ref{htt}.
Since $\Fcc$ decays
slower than $\Ir$ (at least in the high-dilution and weak-coupling regime) , 
\begin{equation}
\label{structf1}
f_1(k)=\frac{4\pi}{\vk^2+\kd^2} g(k),
\end{equation}
where the poles or branch points of $g(k)$ in the upper complex half-plane 
are more distant from the real axis
 than the pole $k=i\kd$ of $1/(\vk^2+\kd^2)$. 
 (For $h^{\rm cc\, {\scriptscriptstyle B}}$, $g(k)=1$, while, for 
 $h^{\rm c-\, {\scriptscriptstyle B}}$, 
 $g(k)$ is given by \eqref{resumhctbis} with $\Irbar(\vk)=0$ 
 and, for $h^{\rm --\, {\scriptscriptstyle B}}$, $g(\vk)$ is given by the second
 term in the r.h.s. of \eqref{resumhttbis} with $\Irbar(\vk)=0$.) Subsequently, the slowest exponential
tail $f_1^{{\rm slow}}(r)$ of $f_1(r)$ is obtained from \eqref{fslow}
by calculating the 
 residue of 
$\exp[ikr]k f_1(k)$ at $k=i\kd$, with the result
\begin{equation}
\label{f1slow}
f_1^{{\rm slow}}(r)=\frac{1}{2\pi r}
\left.\left[e^{ikr}k f_1(k) (k-i\kd)\right]
\right\vert_{k=i\kd} .
\end{equation}
$f_1^{{\rm slow}}(r)$ takes the form
\begin{equation}
\label{valuef1slow}
f_1^{{\rm slow}}(r)=\frac{e^{-\kd r}}{r}F_{1,0}.
\end{equation}
According to  definitions \eqref{defhcc}-\eqref{defhtt} and \eqref{defIrbar}, 
the Fourier transform of the graph  $f_{m}$ 
with $m$ bonds $\Fcc$ ($m\geq 1$)
 reads 
\begin{equation}
\label{relff}
f_{m}(k)=\left[\frac{-4\pi \Irbar(k)}{k^2+\kd^2}\right]^{m-1}f_1(k).
\end{equation}
According to the argument leading to \eqref{f1slow}, $k=i\kd$ is the singular point of 
$f_{m}(k)$ that is the closest one to the real axis in the  complex upper-half
plane, and the slowest exponential
tail $f_{m}^{{\rm slow}}(r)$ of $f_{m}(r)$ is given by the
residue of 
$\exp[ikr]k f_{m}(k)/(2\pi)$ at $k=i\kd$  according to \eqref{fslow}.
Since $k=i\kd$ is a simple pole  for $f_1$, it  is a
multiple  pole of rank $m$ for $f_{m}$, and the latter residue is equal to 
\begin{equation}
\label{residuefnpun}
\frac{1}{2\pi}\times\frac{1}{(m-1)!} \left.\left[\frac{\partial^{m-1}}{\partial k^{m-1}}
\left(e^{ikr}k f_{m}(k) (k-i\kd)^{m}\right)\right]
\right\vert_{k=i\kd}.
\end{equation}
The expression \eqref{residuefnpun} is equal to $\exp[-\kd r]$  
times a polynomial in $r$ of 
rank $m-1$, $\sum_{p=0}^{m-1}F_{m,p}r^p$,  so that
\begin{equation}
\label{polynom}
f_{m}^{{\rm slow}}(r)
= \frac{e^{-\kd r}}{r}
 \sum_{p=0}^{m-1}F_{m,p}\,r^p.
\end{equation}
The leading term $F_{m,m-1}\,r^{m-1} \exp[-\kd r]$ in the residue \eqref{residuefnpun}
arises from the 
$(m-1)^{\rm th}$ derivative
of $\exp[ikr]$.  Relation \eqref{relff}
and  comparison of 
\eqref{f1slow} with \eqref{residuefnpun} imply that
\begin{equation}
\label{valueA}
F_{m,m-1}=\frac{1}{(m-1)!}
\left[-2\pi\frac{\Irbar(i\kd)}{\kd}\right]^{m-1}F_{1,0}.
\end{equation}

The  large-distance behavior  $f^{{\rm as}}(r)$ of $f(r)$ is the sum of the slowest exponential
tails of all graphs $f_m$,
\begin{equation}
\label{deffas}
f^{{\rm as}}(r)=\sum_{m=1}^{+\infty}f_{m}^{{\rm slow}}(r).
\end{equation} 
It reads
\begin{equation}
\label{sumfslow}
f^{{\rm as}}(r)
= \frac{e^{-\kd r}}{r}
 \sum_{p=0}^{+\infty}r^p\sum_{m=p+1}^{+\infty}F_{m,p},
\end{equation}
and the large-distance behavior  $h^{\rm {\scriptscriptstyle B}\, as}_{\as\asp}$ of $h^{\rm \scriptscriptstyle B}_{\as\asp}$ takes the form \eqref{devtail}.

We now turn to $\eps$-expansions of the previous slowest tails.
The $\eps$-expansion of $\Irbar(\vk)$, written as $\kd^2$ times 
a function of $k/\kd$ (see \eqref{valueIrbar1}), generates an $\eps$-expansion for $f_m$ through \eqref{structf1}
 and \eqref{relff}. 
We recall that $f_1(k)$ is the generic notation for the Fourier transform of the graphs with
 only one bond $\Fcc$ in Figs.\ref{hcc}-\ref{htt}.
 Since the $\eps$-expansion of $\Irbar(k)$
 starts at order $\eps$ (see \eqref{valueIrbar1}),
 the $\eps$-expansion of $f_1(k)$ begins at
 order
$\eps^{n_0}$ with $n_0=0$ if $f=\hccb$, $n_0=1$ 
 if $f=\hctb$ or $f=\htcb$, and $n_0=2$ if 
 $f=\httb$. 
For the same reason, 
the $\eps$-expansion of $f_{m}$
 starts at the order $\eps^{n_0+m-1}$, and so does 
the $\eps$-expansion of the 
coefficient $F_{m,p}$ of $r^p$ in \eqref{polynom},
\begin{equation}
F_{m,p}=\eps^{n_0}\sum_{q=m-1}^{+\infty}F_{m,p}^{(n_0+q)} \eps^q.
\end{equation} 
As a consequence, the large-distance  tail $f^{{\rm as}}(r)$ of $f(r)$ defined in \eqref{deffas}
has an $\eps$-expansion of
the form
\begin{eqnarray}
\label{seriesfslow}
f^{{\rm as}}(r)
&=& \frac{e^{-\kd r}}{r}
 \eps^{n_0}\sum_{p=0}^{+\infty} r^p
\sum_{q=p}^{+\infty} a_{p}^{(n_0+ q)} \eps^q\nonumber\\
&=& \frac{e^{-\kd r}}{r}
 \eps^{n_0}\sum_{q=0}^{+\infty} \eps^q\sum_{p=0}^{q}a_{p}^{(n_0+ q)}\,r^p,
\end{eqnarray}
with $a_{p}^{(n_0+ q)}=\sum_{m=p+1}^{q+1}F_{m,p}^{(n_0+q)}$,
and the $\eps$-expansion of $h_{\as\asp}^{\rm {\scriptscriptstyle B}\, as}(r)$ has the structure \eqref{devepshas}.
In \eqref{seriesfslow} the coefficient $a_{q}^{(n_0+q)}$ of 
$\eps^{n_0+q}r^q$ has a simple expression,
\begin{equation}
\label{prop}
a_{q}^{(n_0+q)}=F_{q+1,q}^{(n_0+q)}.
\end{equation}
The leading tail at order $\eps^{n_0+q}$  in the $\eps$-expansion of $f^{{\rm as}}$ 
is proportional to $r^{q}\exp[-\kd r]/r$. It comes from the
leading $r^q$-term  in the slowest tail $f_{q+1}^{{\rm slow}}(r)$ of the graph with $q+1$
bonds calculated at its
lowest order in $\eps$, namely $\eps^{n_0+q}$.
 
The sum $f^{{\rm as}\, \star}(r)$  of the  leading tails at every order in $\eps$ in 
the $\eps$-expansion of  $f^{\rm as}$ 
 reads
\begin{equation}
\label{deffslowstar}
f^{{\rm as}\, \star}(r)\equiv
\frac{e^{-\kd r}}{r} \eps^{n_0}
\sum_{q=0}^{+\infty} a_q^{(n_0+q)}\left(\eps r\right)^{q}.
\end{equation}
 It can be calculated explicitly by virtue of \eqref{prop}, because $F_{q+1,q}^{(n_0+q)}$
 is given by \eqref{valueA}, where $\Irbar(k)$ is 
replaced  by $\Irbar^{(1)}(k)$, while $F_{1,0}$ is replaced by the first term in
its $\eps$-expansion, namely $F_{1,0}^{(n_0)}$.
We get
\begin{equation}
\label{valueF}
\eps^{n_0+q}F_{q+1,q}^{(n_0+q)}=\frac{1}{q!}
\left[-2\pi\frac{\Irbar^{(1)}(i\kd)}{\kd}\right]^{q}\,
\eps^{n_0}F_{1,0}^{(n_0)}.
\end{equation}
\eqref{valueF} implies that the coefficient of $r^{q}$ in the definition \eqref{deffslowstar} of $f^{{\rm as}\, \star}(r)$  is indeed such that
$f^{{\rm as}\, \star}(r)$ coincides with the series of  an exponential,
\begin{equation}
\label{resumslow}
f^{{\rm as}\, \star}(r)
= \frac{1}{r}
e^{-(\kd +\delta\kappa^{\star}_{\rm{\scriptscriptstyle B}})r}\,\eps^{n_0}F_{1,0}^{(n_0)}
\quad\mbox{with}\quad
\delta\kappa^{\star}_{\rm{\scriptscriptstyle B}}=2\pi
\frac{ \Irbar^{(1)}(i\kd)}{\kd}
\end{equation}
According to \eqref{renormkappa1},
$\delta\kappa^{\star}_{\rm{\scriptscriptstyle B}}$ coincides with the  correction
 of order $\eps$ in  $\kappa^{\rm
{\scriptscriptstyle B}}$, $\delta\kappa^{\star}_{\rm{\scriptscriptstyle B}}=\kbun$.
By comparison with 
\eqref{valuef1slow}, the relation 
\eqref{resumslow} can be rewritten as
\begin{equation}
\label{resumslowbis}
f^{{\rm as}\, \star}(r)
=e^{-\kbun r} \,f_1^{{\rm as}\, (n_0)}(r),
\end{equation}
where $f_1^{{\rm as}\, (n_0)}=f_1^{{\rm slow}\, (n_0)}$ is proportional to
$\exp[-\kd r]/r$.
In other words, the sum  $f^{{\rm as}\, \star}(r)$ of the leading tails 
at every order in $\eps$ in the $\eps$-expansion of $f^{\rm as}(r)$ around its
$\exp[-\kd r]/r$ behavior in the vanishing-$\eps$ limit  is
equal to $\exp[- \kbun r]$ times 
the large-distance behavior  of the graph $f_1$ with only one bond $\Fcc$ 
calculated at the first order
$\eps^{n_0}$. 
When $f=\hccb$, $n_0=0$ and $f_1^{{\rm as}\, (n_0)}$ coincides with the diagram ${\widetilde \Pi}_a=\Fcc$ shown in Fig.\ref{diaga}.
 When $f=\hctb$ or $\htcb$, $n_0=1$  and
$f_1^{{\rm as}\, (n_0)}$  
is  the $\exp[-\kd r]/r$ tail 
of diagram ${\widetilde{\Pi}}_b$ or ${\widetilde{\Pi}}_{b^{\star}}$, respectively (see
Fig.\ref{diagbbe}), the amplitude of which is of  order $\eps$ with respect to that of $\Fcc$. When $f=\httb$,  $n_0=2$ and 
$f_1^{{\rm as}\, (n_0)}$ is the $\exp[-\kd r]/r$ tail of the diagram built with
$(1/2)\left[\Fcc\right]^2\ast\Fcc\ast
 (1/2)\left[\Fcc\right]^2$, the amplitude of which is of  order $\eps^2$ with respect to that of $\Fcc$.


\section{ }
\label{AppC}

In the present Appendix, we consider  the limit $\bt \ll 1$ at fixed $\beta e^2/(\epsilon_{\rm solv}b)$ and we calculate the explicit values of 
$C_{\phi}^{(1)}$ and
${\overline G}_{\phi}^{\rm exp\, (1)} (\xt)$ [see definition \eqref{formDunphi}] up to terms of order $\eps$ times  $\ln\bt$ plus a function of  $\beta e^2/(\epsilon_{\rm solv}b)$.
  According to
\eqref{formvalueC1}, \eqref{formvalueGphi1} and \eqref{rapBZ}, the values  are determined from 
$H^+(\xt,\vq^2=0)$ and from  its derivative with respect to $\xt$ at point 
$\xt=\bt$. 

First, we calculate $H^{+(1)}(\xt,\vq^2=0)$.
According to \eqref{HTun}, at first order in $\eps$,
$H^{+(1)}=-\left.{\mathcal{T}}^{+}[1]\right\vert^{(1)}$. In the following, the definition 
\eqref{defTp} of ${\mathcal{T}}^{+}[1]$ is rewritten for $\vq^2=0$, thanks to an integration by parts, as 
\begin{equation}
\label{C1}
{\mathcal{T}}^{+} [1] (\xt;\vq^2=0)=K(\bt)-K(\xt)
\end{equation}
with
\begin{equation}
\label{defK}
K(\xt)=\frac{1}{2\somd} \int_{\xt}^{+\infty} du
\left[1-e^{-2(u-\xt)}\right]
\soma \roab Z_{\as}^2
 \left[\frac{\roa(u/\kd)}{\roab}-1\right].
\end{equation} 
With these definitions
\begin{equation}
\lim_{\xt\rightarrow +\infty}H^{+\,(1)}(\xt,\vq^2=0)
=-K^{(1)}(\bt),
\end{equation} 
and \eqref{formvalueGphi1} is rewritten as
\begin{equation}
\label{relGK}
{\overline G}_{\phi}^{\rm exp\, (1)} (\xt)
=K^{(1)}(\xt).
\end{equation}

From now on, we consider the regime $\bt \ll 1$ at fixed $\beta e^2/(\epsilon_{\rm solv}b)$. 
 $K^{(1)}(\xt)$ is determined from the 
density profiles up to order $\eps$ given in \eqref{profileform}. At leading order in the limit where $\bt$ vanishes,  by virtue of \eqref{valueLb} and \eqref{valuePhi}, 
\begin{equation}
\label{defR}
-Z_{\as}^2 \eps \Lb(\xt;\bt) - Z_\as\beta e
 \Phi^{(1)}\left(x;\kd,\bt,\frac{\beta e^2}{\epsilon_{\rm solv} b}\right)
=
\eps R_\as(\xt)+ \cO(\eps\bt),
\end{equation} 
where $R_\as(\xt)$ is a linear combination of  functions of $\xt$, where one coefficient  involves $\ln \bt$ plus a function of
$\beta e^2/(\epsilon_{\rm solv}b)$ in such a way that the limit of this sum is finite if $\dw\leq 0$ and $\bt=0$. $\cO(\eps\bt)$ is a short notation  for  terms of order $\eps\bt$. Then, 
the expression  of the  density profiles at order $\eps$ can be rewritten at leading order in $\bt$ as
\begin{equation}
\label{profilestruct}
\frac{\roa(x)}{\roab}-1 =
\left\{\exp\left[- Z_{\as}^2 \frac{ \beta e^2}{\epsilon_{\rm solv}} \Vimsc(x)\right]-1\right\}
+\eps R_\as(\xt) 
+  \left\{\exp\left[-Z_{\as}^2 \frac{ \beta  e^2}{\epsilon_{\rm solv}} \Vimsc(x)\right]-1\right\} 
\eps R_\as(\xt)+\cO(\eps\bt,\eps^2),
\end{equation}
where $\cO(\eps^2)$ stands  both for terms of orders written in \eqref{orders} with $\Gamma\propto\eps^{2/3}$ (as in \eqref{profileform}), and for terms of order $\eps^2$ times a possible sum of a $\ln\bt$ term  and a function of
$(\beta e^2/\epsilon_{\rm solv})$, which is  similar to the coefficient in  $R_\as(\xt) $ (see the comment after \eqref{defR}).
$\eps R_\as(\xt)$ is a bounded integrable function  of  only $\xt=\kd x$, while 
$(\beta e^2/\epsilon_{\rm solv})\Vimsc(x)$ is a function of both $x/(\beta e^2/\epsilon_{\rm solv})$ and $\kd x$.  As  
already noticed in Section 3.3 of Ref.\cite{AquaCor01I},  an 
integral  where $\left\{\exp\left[- Z_{\as}^2 (\beta e^2/\epsilon_{\rm solv}) \Vimsc(x)\right]-1\right\}$ is multiplied by $\eps$ times a bounded integrable function of $\xt$
is of order $\eps^2$ times a function of $\bt$ and 
$\beta e^2/(\epsilon_{\rm solv}b)$, which has a structure similar to $R_\as(\xt) $ in \eqref{profilestruct}. Thus, according to \eqref{defK} and \eqref{profilestruct}, 
the expression of  $K(\xt)$ at order $\eps$, $K^{(1)}(\xt)$, can be written at leading order in $\bt$ 
  as the sum of only two
contributions  
\begin{equation}
  \label{devT}
 K^{(1)}(\xt)
 =K_{\rm im}^{(1)}(\xt)
+ \eps K_{R}(\xt)+\cO(\eps\bt),
\end{equation}
where  $K_{\rm im}(\xt)$ and $K_{R}(\xt) $ are defined as 
$K(\xt)$ by replacing    $[\roa (u/\kd)/\roab]-1$ in  \eqref{defK} by
\linebreak $\left\{\exp \left[-   Z_{\as}^2 (\beta e^2/\epsilon_{\rm solv})\Vimsc (u/\kd )\right]-1\right\}$
and $ R_{\as}(u)$, respectively.
 $ R_\as(u)$  defined in \eqref{defR} is given by 
the explicit expressions
  \eqref{valueLb} and \eqref{valuePhi}. 
$K_{R}(\xt)$ is calculated by reversing the order of the
 integrations $\int du$ from the definition of $K_R$ and $ \int dt$  from the expressions of $\Lb$ and $\Phi^{(1)}$ in $R_{\as}(u/\kd)$. Eventually,
\begin{equation}
  \label{relHK}
  K_{R} (\xt)= K_{\Lb}(\xt) + K_{\Mb} (\xt) + K_{\rm Exp}
  (\xt) + K_{S_-}  (\xt),
\end{equation}
where the four contributions arising from
the terms in the density profiles involving either
$\Lb (x)$, $\Mb (x)$, $\exp(-\xt)$ or $S_{-}$ are written 
in \eqref{bbis}, \eqref{cbis}, \eqref{ebis} and \eqref{dbis}, respectively, by virtue of \eqref{relGK}.

Now we  show that 
\begin{equation}
\label{lin}
K_{\rm im}^{(1)}(\xt)
=K^{\rm lin}_{\rm im}(\xt)+\cO(\eps\bt),
\end{equation} 
where $K^{\rm lin}_{\rm im}(\xt)$ is deduced from $K_{\rm im}(\xt)$, defined after \eqref{devT}, by linearizing  the exponential that contains the bulk-screened self-image interaction  $Z_{\as}^2 ( e^2/\epsilon_{\rm solv})\Vimsc $. Indeed,
$K_{\rm im}(\xt)-K^{\rm lin}_{\rm im}(\xt)=Q(\xt)$ with
\begin{equation}
\label{defFK}
Q(\xt)\equiv\frac{1}{2\somd} \int_{\xt}^{+\infty} du
\left[1-e^{-2(u-\xt)}\right]
\soma \roab Z_\as^2\left[\exp\left(\frac{\dw}{2}
 Z_\as^2\frac{\eps}{u}e^{-2u}\right)-1
-\frac{\dw}{2}Z_\as^2\frac{\eps}{u}e^{-2u}\right].
\end{equation} 
Since the functions in the square brackets are positive and  $1-\exp[-2(u-\xt)]\leq 1-\exp[-2(u-\bt)]$ for $\xt\geq \bt$,  
\begin{equation}
\label{borne}
0\leq Q(\xt)\leq Q(\bt)=Q\left(\bt,\eps;\frac{\beta e^2}{\epsilon_{\rm solv}b}\right).
\end{equation} 
The double expansion of $Q(\bt,\eps; \beta e^2/(\epsilon_{\rm solv} b))$ in powers of $\eps$ and $\bt$ at fixed $\beta e^2/(\epsilon_{\rm solv} b)$ can be calculated thanks to the following formula (already used in Ref.\cite{AquaCor01II}). We set $\eps_{\as}\equiv 
 Z_\as^2\eps/2$. The function $f$ in the integrand of \eqref{defFK} is a function of $u$ that depends on $\eps_{\as}$ as if $f$ were a function of the two independent variables $u$ and $u_1=u/\eps_{\as}$. We write it as
\begin{equation}
f\left(u,\frac{u}{\eps_{\as}}\right)=
g(\eps_{\as}u_1,u_1).
\end{equation} 
Since $\bt\ll 1$ the integral $\int_{\bt}^{+\infty}$ can be split into the sum of integrals $\int_{\bt}^{\lt}$ and 
$\int_{\lt}^{+\infty}$ with $\bt<\lt$ and $\eps_{\as}\ll \lt \ll 1$. Then
\begin{eqnarray}
  \label{decompint}
 && \underset{\eps_{\as} \rightarrow 0}{{\rm Exp}}
\left[  \int_{\bt}^{+\infty} du \,
 f \left( u, \frac{u}{\eps_{\as}}\right) \right]=\\ 
&&\qquad\qquad \qquad\qquad \qquad\qquad 
\underset{(\lt/\eps_{\as})\rightarrow  +\infty }{{\rm Exp}} \, \eps_{\as}\int_{\bt/\eps_{\as}}^{\lt/\eps_{\as}} d u_1 \, \underset{\eps_{\as} 
\rightarrow 0}{{\rm Exp}} \, \, g( \eps_{\as}  u_1,  u_1 )  +
\underset{\lt\rightarrow  0 }{{\rm Exp}}
 \int_{\lt}^{+\infty} du \, \underset{\eps_{\as} \rightarrow 0}{
{\rm Exp}} \, \, f\left(u , \frac{u}{\eps_{\as}}\right), \nonumber
\end{eqnarray}
where 
 $\underset{\delta \rightarrow 0}{{\rm Exp}}$ denotes an $\delta$-expansion.
The identity holds, because when $u_1<\lt/\eps_{\as}$ then
$\eps_{\as} u_1\ll 1$, and when $u>\lt$ 
then $(\eps_{\as}/u)\ll 1$. When  \eqref{decompint} is applied to  the calculation of $Q(\bt,\eps; \beta e^2/(\epsilon_{\rm solv} b))$, the second integral in \eqref{decompint} gives a term of order $\eps_{\as}^2\propto\eps^2$, while
$ \underset{\eps_{\as} \rightarrow 0}{{\rm Exp}} \left\{1-\exp\left[-2(\eps_{\as}u_1-\bt\right]\right\}$ behaves as $-2\bt$, so that  the first integral provides a contribution which starts at  order $\eps\bt$; more precisely,
\begin{equation}
\label{valueQb}
Q\left(\bt,\eps;\frac{\beta e^2}{\epsilon_{\rm solv}b}\right)=  \eps \bt \frac{\dw}{2}
\frac{\soma \roab Z_{\as}^4 g_{\as}}{\somd}
+ \cO(\eps^2),
\end{equation} 
where $g_{\alpha}$ has been defined in \eqref{defggamma}--\eqref{defgbis}, and $ \cO(\eps^2)$ is equal to $\eps^2$ times a function with a structure similar to that of $R_{\as}(x)$ in \eqref{defR}. 
The result \eqref{valueQb} combined with   inequalities \eqref{borne} leads to \eqref{lin}. Eventually \eqref{devT} can be written as 
\begin{equation}
 K^{(1)}(\xt) \equiv K^{\rm lin}_{\rm im}(\xt)+\eps K_{R}  (\xt)
+\cO(\eps\bt).
 \end{equation}
$K_{R}  (\xt)$ is given in \eqref{relHK} and, according to its definition,
\begin{equation}
  \label{valueKself}
  K_{\rm im}^{\rm lin}  (\xt)=  \eps \,\frac{\dw}{4}\, \frac{\somq}{\somd} \left[ 
 e^{2\xt} \Ei(-4\xt) - \Ei(-2\xt) \right],
\end{equation}
where ${\rm Ei}(u)$ is the Exponential-Integral function defined in \eqref{defEi}.

The derivative  $\partial H^+(\xt,\vq^2=0)/\partial \xt$ at first order in 
$\eps$ must be performed more
carefully. The reason that leads to $H^{+(1)}=-{\mathcal{T}}^{+(1)}[1]$ also implies that
\begin{equation}
\left.\frac{\partial H^{+}}{\partial \xt}
\right\vert^{(1)}
=\left.\frac{d K}{d\xt}
\right\vert^{(1)},
\end{equation}
and, similarly to \eqref{devT},
\begin{equation}
  \label{devpartialT}
\left.\frac{d K}{d \xt}\right\vert^{(1)}=
\left.\frac{d K_{\rm im}}{d \xt}\right\vert^{(1)}+
 \eps \frac{d K_{R}}{d \xt}+\cO(\eps\bt).
\end{equation}
The decomposition \eqref{decompint} leads
to
\begin{equation}
\left.\frac{d K_{\rm im}}{d \xt}
\right\vert^{(1)}_{\xt=\bt}=
\frac{\dw}{2} \, \eps \,  \left\{
\frac{\somg \rogb Z_{\gamma}^4 g_{\gamma}}{\somd}
-\frac{\somq}{\somd}\left[C+\ln\left(4\bt\right)\right]\right\}.
\end{equation}
We notice that
\begin{equation}
\label{nonid}
\left.\frac{d K_{\rm im}}{d \xt}
\right\vert^{(1)}
\not=\frac{d K_{\rm im}^{(1)}}{d \xt}.
\end{equation}
The reason is that, though $K_{\rm im}^{(1)}$ is only
a function of $\xt$, $K_{\rm im}^{(2)}$ involves a contribution that is equal to $\eps^2$ times a function  of the two variables $\xt$ and 
$\xt/\eps$, and the  derivative of the latter contribution  with respect 
to the second argument $\xt/\eps$ is of order $\eps$. 
The existence of such a contribution in  $K_{\rm im}^{(2)}$  is due
to the fact that the function $ Z_{\as}^2 (\beta e^2/\epsilon_{\rm solv})\Vimsc (x;\kd)$ in the expression 
 \eqref{valueVselfscreened}
 of the density profile 
  varies both
over the Bjerrum length $\beta e^2/\epsilon_{\rm solv}$ and over the screening length $\xid$.  (This structure arises directly when the equation obeyed by $H^+$ 
is solved by a multi-scale expansion method.) 

\section{ }
\label{AppB}

\begin{figure*}[H]
\includegraphics[width=0.96 \textwidth]{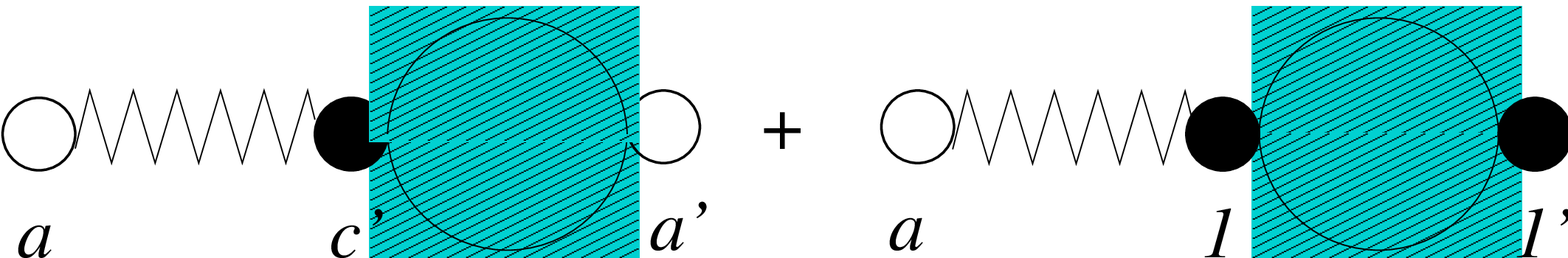}
\caption{\label{hct}}
\end{figure*}

\begin{figure*}[H]
\includegraphics[width=0.96 \textwidth]{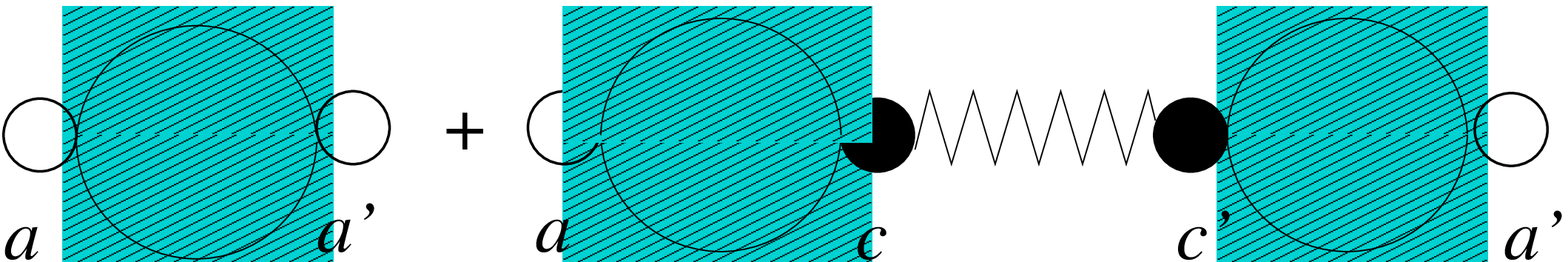}
\caption{\label{htt}}
\end{figure*}

In the present Appendix we consider the large-$x$ behavior $D^{\rm c\,as}_{\as} (x)$ of the dipole
$D^{\rm c}_{\as} (x)$ that appears in the large-$y$
 tail $D^{\rm c}_{\as} (x)D^{\rm c}_{\asp} (x')/y^3$ of $\hcc_{\as\asp}$. Calculations are not as straightforwrd as for $\hccb_{\as\asp}$ in the bulk, and  we calculate only the sum  $D^{\rm c\,as\,\star}_{\as} (x)$ of the leading large-$x$ terms  at every order $\eps^q$ in
the $\eps$-expansion of   
$D^{\rm c\, as}_{\as} (x)$ around its infinite-dilution and vanishing-coupling  limit
$D^{\rm c\,as\,(0)}_{\as} (x)$. According to 
\eqref{tailhcc},
\begin{equation}
\label{defDc}
D^{\rm c}_{\as} (x)\equiv \frac{e}{\sqrt{\epsilon_{\rm solv}}}
Z_\as \left[{\overline D}_{\phi}(x) + \overline {C}^{{\rm c}-}(x)\right].
\end{equation}
After insertion of the graphic representation \eqref{defhct} of $\hct$ in the definition \eqref{valueCct} of $\overline {C}^{{\rm c}-}(x)$,  the latter can be written as
\begin{equation}
\overline {C}^{{\rm c}-}(x)=\sum_{m=1}^{+\infty} J_m(x),
\end{equation}
where $J_m(x)$ is the contribution to $\overline {C}^{{\rm c} -}(x)$ from the graph in $\hct_{\as\gs''}$  with $m$ bonds $\Fcc$(see Fig.\ref{hct}).

The graph with $m$ bonds $\Fcc$ in $\hct_{\as\gs''}$ also contains $m$ bonds $\Ir$. Therefore,
according to the scaling analysis of Section \ref{section42}, the $\eps$-expansion of 
$J_m(x)$ starts at order $\eps^m$,  $J_m(x)=\sum_{q=m}^{+\infty}J_m^{(q)}(x)$, where $J_m^{(q)}(x)$ denotes the term of order $\eps^q$.  Therefore, the leading tail  at order $\eps^q$ in the large-$x$ behavior  $\overline {C}^{\rm c-\, as}(x)$ of $\overline {C}^{{\rm c}-}(x)$ can arise only from the leading tails of the $J_m(x)$'s  with $m\leq q$.
Though we are not able to derive the leading tail of $J_m(x)$ systematically, we  expect, by analogy with the $F_m$'s in the bulk case,   that the leading tail of $J_m(x)$ has the same $x$-dependence as  the  leading tail of  the first term $J_m^{(m)}(x)$ in the 
$\eps$-expansion of   $J_m(x)$.
As shown herefater, the leading tail  $J_m^{(m)\, {\rm as}}(x)$ of $J_m^{(m)}(x)$ is proportional to $\eps^m(\xt-\bt)^m\exp[-(\xt-\bt)]$.  As a consequence,  the leading tail  at order $\eps^q$ in the large-$x$ behavior  $\overline {C}^{\rm c-\, as}(x)$ of 
$\overline {C}^{{\rm c}-}(x)$ coincides with the leading tail $J^{(q)\, {\rm as}}_q(x)$ of $J_q^{(q)}(x)$, and the sum $\overline {C}^{\rm c-\, as\,\star}(x)$ of the leading tails at every order $\eps^q$ in
$\overline {C}^{\rm c-\, as}(x)$ is 
$\overline {C}^{\rm c-\, as\,\star}(x)=\sum_{q=1}^{+\infty}  J^{(q)\, {\rm as}}_q(x)$.
Similarly, for $D^{\rm c}_{\as} (x)$ defined in \eqref{defDc} 
\begin{equation}
\label{valueDcstar}
D^{\rm c\,as\,\star}_{\as} (x)
=\frac{e}{\sqrt{\epsilon_{\rm solv}}}
Z_\as \left[{\overline D}^{(0)}_{\phi}(x) + \sum_{q=1}^{+\infty}  J^{(q)\, {\rm as}}_q(x)\right].
\end{equation}

The term $J_q^{(q)}(x)$ of order $\eps^q$ in the $\eps$-expansion of $J_q(x)$
is obtained by replacing  every bond $\Fcc$
by its zeroth-order expresssion $F^{\rm c \, c\,(0)}$, every $\Ir$  by its lowest-order value 
  $[F^{\rm c \, c\,(0)}]^2/2$, and 
  every weight $\roa(x_n)$ by its  bulk 
  value $\roab$. 
 Inspection of $J_q^{(q)}(x)$ for small values of $q$
 shows that only the part $\phi_{\rm sing}^{(0)}=\phid$ of $\phi^{(0)}$ does
 contribute to the leading tail  $J^{(q)\, {\rm as}}_q(x)$.
  Let us denote by
 ${\widehat J}_q^{(q)}(x)$ the corresponding part in 
 $J_q^{(q)}(x)$.
 For the sake of simplicity, we relabel point pairs $\{p,p'\}\equiv\{(\vecr_p,\gamma_p),(\vecr'_p,\gamma'_p)\}$, with $p=1,\ldots, q$, $x_{q}=x_{c'}$ and $x'_q=x''$,  in
 the opposite sense and we set $u_p\equiv \xt_{q-(p-1)}$ and
 $u'_p\equiv \xt'_{q-(p-1)}$. 
  By using \eqref{valueDzero}, \eqref{phizerobulk} and the change of 
  variable
  $t=2\sqrt{1+q^2}$, we get
  \begin{eqnarray}
  \label{defJhatqq}
  {\widehat J}_q^{(q)}(x)=
   -\frac{1}{\kd}\sqrt{\frac{2\ew}{\epsilon_{\rm solv}}}
\left(-\eps \frac{1}{4} \left(\frac{\somt}{\somd}\right)^2\right)^q &&\int_2^{+\infty} \frac{dt_q}{t_q}
  \int_{\bt}^{+\infty}du_q \, e^{-|\xt-u_q|}
  \int_{\bt}^{+\infty}du'_q \,e^{-t_q|u_q-u'_q|}\\
  &&\cdots
  \int_2^{+\infty} \frac{dt_1}{t_1}
  \int_{\bt}^{+\infty}du_1 \, e^{-|u'_2-u_1|}
  \int_{\bt}^{+\infty}du'_1 \,e^{-t_1|u_1-u'_1|}e^{-(u'_1-\bt)}\nonumber.
\end{eqnarray} 
Next steps of the calculations involve the following formulae,
\begin{equation}
I(u_1)\equiv 
\int_{\bt}^{+\infty}du'_1 \,e^{-t_1|u_1-u'_1|}e^{-(u'_1-\bt)}
= \frac{2t_1}{t_1^2-1} e^{-(u_1-\bt)}-
\frac{1}{t_1-1} e^{-t_1(u_1-\bt)}
\end{equation}
and
\begin{equation}
 \int_{\bt}^{+\infty}du_1 \, e^{-|u'_2-u_1|} I(u_1)
 = \frac{2t_1}{t_1^2-1} (u'_2-\bt)e^{-(u'_2-\bt)}+R_0(u'_2-\bt),
 \end{equation}
 where $R_p(u)$ denotes a function whose slowest exponential tail 
is equal to $\exp(-u)$ times a polynomial of rank $p$ in the variable $x$.
 More generally, we find 
 \begin{equation}
 \int_{\bt}^{+\infty}du'_p \, 
 e^{-t_p|u_p-u'_p|} (u'_p-\bt)^{p-1}e^{-(u'_p-\bt)}
 = \frac{2t_p}{t_p^2-1} (u_p-\bt)^{p-1}e^{-(u_p-\bt)}+R_{p-2}(u_p-\bt)
 \end{equation}
 and 
 \begin{equation}
 \int_{\bt}^{+\infty}du_p \, e^{-|u'_{p+1}-u_p|} (u_p-\bt)^{p-1}
 e^{-(u_p-\bt)}
 = \frac{1}{p} (u'_{p+1}-\bt)^{p}e^{-(u'_{p+1}-\bt)}+R_{p-1}(u'_{p+1}-\bt).
 \end{equation}
 Eventually, the multiple integral in \eqref{defJhatqq} is equal to
 \begin{equation}
 \label{valueJbar}
   \left(\int_2^{+\infty} dt \frac{2}{t^2-1}\right)^q
 \frac{(\xt-\bt)^q}{q!} e^{-(\xt-\bt)} +R_{q-1}(\xt-\bt)
  \end{equation} 
where $\int_2^{+\infty} dt\, 2/(t^2-1)=\ln 3$, and 
\begin{equation}
J^{(q)\, {\rm as}}_q(x)=
 -\frac{1}{\kd}\sqrt{\frac{2\ew}{\epsilon_{\rm solv}}}
\left(-\eps\frac{\ln 3}{4}
\left(\frac{\somt}{\somd}\right)^2\right)^q \frac{(\xt-\bt)^q}{q!} e^{-(\xt-\bt)}.
\end{equation}
Therefore, $D^{\rm c\, as\, \star}_{\as} (x)$, given by \eqref{valueDcstar} with ${\overline D}^{(0)}_{\phi}(x)$ written in \eqref{valueDzero}, proves 
 to be  the series of an exponential
whose argument is proportional to $\eps (\xt-\bt)$,
\begin{equation}
\label{totu}
D^{\rm c\, as\, \star}_{\as} (x)=
-\frac{ e}{\sqrt{\epsilon_{\rm solv}}}\,
  Z_\as 
\sqrt{\frac{2\ew}{\epsilon_{\rm solv}}}
\frac{e^{-(\kd+\delta\kappa^{\star})(x-b)}}{\kd}
\end{equation}
with
\begin{equation}
\frac{\delta\kappa^{\star}}{\kd}=\eps\frac{\ln 3}{4}
\left(\frac{\somt}{\somd}\right)^2.
\end{equation}
By virtue of \eqref{valuecorkappabulk},  $\delta\kappa^{\star}$  
coincides with the first-order correction $\kbun$ to the screening length in the bulk.
We notice that \eqref{totu} can be rewritten as 
\begin{equation}
\label{tito}
D^{\rm c\, as\, \star}_{\as} (x)=D_{\as}^{(0)}(x)\,
e^{-\delta\kappa^{\star}(x-b)}.
\end{equation}
\eqref{tito} corresponds to the relation \eqref{relhstarh0}  in the bulk case.


\vskip 2cm

{\bf FIGURE CAPTIONS}

{\bf Fig.1} Representation of $h^{cc}_{\as\asp}(\vecr,\vecr')$ as the graph series defined in \eqref{defhcc}. A wavy line represents a bond $\Fcc$ and a grey disk stands for a bond $I$. A couple of variables $(\vecr_i,\gamma_i)$ is associated with every circle. For a white circle $a=(\vecr,\alpha)$ [or $a'=(\vecr',\alpha')$],
$\vecr$ and $\alpha$ are fixed, whereas,  for a black circle $i=(\vecr_i,\gamma_i)$, $\vecr_i$ and $\gamma_i$ are integrated over with the measure $\int d{\vecr}_i \sum_{\alpha_i} \rho_{\alpha_i}(\vecr_i)$.

{\bf Fig.2} Graphic representation of definition \eqref{defhct} for $h^{c-}_{\as\asp}(\vecr,\vecr')$.

{\bf Fig.3} Graphic representation of definition \eqref{defhtt} for $h^{--}_{\as\asp}(\vecr,\vecr')$.

{\bf Fig.4} Diagrams in $h^{\rm cc}_{\as\asp}(\vecr,\vecr')$ that contribute to the correction of  order $\eps$ in the screening length. A double wavy line denotes  a bond $(1/2)\left[\Fcc\right]^2$.

{\bf Fig.5} Diagrams in $h^{\rm c-}_{\as\asp}(\vecr,\vecr')$ that contribute to the correction  of  order $\eps$ in the screening length.

{\bf Fig.6} Diagrams in $h^{\rm --}_{\as\asp}(\vecr,\vecr')$ that contribute to the correction  of  order $\eps$ in the screening length. 

{\bf Fig.7} Diagram ${\widetilde \Pi}_a$.

{\bf Fig.8} Diagrams ${\widetilde \Pi}_b$ (on the left) and 
${\widetilde \Pi}_{b \star}$ (on the right).

{\bf Fig.9} Diagram ${\widetilde \Pi}_c$.


\newpage

\begin{figure*}[H]
\includegraphics[width=0.96 \textwidth]{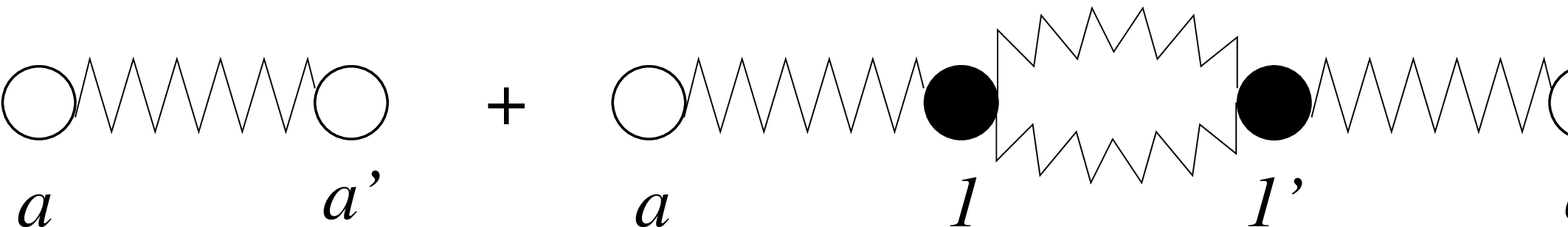}
\caption{\label{hccun}}
\end{figure*}

\begin{figure*}[H]
\includegraphics[width=0.96 \textwidth]{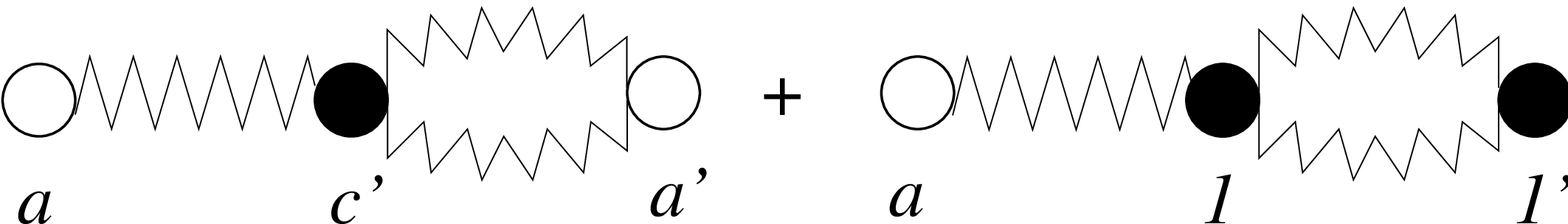}
\caption{\label{hctun}}
\end{figure*}

\begin{figure*}[H]
\includegraphics[width=0.96 \textwidth]{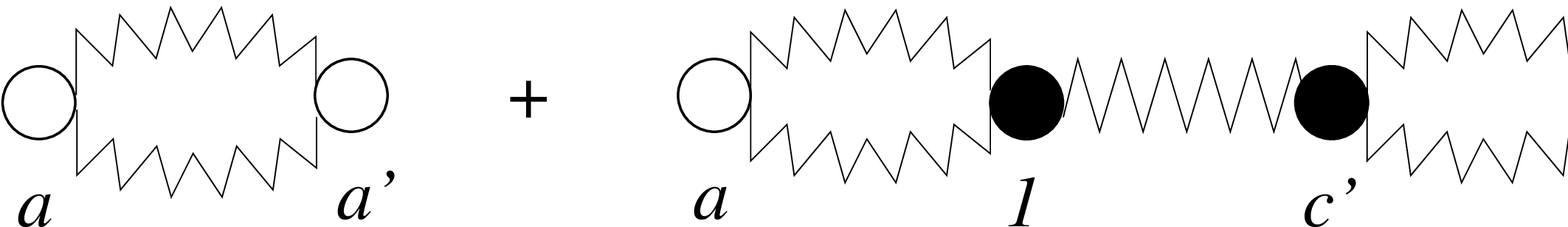}
\caption{\label{httun}}
\end{figure*}

\begin{figure*}[H]
\includegraphics[width=2cm]{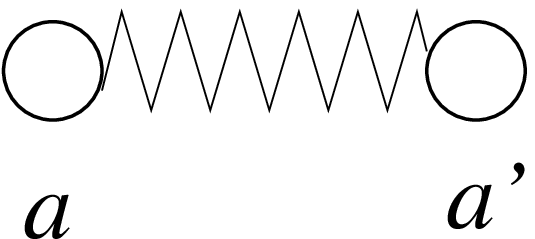}
\caption{\label{diaga}}
\end{figure*}

\begin{figure*}[H]
\includegraphics[width=3cm]{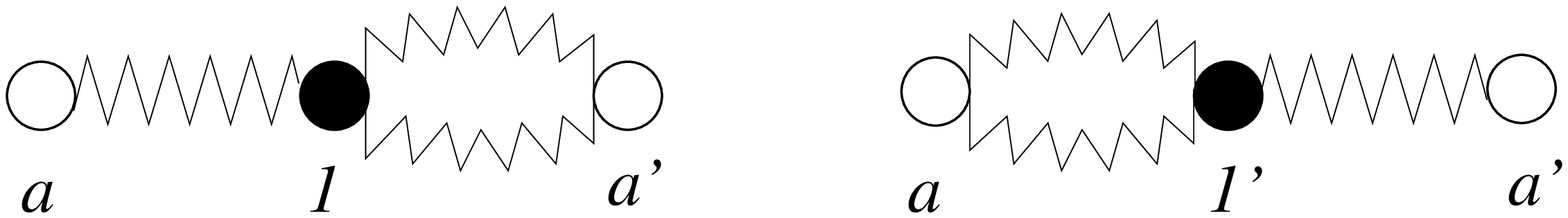}
\caption{\label{diagbbe}}
\end{figure*}

\begin{figure*}[H]
\includegraphics[width=3cm]{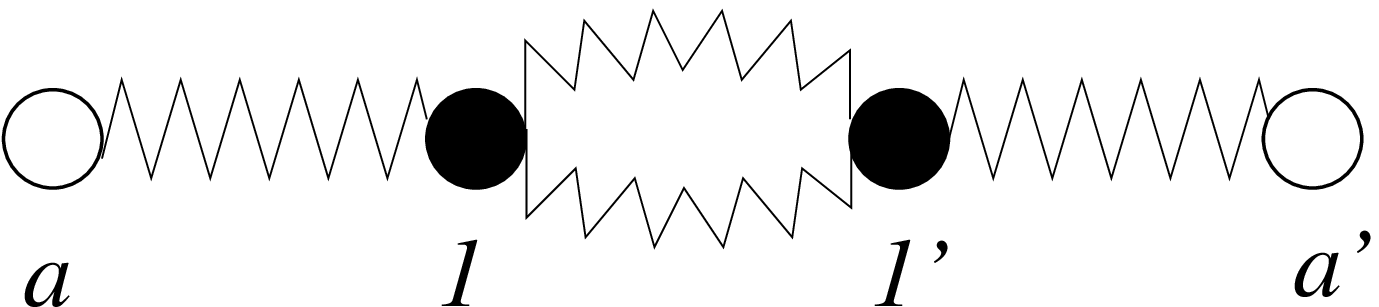}
\caption{\label{diagc}}
\end{figure*}

\end{document}